\documentclass[useAMS,usenatbib,usegraphicx]{mn2e}

\usepackage[usenames,dvipsnames]{color}
\usepackage{amsmath}
\usepackage{amssymb}
\usepackage{amsfonts}
\usepackage[xindy]{glossaries}
\usepackage{graphicx}
\usepackage{multirow}
\usepackage{subfigure}
\usepackage{aas_macros}
\makeglossaries

\newacronym{aa}{AA}{aperture array}
\newacronym{db}{dB}{decibel}
\newacronym{fov}{FoV}{field of view}
\newacronym{ixr}{IXR}{polarimeter intrinsic cross-polarization ratio}
\newacronym{mjd}{MJD}{modified Julian Date}
\newacronym{msp}{MSP}{millisecond pulsar}
\newacronym{paf}{PAF}{phased-array feed}
\newacronym{rime}{RIME}{radio interferometer measurement equation}
\newacronym{rms}{rms}{root mean square}
\newacronym{ska}{SKA}{Square Kilometre Array}
\newacronym{snr}{S/N}{signal-to-noise ratio}
\newacronym{svd}{SVD}{singular value decomposition}
\newacronym{toa}{ToA}{time of arrival}
\newacronym{xpr}{XPR}{cross-polarization ratio}

\title[Intrinsic Instrumental Polarization and Pulsar Timing]{Intrinsic Instrumental Polarization and High-Precision Pulsar Timing}
\author[G. Foster, et al. ]{G. Foster$^{1,2}$, A. Karastergiou$^{2,1,3}$, R. Paulin$^{1}$, T.D. Carozzi$^{4}$, S. Johnston$^{5}$, and \newauthor W. van Straten$^{6}$ \\
  $^{1}$Department of Physics and Electronics, Rhodes University, P.O. Box 94, Grahamstown 6140, South Africa\\
  $^{2}$University of Oxford, Sub-Department of Astrophysics, Denys Wilkinson Building, Keble Road, Oxford, OX1 3RH, United Kingdom\\
  $^{3}$Physics Department, University of the Western Cape, Cape Town 7535, South Africa\\
  $^{4}$Onsala Space Observatory, Department of Earth \& Space Physics, Chalmers University, 43992 Onsala Sweden\\
  $^{5}$CSIRO Astronomy and Space Science, PO Box 76, Epping, NSW 1710, Australia\\
  $^{6}$Centre for Astrophysics and Supercomputing, Swinburne University of Technology, Hawthorn, VIC 3122, Australia}

\begin{document}

\date{\today}

\pagerange{\pageref{firstpage}--\pageref{lastpage}} \pubyear{2015}

\maketitle

\begin{abstract}
  Radio telescopes are used to accurately measure the \gls{toa} of radio pulses in
  pulsar timing experiments that target mostly \glspl{msp} due to their high rotational
  stability. This allows for detailed study of \glspl{msp} and forms the basis of
  experiments to detect gravitational waves. Apart from intrinsic and propagation effects,
  such as pulse-to-pulse jitter and dispersion variations in the interstellar medium,
  timing precision is limited in part by the following: polarization purity
  of the telescope's orthogonally polarized receptors, the \gls{snr} of the pulsar
  profile, and the polarization fidelity of the system. Using simulations, we present
  how fundamental limitations in recovering the true polarization reduce the precision
  of \gls{toa} measurements. Any real system will respond differently to each source
  observed depending on the unique pulsar polarization profile. Using the profiles
  of known \glspl{msp} we quantify the limits of observing system specifications that
  yield satisfactory \gls{toa} measurements, and we place a practical design limit
  beyond which improvement of the system results in diminishing returns. Our aim is
  to justify limits for the front-end polarization characteristics of next generation
  radio telescopes, leading to the \gls{ska}.
\end{abstract}


\glsresetall
\label{firstpage}

\section{Introduction}

Any dual-polarization polarimeter is characterized by a degree of
polarization purity, i.e. the cross-polarization between orthogonal
feeds, and the extent to which calibration can be used to retrieve
accurate polarization information.  In this paper we examine how both of
these limitations affect pulsar \gls{toa} measurements, especially in
the case of \glspl{msp}. We do this by simulating \gls{toa}
measurements through a sampling of the \gls{snr}, calibration error,
and intrinsic polarization leakage parameter space. Intrinsic
polarization leakage includes the apparent leakage between orthogonal
receptors due to differential receptor gains, in addition to the
cross-coupling between receptors, which is typically thought of
as `polarization leakage'.

Simulations are used, as it is difficult to analytically quantify the effects of
calibration error, integration time, and intrinsic polarization leakage in a general
form. The \gls{toa} measurement error depends on the ability to observe pulsar
profiles with high fidelity. By profile, we mean the stable average shape of the
radio pulse of a given pulsar, and its polarization properties. We perform our
analysis using profiles from the 20 \glspl{msp} in \cite{2013PASA...30...17M}.

A fundamental limit to any \gls{toa} measurement is the design of the
polarimeter feeds, which is set by the telescope specifications.  As
pulsar timing is a key science project for \gls{ska}, see \cite
{2015arXiv150100127J} and \cite{2004NewAR..48.1413C}, it is important
to consider the science limitations set during the design process.
The decimetre wavelength band, where pulsars are typically observed
for timing, will be covered by both dishes and aperture arrays.  The
analysis presented here applies to both telescope types.

For the design of the feeds, we need to consider the capacity of any
dish or aperture array to produce data from which the true
polarization of a signal can be recovered.  For example, the full
polarization description of an incoming signal can never be recovered
by a single dipole, no matter how good the calibration procedure
is. On more realistic systems, there is a fundamental limitation in
recovering the polarization state due to differential gains between
orthogonal receptors. These differential gains coupled with noise in
the system result in measurement errors which can not be corrected via
any currently used calibration procedure. This affects all \gls{toa} measurement
methods. For total intensity (e.g. \cite{1992RSPTA.341..117T}) timing
methods, i.e. the determination of a \gls{toa} through
cross-correlation of a total power template, this is an effect in
addition to calibration error. Techniques such as the invariant
interval \citep{2000ApJ...532.1240B} and matrix template matching
\citep{2006ApJ...642.1004V} methods, despite being largely independent
of polarization calibration error, are also affected.

The remainder of this introduction covers the relevant Jones and
Mueller mathematical formalisms necessary to describe intrinsic
polarization leakage. In Section \ref{section:sims} we describe the
simulation setup and the strategy for exploring the relevant parameter
space, how the simulated observations are generated, and the methods
for determining the \gls{toa}.  Results are presented in Section
\ref{section:results}. Discussion of the results and the implications
for current and future telescopes are presented in Section
\ref{section:discussion}.

\subsection{Intrinsic polarization leakage in Jones and Mueller formalism}

We are interested in describing the intrinsic polarization leakage of a linear
feed, dual-polarization system as this is a typical design for single pixel feeds
and phased arrays such as \glspl{paf} and \glspl{aa} used for pulsar timing.
Intrinsic polarization leakage can be described with the mathematical structure
developed for the \gls{rime} presented in \citet{1996A&AS..117..137H} and \citet{2011A&A...527A.106S}.
Additionally, the \gls{rime} can be extended to phased arrays where, to first order,
the formed beam is a linear combination of the individual element beams; a full
description would also include element mutual coupling terms. Jones matrix formalism
is useful to frame the \gls{rime} in terms of instrumentation and environmental
effects on an electromagnetic signal.  The Mueller matrix formalism, which is used
in our simulations, is useful in interpreting the \gls{rime} in terms of detected
power, represented by the Stokes parameters of the signal.

In Jones formalism, transformations are applied to an input electromagnetic signal
to produce the observed signal.  The transformation from the complex electromagnetic
sky Jones vector $\mathbfit{e}$ to the observed voltage Jones vector $\mathbfit{v}$
is
\begin{equation}
\label{jones_rime}
\mathbfit{v} = \mathbfss{J}_{\text{sys}} \mathbfit{e}
\end{equation}
\noindent where $\mathbfss{J}_{\text{sys}}$ is the total system Jones
matrix representation which is constructed out of multiple linear
Jones transformations, each of which can have dependence on time,
observing frequency, and source direction \citep{2011A&A...527A.107S}.
In the scope of this paper we are interested in the effect of
intrinsic polarization leakage. The polarization leakage matrix
$\mathbfss{D}$ is usually defined as a direction-independent Jones
matrix with the direction-dependent polarization leakage components
incorporated into the primary beam matrix $\mathbfss{E}$. For this
work we are not focusing on the primary beam direction-dependent
sensitivity, but only the potentially direction-dependent polarization
leakage, thus for phased arrays we are defining $\mathbfss{D}$ to also
include the direction-dependent polarization leakage
\begin{equation}
\label{eq:d_matrix}
\mathbfss{D}= 
	\begin{pmatrix}
	1	                    &	d_{p \rightarrow q}(\nu,\theta,\phi) \\
	-d_{q \rightarrow p}(\nu,\theta,\phi)	&	1
	\end{pmatrix}
\end{equation}
\noindent where $(\theta,\phi)$ are reference frame dependent position
angles. In Equation \ref{eq:d_matrix}, $d_{p \rightarrow q}$ is the
intrinsic leakage of feed $p$ into feed $q$. In feed design, the
off-diagonal terms are minimized. Ideally they are zero. For phased
arrays, however, projection effects will cause $\mathbfss{D}$ to vary
with observing direction.

We can then define an explicit \gls{rime} for our dish and phased array systems as
\begin{subequations}
\label{eq:rimes}

\begin{equation}
\label{explicit_dish_rime}
\mathbfss{J}_{\text{sys,dish}} = \mathbfss{B}(\nu) \mathbfss{G}(t) \mathbfss{C} \mathbfss{D}(\nu)
\end{equation}

\begin{equation}
\label{explicit_pa_rime}
\mathbfss{J}_{\text{sys,PA}} = \sum^{n}_{i=1} \mathbfss{W}_i(\nu,\theta,\phi) \mathbfss{B}_i(\nu) \mathbfss{G}_i(t) \mathbfss{C}_i \mathbfss{D}_i(\nu,\theta,\phi)
\end{equation}

\end{subequations}
\noindent where $\mathbfss{B}$ is the frequency-dependent bandpass
structure, and $\mathbfss{G}$ is the time-dependent electronic
gain. $\mathbfss{B}$ and $\mathbfss{G}$ are diagonal matrices. The
idealized nominal feed configuration $\mathbfss{C}$ is a coordinate
transform from the sky to observing frame. For a single pixel feed, on
axis observation, $\mathbfss{D}$ has no direction dependence and can
be simplified to $\mathbfss{D}(\nu)$.

For a phased array, the $\mathbfss{J}_{\text{sys}}$ is the weighted
sum over all $n$ elements of the array, where $\mathbfss{W}_i$ are the
complex weights to shape the array beam pattern to `point' in a
direction. For a \gls{paf}, the direction dependence of $\mathbfss{D}$
will be relative to the dish pointing centre. For an aperture array,
the direction dependence of $\mathbfss{D}$ will be relative to the
direction of boresight, usually zenith.

Jones formalism is useful for understanding the instrumental effects
on a signal.  For our simulation, however, we use pulsar profiles
described as Stokes vectors, for which Mueller matrices are used to
perform operations. Using the notation from
\cite{1996A&AS..117..137H}, any Jones matrix $\mathbfss{J}$ can be
transformed to a corresponding Mueller matrix $\mathbfss{M}$ by use of
the Kronecker product,
\begin{equation}
\label{eq:jones2mueller}
\mathbfss{M} = \mathbfss{S}^{-1} \left ( \mathbfss{J} \otimes \mathbfss{J}^{*} \right ) \mathbfss{S}
\end{equation}
where $\mathbfss{S}$ and $\mathbfss{S}^{-1}$ are the conversion
matrices to transform between Stokes parameters and the brightness
coherency vector. For reference they are presented below:

\begin{equation}
\label{eq:s_transform}
\mathbfss{S} = \frac{1}{2}
    \begin{pmatrix}
    1 & 1 & 0 & 0 \\
    0 & 0 & 1 & i \\
    0 & 0 & 1 & -i \\
    1 & -1 & 0 & 0 \\
    \end{pmatrix}
\end{equation}

\begin{equation}
\label{eq:s_inv_transform}
\mathbfss{S}^{-1} =
    \begin{pmatrix}
    1 & 0 & 0 & 1 \\
    1 & 0 & 0 & -1 \\
    0 & 1 & 1 & 0 \\
    0 & -i & i & 0 \\
    \end{pmatrix}
\end{equation}

Using equation \ref{eq:jones2mueller} the feed-error matrix
$\mathbfss{D}$ can be converted to a Mueller form
$\mathbfss{D}_{\text{M}}$ (Eq.
\ref{eq:d_matrix_mueller}). 
Each element of which can be directly computed using $\mathbfss{M}_{i,j} = \frac{1}{2} \; \textrm{tr} (\pmb{\sigma}_i \, \mathbfss{J} \, \pmb{\sigma}_j \, \mathbfss{J}^{\dag})$, where $\pmb{\sigma}_i$ is the $i^{\textrm{th}}$ Pauli matrix.
In this form, the $(i,j)^{\textrm{th}}$ element can be understood as the response of the $i^{\textrm{th}}$ output to the $j^{\textrm{th}}$ input.

\begin{figure*}
\begin{equation}
\label{eq:d_matrix_mueller}
\mathbfss{D}_{\text{M}} =
    \begin{pmatrix}
    1+ \frac{1}{2}(|d_{p \rightarrow q}|^2 + |d_{q \rightarrow p}|^2)   & \frac{1}{2}(|d_{p \rightarrow q}|^2 + |d_{q \rightarrow p}|^2)    & \text{Re}[d_{p \rightarrow q} -d_{q \rightarrow p}]       & \text{Im}[d_{p \rightarrow q} +d_{q \rightarrow p}] \\
    \frac{1}{2}(|d_{p \rightarrow q}|^2 - |d_{q \rightarrow p}|^2)      & 1- \frac{1}{2}(|d_{p \rightarrow q}|^2 - |d_{q \rightarrow p}|^2) & \text{Re}[d_{p \rightarrow q} +d_{q \rightarrow p}]       & -\text{Im}[d_{p \rightarrow q} -d_{q \rightarrow p}] \\
    \text{Re}[d_{p \rightarrow q} -d_{q \rightarrow p}]                 & -\text{Re}[d_{p \rightarrow q} +d_{q \rightarrow p}]              & 1 -\text{Re}[d_{p \rightarrow q} d_{q \rightarrow p}^*]   & \text{Im}[d_{p \rightarrow q} d_{q \rightarrow p}^*] \\
    \text{Im}[d_{p \rightarrow q} +d_{q \rightarrow p}]                 & -\text{Im}[d_{p \rightarrow q} -d_{q \rightarrow p}]              & \text{Im}[d_{p \rightarrow q} d_{q \rightarrow p}^*]      & 1 +\text{Re}[d_{p \rightarrow q} d_{q \rightarrow p}^*] \\
    \end{pmatrix}
\end{equation}
\end{figure*}

\subsection{Polarimeter performance metric}
\glsreset{ixr}

The polarization leakage of a dual-feed receiver is
quantified by using \gls{xpr} metrics \citep{705931}.  In \cite{2011ITAP...59.2058C},
a new \gls{xpr}, the \gls{ixr}, was introduced.  \gls{xpr}s are used
as metrics for radio communication feeds where the polarization of
both the source and receiver is known.  Thus an \gls{xpr} can vary by
choice of coordinate system \citep{2011ITAP...59.2058C}.  The \gls{ixr} is an \gls{xpr} which is
invariant under coordinate transform.  This makes the \gls{ixr} ideal
for a radio astronomical polarimeter as there is no preferred sky
coordinate frame. The \gls{ixr} in Jones formalism is defined as
\begin{equation}
\label{eq:ixr_def}
\textrm{IXR}_{\text{J}} \equiv \left ( \frac{g_{\text{max}}+g_{\text{min}}}{g_{\text{max}}-g_{\text{min}}} \right )^2
\end{equation}
where $g_{\text{max}}$ and $g_{\text{min}}$ are the maximum and
minimum amplitude gains of the polarimeter when using \gls{svd} to
decompose the system Jones matrix $\mathbfss{J}_{\text{sys}}$. The
\gls{svd} theorem (Eq. \ref{eq:svd}) states that any Jones matrix
$\mathbfss{J}$ can be decomposed into two unitary transforms
$\mathbfss{U}, \mathbfss{V}^{\dagger}$ and one diagonal transform
matrix $\mathbf{\Sigma}$.

\begin{equation}
\label{eq:svd}
\mathbfss{J} = \mathbfss{U} \mathbf{\Sigma} \mathbfss{V}^{\dagger} = \mathbfss{U}
\begin{pmatrix}
    \sigma_{\text{max}} & 0\\
    0 & \sigma_{\text{min}}\\
\end{pmatrix}
\mathbfss{V}^{\dagger}
\end{equation}

Given a noise input signal $\mathbfit{e}$, the sensitivity to change
in $\mathbfit{v}$ from Equation \ref{jones_rime} is measured by the
\emph{matrix condition number} $\textrm{cond}_2(\mathbfss{J}) \equiv
\kappa(\mathbfss{J})=\frac{\sigma_{\text{max}}}{\sigma_{\text{min}}}$,
where $\sigma_{\text{max}}$ and $\sigma_{\text{min}}$ are the maximum
and minimum singular values. An \emph{ill-conditioned} matrix, one
with a condition number much larger than 1, will cause an increase in
the error of $\mathbfit{v}$ with respect to the error of
$\mathbfit{e}$. Conversely, a \emph{well-conditioned} matrix, one with
a condition number close to 1, will transform $\mathbfit{e}$ into
$\mathbfit{v}$ with minimal effect on the error.

\cite{2011ITAP...59.2058C} show that by setting the maximum and
minimum amplitude gains to be equal to the maximum and minimum
singular values ($\sigma_{\text{max}}=g_{\text{max}}$ and
$\sigma_{\text{min}}=g_{\text{min}}$), there is always an orthonormal
choice of coordinates systems for the sky and the channels that gives
$\mathbfss{J'}$ from $\mathbfss{J}$ such that the feed error matrix
takes the form
\begin{equation}
\label{eq:ixr_matrix}
\mathbfss{J'} = \frac{g_{\text{max}}+g_{\text{min}}}{2}
\begin{pmatrix}
    1 & 1/\sqrt{\textrm{IXR}_{\text{J}}}\\
    1/\sqrt{\textrm{IXR}_{\text{J}}} & 1\\
\end{pmatrix}
\end{equation}
The $\textrm{IXR}_{\text{J}}$ is in units of power and Equation
\ref{eq:ixr_matrix} has components of $\sqrt{\textrm{IXR}_{\text{J}}}$
as a Jones matrix acts as an operation on an electric field
vector. Intrinsic polarization leakage can be seen as differential
gains or `canonical' polarization leakage depending on the basis.  The
condition number can thus be related back to the intrinsic
polarization leakage.  That is $d_{p \rightarrow q} = -d_{q
  \rightarrow p} =
\frac{\kappa(\mathbfss{J})+1}{\kappa(\mathbfss{J})-1}$ with a
normalization factor of
$\frac{\kappa(\mathbfss{J})+1}{2\kappa(\mathbfss{J})}$.  Redefining
the $\textrm{IXR}_{\text{J}}$ in terms of the condition number,
Equation \ref{eq:ixr_def} becomes
\begin{equation}
\label{eq:ixr_def_kappa}
\textrm{IXR}_{\text{J}} = \left ( \frac{\kappa(\mathbfss{J}) + 1}{\kappa(\mathbfss{J}) -1} \right )^2
\end{equation}
This is a crucial quantity which represents a fundamental limit in our
ability to recover the true signal. This limit is independent of the
polarization calibration.

The \gls{ixr} is conceptually equivalent to the \emph{polconversion}
\citep{2000A&AS..143..515H} and Lorentz boost
\citep{2000ApJ...532.1240B} transformations that have been employed in
previous works based on quaternions and geometric algebra,
respectively.  For example, where $\beta$ is the velocity parameter
that describes a Hermitian Jones matrix in Equation 11 of
\cite{2000ApJ...532.1240B}, $\kappa(\mathbfss{J}) = e^{2\beta}$ and
$\textrm{IXR}_{\text{J}} = \coth^2(\beta$). These equations enable
meaningful comparisons between the results presented in this work and
the notation employed in some previous studies (e.g. \cite{2013ApJS..204...13V}).

To understand how the condition number, and by extension the
\gls{ixr}, affects an observation, we can look at how the true sky
vector $\mathbfit{e}$ is determined.  To obtain an estimate of the
true Jones sky vector $\hat{\mathbfit{e}}$ from the observed Jones
vector $\mathbfit{v}$, the system Jones matrix
$\mathbfss{J}_{\text{sys}}$ must be determined and inverted
(Eq. \ref{invert_jones_rime}) via calibration
\begin{equation}
\label{invert_jones_rime}
\hat{\mathbfit{e}} = \mathbfss{J}_{\text{sys}}^{-1} \mathbfit{v}
\end{equation}
where \noindent $\hat{\mathbfit{e}}$ is an estimate of $\mathbfit{e}$
due to multiple compounding effects. First, in the measurement of
$\mathbfit{v}$ there is a limit in precision due to noise. Second,
$\mathbfss{J}_{\text{sys}}$ is not perfectly known, but is an estimate
based on modeling and calibration. Finally, the condition of the
components of $\mathbfss{J}_{\text{sys}}$ determine how the errors in
measurement affect the estimation of $\hat{\mathbfit{e}}$. For an
ill-conditioned $\mathbfss{J}_{\text{sys}}$, a small error in
$\mathbfit{v}$ will result in a large error in estimated sky vector
$\hat{\mathbfit{e}}$ compared to the true sky vector
$\mathbfit{e}$. An ill-conditioned $\mathbfss{J}_{\text{sys}}$ matrix
will lead to a noisy estimate of $\mathbfit{e}$, no matter how well
known $\mathbfss{J}_{\text{sys}}$ is, due to the inherent noise in the
measurement of $\mathbfit{v}$. As the conditioning of the
$\mathbfss{J}_{\text{sys}}$ improves, so too does $\hat{\mathbfit{e}}$
more accurately describe $\mathbfit{e}$.

By definition $\kappa(\mathbfss{J}) \geq 1$. Ideally there is no
intrinsic polarization leakage between feeds, i.e. the matrix is
perfectly conditioned $g_{\text{max}}=g_{\text{min}} \Rightarrow
\kappa(\mathbfss{J})=1$ and the $\textrm{IXR}_{\text{J}} \rightarrow
\infty$.  That is, the two receptors are completely orthonormal.  If
there is intrinsic leakage between the two feeds then $g_{\text{max}}
> g_{\text{min}} \Rightarrow \kappa(\mathbfss{J})$ increases and the
$\textrm{IXR}_{\text{J}}$ decreases.  In the worst case (e.g. where the two
feeds are perfectly coupled, or one receptor's sensitivity goes to 0),
then $g_{\text{min}} \rightarrow 0 \Rightarrow \kappa(\mathbfss{J})
\rightarrow \infty$ and the $\textrm{IXR}_{\text{J}} \rightarrow 1$.
Since the $\textrm{IXR}_{\text{J}}$ is a measure of feed response, it
is common to use \gls{db} units, $\textrm{IXR}_{\text{J,dB}}= 10
\log_{10} (\textrm{IXR}_{\text{J}})$.

We can also consider the \gls{ixr} in terms of Mueller
matrices. \cite{2011ITAP...59.2058C} connects the \gls{ixr} to Mueller
matrices by showing $\kappa(\mathbfss{M}) = \kappa^2(\mathbfss{J})$.
This relation is used to show the \gls{ixr} in Mueller formalism is
\begin{equation}
\label{eq:ixr_m_ixr_rel}
\textrm{IXR}_{\text{M}} = \frac{\kappa(\mathbfss{M})+1}{\kappa(\mathbfss{M})-1} = \frac{1 + \textrm{IXR}_{\text{J}}}{2 \sqrt{\textrm{IXR}_{\text{J}}}}
\end{equation}
which provides a useful metric for measuring the impact of instrumental polarization on the Stokes parameters, especially in the case of impure transformations with no corresponding Jones matrix.

\begin{figure}
    \includegraphics[width=1.0\linewidth]{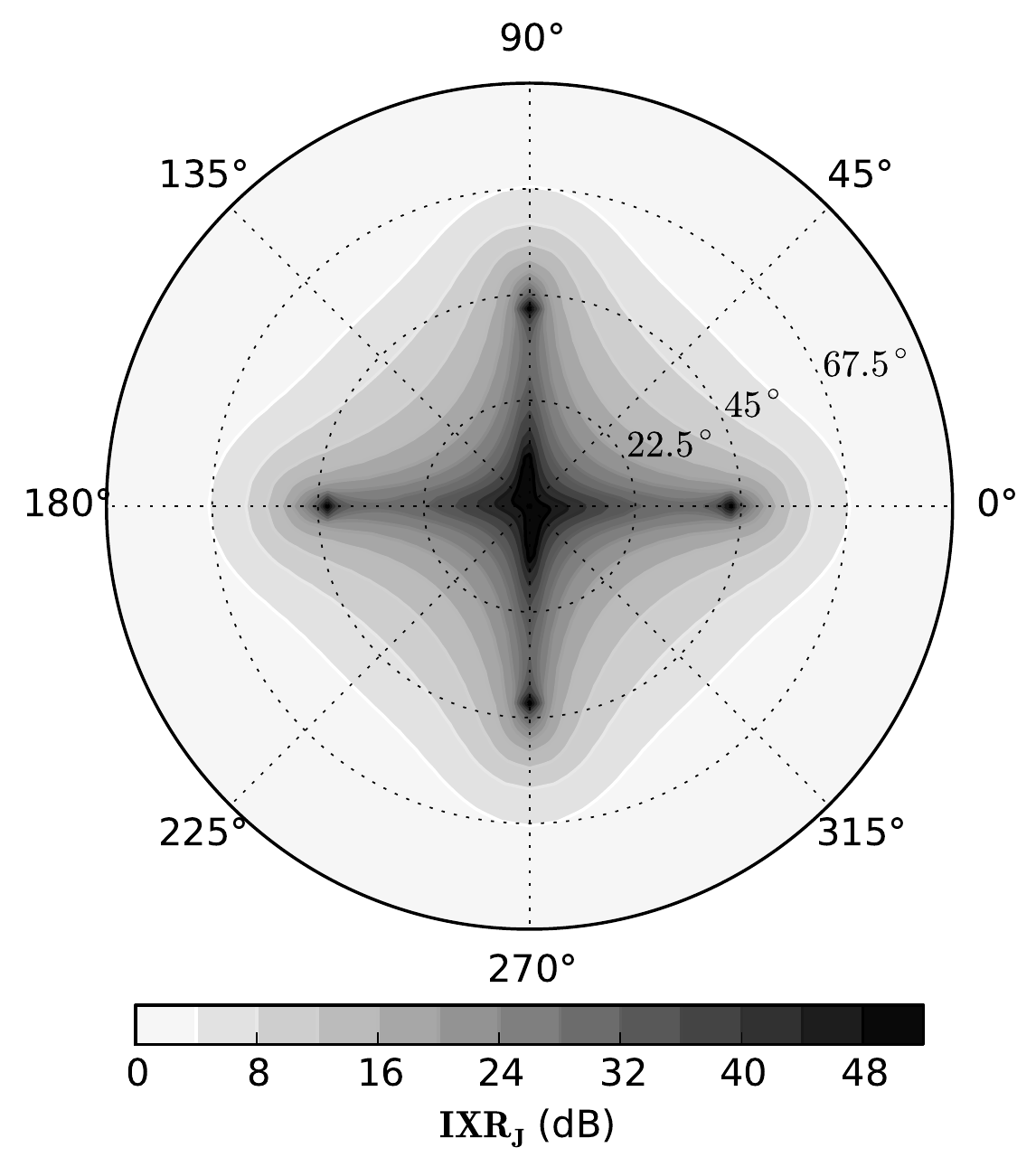}
    \caption{$\textrm{IXR}_{\text{J}}$ of a simple `all-sky' element based on the
    Precision Array for Probing the Epoch of Reionization (PAPER) dipole beam model
    at 130 MHz. The orthogonal dipoles are oriented along the $(45^{\circ},225^{\circ})$
    and $(135^{\circ},315^{\circ})$ axes.
    }
    \label{fig:paper_ixr}
\end{figure}

An example of the variation in $\textrm{IXR}_{\text{J}}$ across the
field of view of a simple dipole element is shown in Figure
\ref{fig:paper_ixr}. The $\textrm{IXR}_{\text{J}}$ is maximized ($70$
dB) in the direction of zenith, but rises when observing away from
boresight. The low $\textrm{IXR}_{\text{J}}$ structure
($(0^{\circ},180^{\circ})$ and $(90^{\circ},270^{\circ})$ axes) is
along the $45^{\circ}$ line between the two orthogonal receptors. The
variation in $\textrm{IXR}_{\text{J}}$ across the field of view also
depends on observing frequency.

\cite{2009wska.confE..17C} and \cite{2013ITAP...61.2852S} show the
$\textrm{IXR}_{\text{J,dB}}$ can vary between $0$ dB and $66$ dB
across an aperture array depending on pointing direction and element
design.  For an idealized short dipole the $\textrm{IXR}_{\text{J}}$
varies smoothly over the observable hemisphere.  But for many feeds
--- such as Vivaldi-type, bow-tie, and narrow-band half-wavelength
dipoles --- sharp intrinsic polarization leakage structures form
across the hemisphere.  Figures $3$ and $5$ in
\cite{2009wska.confE..17C} show the $\textrm{IXR}_{\text{J}}$ over a
hemisphere for short dipoles and Vivaldi-type elements.  Figures $2$
and $3$ in \cite{2013ITAP...61.2852S} show the
$\textrm{IXR}_{\text{J}}$ across the \gls{fov} of an MWA bow-tie
element.  These published values and maps provide insight into what
range of $\textrm{IXR}_{\text{J}}$ to use in our simulations.

\subsection{IXR and signal-to-noise ratio}
\label{sec:ixr_to_snr}

\begin{figure}
    \includegraphics[width=1.0\linewidth]{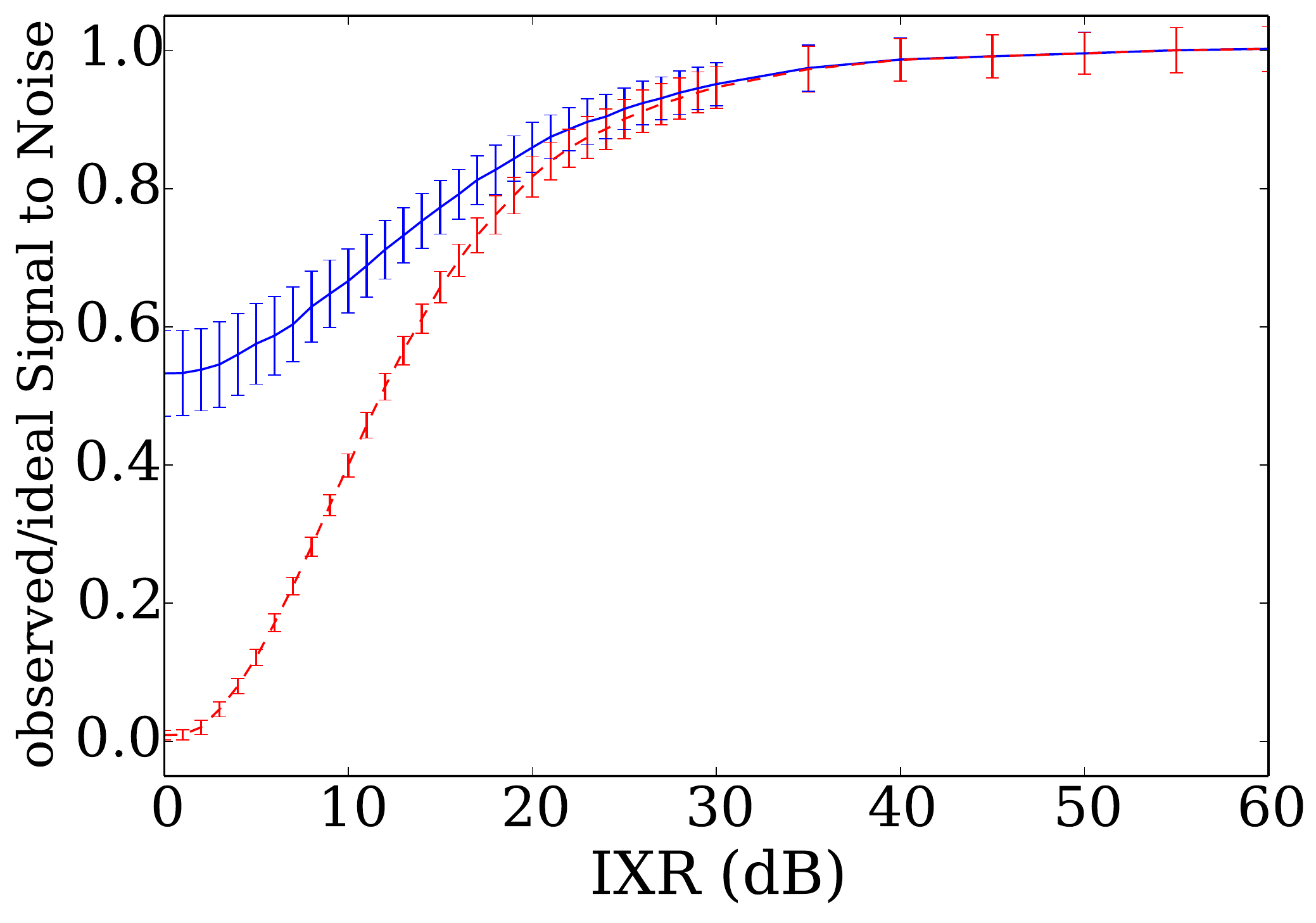}
    \caption{The average observed S/N of J1603$-$7202 as a fraction of the ideal
    S/N for a range of IXR values using `gain' (blue/solid) and `full' (red/dashed) calibration
    methods described in Section \ref{sec:sim_obs}.
    }
    \label{fig:actual_snr_vs_ixr}
\end{figure}

The error on a \gls{toa} measurement is, in general, a function of the
\gls{snr} of a given observation. By \gls{snr}, we hereby refer to the
peak pulse value in the Stokes \emph{I} profile to the standard deviation
of the off-pulse signal. Under ideal circumstances, the \gls{snr} increases
with the square root of integration time. For a polarized
source, instrumental intrinsic polarization leakage will result in a
lower observed \gls{snr} compared to the ideal \gls{snr}, i.e. that
obtained from a system with no intrinsic polarization leakage, for a
given amount of integration. The blue/solid line in Figure \ref{fig:actual_snr_vs_ixr} shows
the fractional observed \gls{snr} compared to the ideal \gls{snr} as a
function of the \gls{ixr} for J1603$-$7202. The red/dashed line shows how using
the inverse of a poorly conditioned matrix for calibration amplifies the
noise in the measured profile. This will be discussed further in Section
\ref{section:sims}. All the simulated pulsars
have a similar response. An effect of a low \gls{ixr} is the
introduction of a differential gain between feeds. As the \gls{ixr}
goes to $0$ the receiver system becomes effectively blind to one
polarization, thus the observed \gls{snr} is approximately half (blue/solid line) that
of the ideal \gls{snr} in the limit $\textrm{IXR} \rightarrow 0$.
When calibration is applied, not only is there a differential gain effect,
but the inversion of the ill-conditioned matrix will significantly degrade
the \gls{snr} of any profile (red/dashed line).
The ideal \gls{snr} is achieved only in the limit as $\textrm{IXR} \rightarrow \infty$,
and there is a one-to-one correspondence with integration
time. A reference integration time of $\tau_{\text{int}}=1$ is defined
as the time it would take to build up an ideal \gls{snr} of 1000. All
integration time values quoted in this paper are a fraction of this
reference integration time. The relationship between
$\tau_{\text{int}}$ and ideal \gls{snr} $\textrm{SN}_{\text{I}}$ is
\begin{equation*}
\tau_{\text{int}} = \left ( \frac{\textrm{SN}_{\text{I}}}{1000} \right )^2
\end{equation*}
%

\section{The simulations}
\label{section:sims}

We have performed simulations with the goal to quantify the effect of
intrinsic polarization leakage on pulsar \gls{toa} measurements.  We
have sampled the three-dimensional parameter space that includes the
$\textrm{IXR}_{\text{J}}$, ideal pulse profile \gls{snr}, and
calibration error, which covers current telescope measurements and
future telescope specifications.  For each point in the sampled
parameter space, `observed' profiles are generated for 500 epochs, by
modifying a template profile with the appropriate Mueller matrices.
For every epoch, we stochastically generate the
$\mathbfss{J}_{\text{sys}}$ Jones matrix, and the corresponding
Mueller matrix $\mathbfss{M}_{\text{sys}}$, for a given intrinsic
polarization leakage. The form of this matrix is described in \S
\ref{sec:sim_obs}. The observed profile is then calibrated by
multiplying by the inverse of the system Mueller matrix with additive
random calibration errors.

A \gls{toa} is determined at each of the 500 epochs, using a standard
timing method (section \ref{section:timing}) included in
\texttt{PSRCHIVE} \citep{2004PASA...21..302H,2012AR&T....9..237V}.  In a normal pulsar
timing experiment, a model would then be fit to the \gls{toa}
measurements using \texttt{TEMPO2} \citep{2006MNRAS.369..655H} and, the
goodness of the model would be measured by the \gls{rms} of the timing
residuals. Since we are using a simple model of an isolated, stable
pulsar of constant period, this \gls{rms} represents the ideal
\gls{rms} for a given set of parameters.

In the following we give further details of the steps of the
simulation.

\subsection{Pulse profiles}

We performed our simulations using profiles based on the mean pulse profiles
of 20 well studied \glspl{msp} from the Parkes Pulsar Timing Array
\cite{2013PASA...30...17M}. It is not our intention to reproduce the
results of that work. We use these \glspl{msp} as they cover a wide
range of profile and polarization structures, and perform our
simulations to much higher \glspl{snr} than is possible with current
experiments.

A high \gls{snr} version of each reference profile was generated by
applying a rolling Hann filter with a width of $\sim 1 \%$ of the
profile. This has the effect of removing high-frequency components and
reduces the amplitude (up to a few percent) of narrow profile
structures. This was done in order to avoid very low uncertainties in
our TOA measurements due to the presence of the same high frequency
structures in the templates and the simulated noisy data. Though we
retain the pulsar names, by applying this filter, the new profiles are
approximations to the original profiles and are meant only to be a
representative set of profile structures. The low frequency components
that remain, may cause our uncertainties to be systematically lower
than real timing experiments for these pulsars, however this effect is
consistent across our simulations and therefore does not affect our
final result.

The ideal profiles are then used to generate imperfect profiles at each
epoch using the process described in Section \ref{sec:sim_obs}. These
profiles were also used as the template profiles when performing timing.
The first and third columns of Figures \ref{fig:psr_plots0} --
\ref{fig:psr_plots1} at the end of this paper are the Stokes
parameters of the ideal profiles.

\subsection{Simulated observations}
\label{sec:sim_obs}

\begin{figure}
\centering
\subfigure[]{
    \includegraphics[width=.22\textwidth]{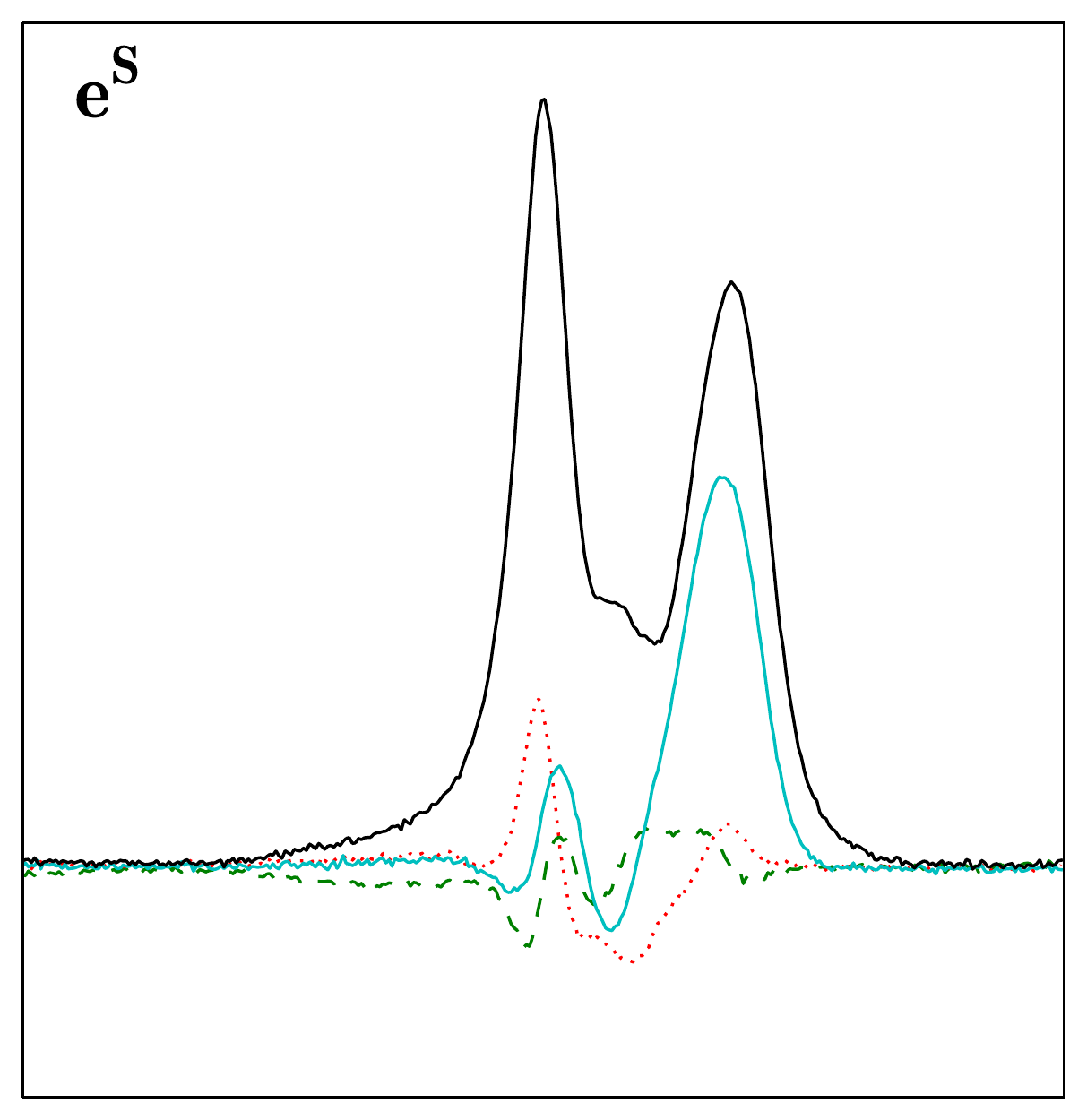}
    \label{fig:simoriginal}
}

\subfigure[]{
    \includegraphics[width=.22\textwidth]{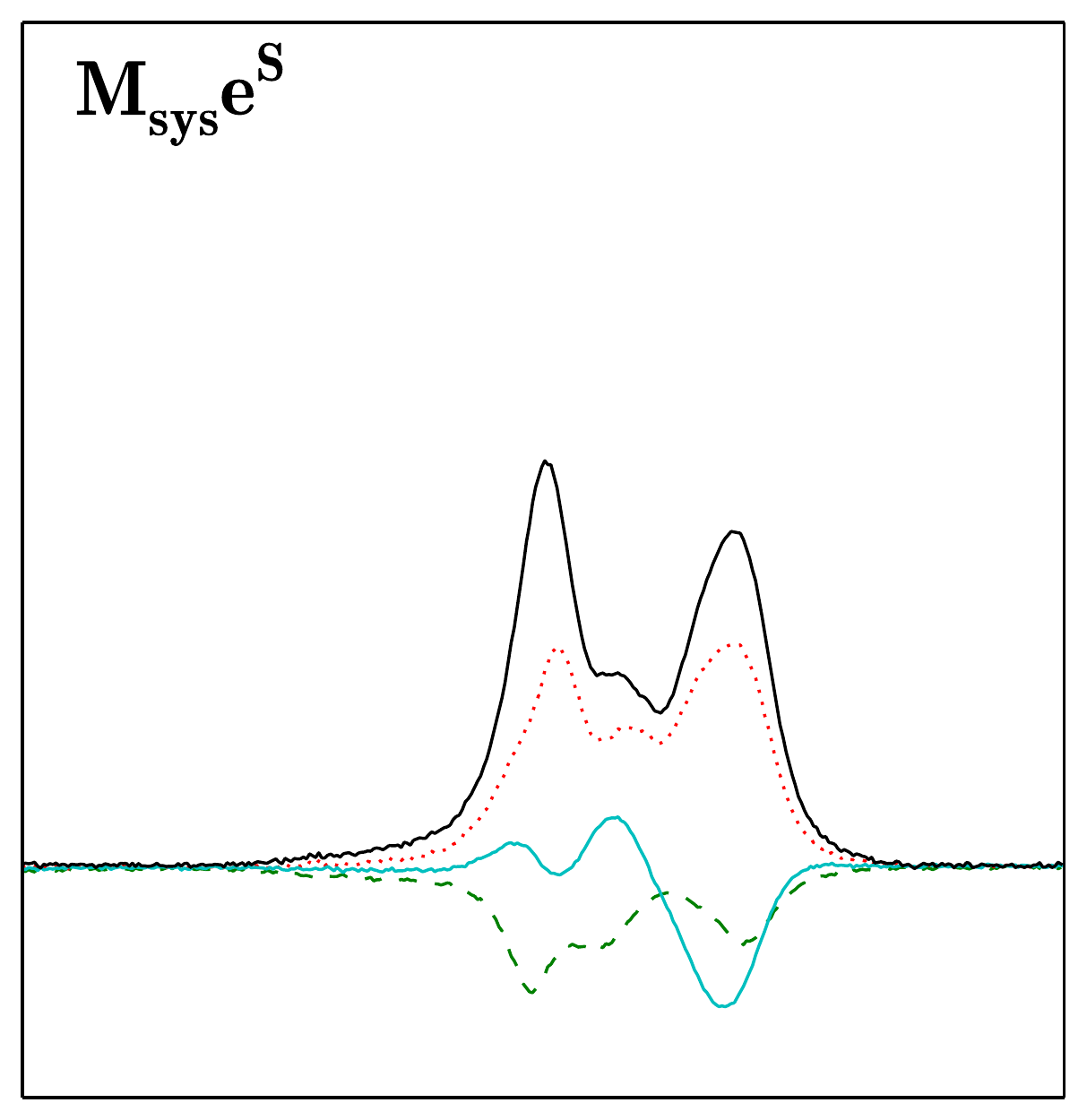}
    \label{fig:sim0}
}
\subfigure[]{
    \includegraphics[width=.22\textwidth]{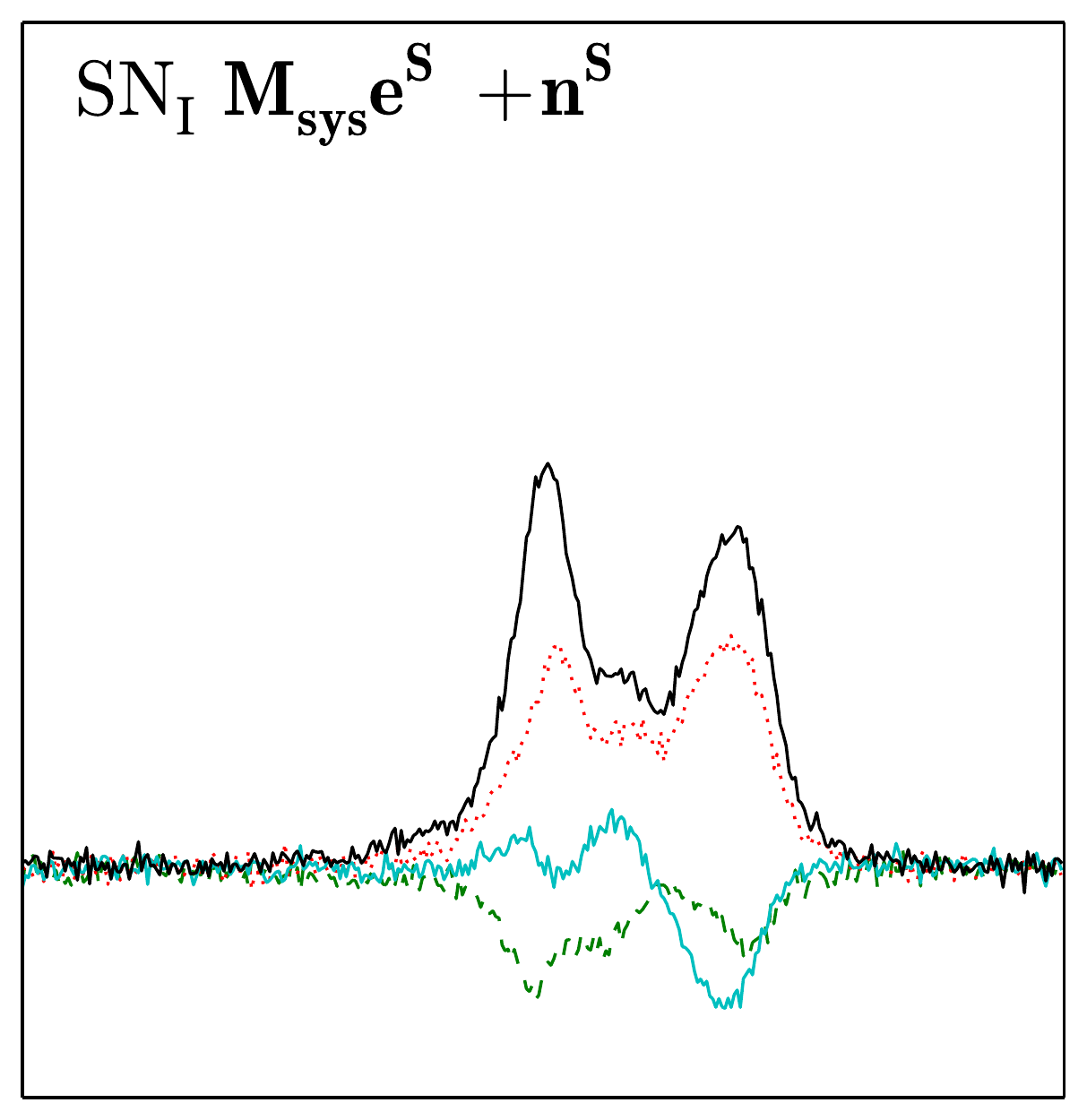}
    \label{fig:sim2}
}
\subfigure[]{
    \includegraphics[width=.22\textwidth]{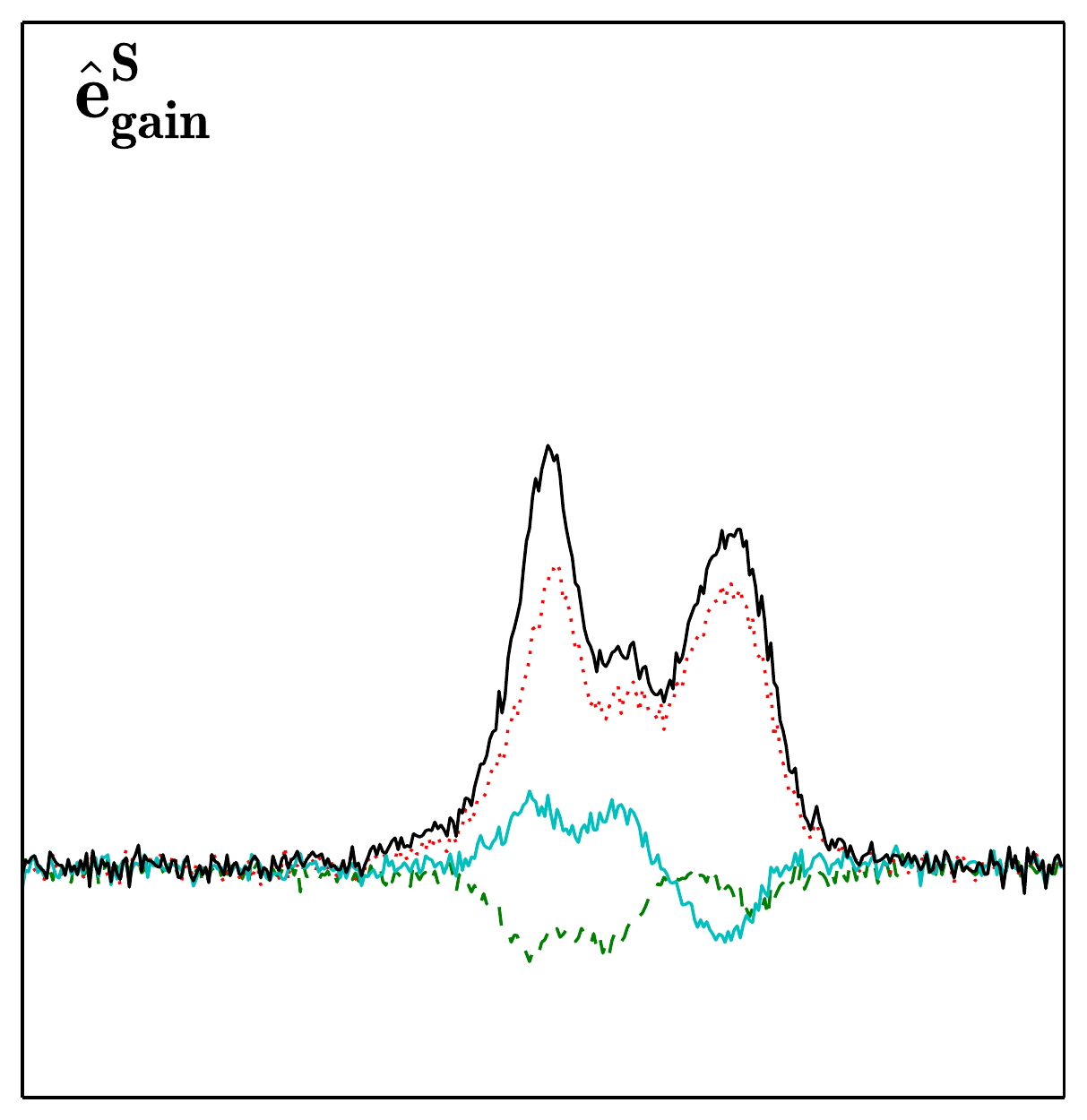}
    \label{fig:sim3}
}
\subfigure[]{
    \includegraphics[width=.22\textwidth]{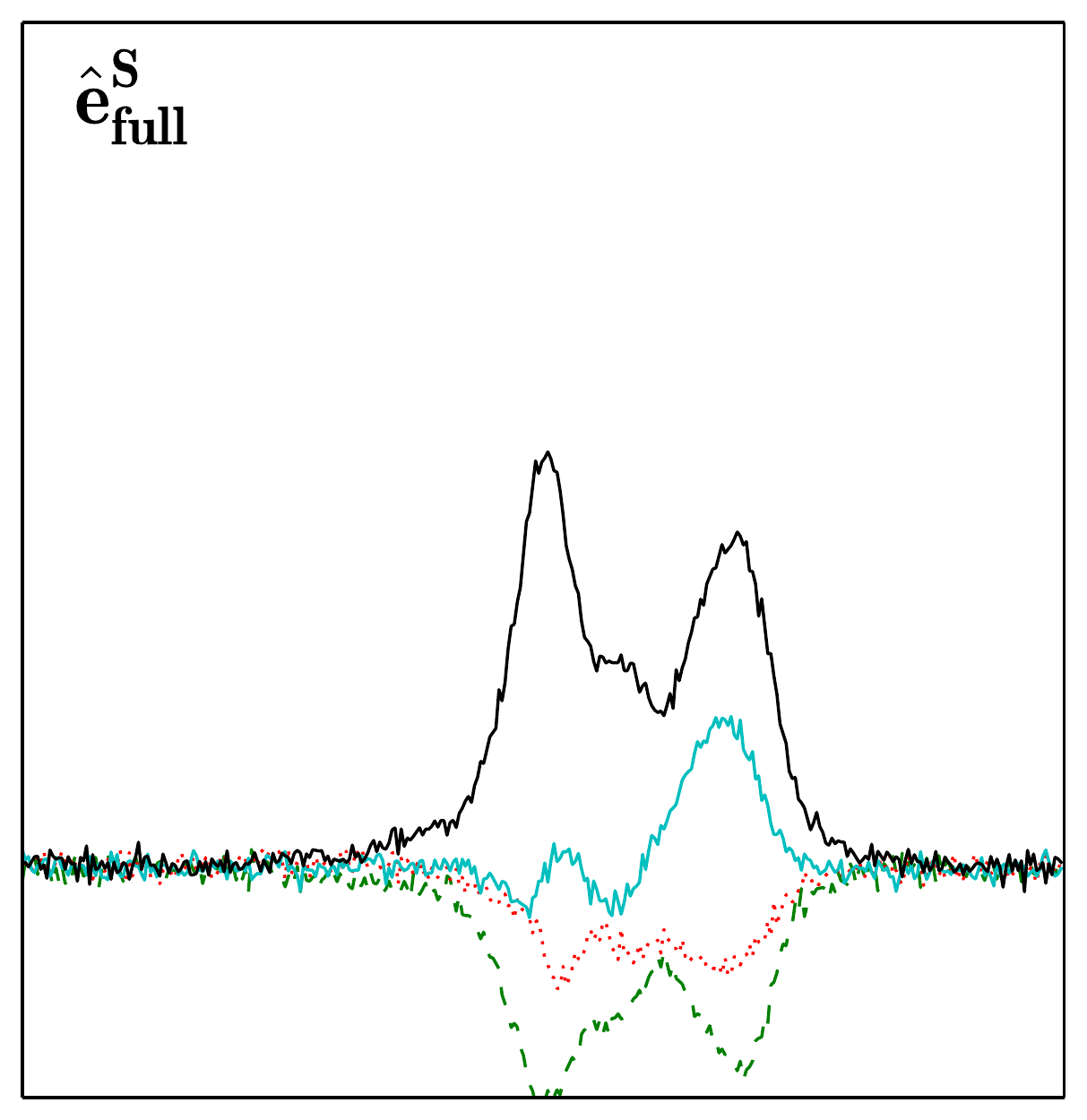}
    \label{fig:sim4}
}

\caption{
    Stages of simulation for J1603$-$7202. Central third of pulse profile plotted.
    Stokes Parameters: I (black/solid), Q (green/dashed), U (red/dotted), and V
    (cyan/solid)
    \subref{fig:simoriginal} True sky Stokes vector $\mathbfit{e}^S$ of the pulsar profile.
    \subref{fig:sim0} True system Mueller matrix $\mathbfss{M}_{\text{sys}}$ with the condition number $\kappa{(\mathbfss{J})}$ set for $\textrm{IXR}_{\text{J,dB}}=7$ dB applied to $\mathbfit{e}^S$.
    \subref{fig:sim2} Added receiver noise for an $\textrm{SN}_{\text{I}}$ of 100.
    \subref{fig:sim3} Estimated Stokes vector $\hat{\mathbfit{e}}^S$ of the pulse profile used in timing experiment using `gain' calibration with $20\%$ calibration error.
    \subref{fig:sim4} Estimated Stokes vector $\hat{\mathbfit{e}}^S$ of the pulse profile used in timing experiment using `full' calibration with $20\%$ calibration error.
    }
\label{fig:sims}
\end{figure}

Our simulated observations produce an estimated sky Stokes vector
$\hat{\mathbfit{e}}^S$ for a given pulsar per epoch by estimating the
system Mueller matrix, including calibration error, intrinsic
polarization leakage, and \gls{snr} parameters.  Figure \ref{fig:sims}
shows the stages to arrive at an example $\hat{\mathbfit{e}}^S$.

The explicit \glspl{rime} defined in Equation \ref{eq:rimes} can be
simplified for the simulations. By using the mean profile we are
making the simplification that the bandpass $\mathbfss{B}$ and time
varying $\mathbfss{G}$ system gains have been solved for and applied
to the observed signal. Thus, $\mathbfss{B}$ and $\mathbfss{G}$ are
unity diagonal matrices. We include a polarization calibration error
term $\Delta \mathbfss{J}$ into our system to simulate the effect of
imperfect calibration of $\mathbfss{B}$ and $\mathbfss{G}$. The
nominal feed matrix $\mathbfss{C}$ is a telescope-specific basis
transform, and will not affect the \gls{ixr} as it is a coordinate
independent metric (i.e. $\textrm{IXR}_{\text{C}} = \infty$). Thus, the polarization leakage matrix
$\mathbfss{D}$ is the only matrix which will vary in our parameter
space. In practice $\mathbfss{D}$ is a function of frequency, but as
the profile is an average across a frequency band, so too is the
\gls{ixr} a frequency averaged intrinsic polarization leakage.

To generate a simulated observed profile we start by creating a
intrinsic polarization leakage Jones matrix representation.
We define the \emph{polarization leakage} to be
$1/\sqrt{\mathrm{IXR}_{\text{J}}}$, in decibel units the intrinsic
polarization leakage is $-\mathrm{IXR}_{\text{J,dB}}$. We have chosen
this definition as polarization leakage is a common concept within the
community. The \gls{ixr} is a measure of both the cross-coupling
between receiver feeds, which is typically thought of as `polarization
leakage', and the apparent leakage due to differential feed gains and
thus is a more complete metric for `polarization leakage' over
previous definitions. A higher intrinsic polarization leakage implies
the two feeds are more coupled together than a lower intrinsic
polarization leakage. To sample a broad range of intrinsic
polarization leakage values we sample the space $ 0 \, \textrm{dB}$ to
$-30 \, \textrm{dB}$, where the upper limit of 0~dB is effectively
blind to one polarization, such as a single polarization receiver. The
intrinsic polarization leakage sample space in \gls{ixr} notation is $
1 \geq 1/\sqrt{\textrm{IXR}_{\text{J}}} \geq 0.031$. By inverting
equation \ref{eq:ixr_def_kappa}, the condition number is related to
the $\textrm{IXR}_{\text{J}}$ is
\begin{equation}
\label{eq:kappa_ixr_rel}
\kappa(\mathbfss{J}) = \frac{\sqrt{\textrm{IXR}_{\text{J}}}+1}{\sqrt{\textrm{IXR}_{\text{J}}}-1}
\end{equation}
For each run of the simulation with a given set of parameters, we
construct a system Jones matrix $\mathbfss{J}_{\text{sys}}$
(Eq. \ref{jones_rime}) by generating a random complex matrix from a
normal distribution ($\mu=0, \sigma=1$) for $\mathbfss{D}$.  All other
matrices in $\mathbfss{J}_{\text{sys}}$ are simplified to unity
matrices as discussed earlier in the section. The random matrix is
decomposed using \gls{svd} (Eq. \ref{eq:svd}), and the singular values
in $\mathbf{\Sigma}$ are replaced with those for the simulated
\gls{ixr} parameter. Without loss of generality we normalize
$\mathbf{\Sigma}$ using $\sigma_{\text{max}}=g_{\text{max}}=1$.  The
normalized condition number becomes
$\kappa(\mathbfss{J})=1/g_{\text{min}}$ and the system Jones matrix
due to intrinsic polarization leakage becomes
\begin{equation}
\label{eq:exp_jones}
\mathbfss{J}_{\text{sys}} = \mathbfss{U} \mathbf{\Sigma} \mathbfss{V}^{\dagger} = \mathbfss{U}
\begin{pmatrix}
    1 & 0\\
    0 & \frac{1}{\kappa(\mathbfss{J})}\\
\end{pmatrix}
\mathbfss{V}^{\dagger}
\end{equation}
We use a random matrix as there is an infinite set of Jones matrices
for a given \gls{ixr}. The decomposition by \gls{svd}, and
reconstruction steps are to maintain the same scaling as with the
calibration error Jones matrix.

For error in the polarization calibration, we use a sample space
starting at the ideal perfect calibration ($0\%$) up to $15\%$
calibration error. This calibration error represents the imperfect
measurement of \mathbfss{B} and \mathbfss{G}, the inverse of which are
applied to the observed signal to integrate the profile in both time
and frequency. A calibration error matrix $\Delta \mathbfss{J}$ is
generated from a random normal complex distribution ($\mu=0,
\sigma=\textrm{precent error}/100$).  The estimated system Jones
matrix is $\hat{\mathbfss{J}}_{\text{sys}} = \mathbfss{J}_{\text{sys}}
+ \Delta \mathbfss{J}$. This error parameter space covers a wide range
of polarization calibration situations, although the typical
calibration error is of the order of a few percent
\citep{2009ApJS..181..557H}. In practice, the polarization calibration
solution will also be affected by the \gls{ixr}. We simulate this
effect by using two different calibration techniques discussed later
in this section.
Additionally, this error does not include the
potential error from approximating \mathbfss{D} based on beam
modelling and observations that would be done in a real system.
Using equation 
\ref{eq:jones2mueller}, $\mathbfss{J}_{\text{sys}}$ and 
$\hat{\mathbfss{J}}_{\text{sys}}$ are converted to their Mueller forms
$\mathbfss{M}_{\text{sys}}$ and $\hat{\mathbfss{M}}_{\text{sys}}$.

To set the simulation \gls{snr} we add a system noise component
appropriate for the integration time of the simulated observation as
described in \S \ref{sec:ixr_to_snr}.  This is set by scaling
the ideal profile $\mathbfit{e}^{S}$, which has been normalized
such that the Stokes \emph{I} peak is unity, by a scalar $\mathrm{SN}_{\text{I}}$
value and adding a system noise Stokes vector $\mathbfit{n}^{S}$ which
is a set of real random values from a normal distribution ($\mu=0,
\sigma=1$).
The noise component $\mathbfit{n}^{S}$ is a direction-independent
effect.  This is a practical approximation when the system noise is
receiver noise dominated or when the sky noise is isotropic on the
scale of the beam primary lobe. In the extreme case where all the
system noise is direction-dependent in the direction of the source the
effect on \gls{rms} timing will only be a loss in \gls{snr} as seen in
Figure \ref{fig:actual_snr_vs_ixr}. As we will see in the following
sections, the effect of direction-independent noise is to further
degrade the timing solutions beyond the loss in \gls{snr}. For
simplicity we are using only direction-independent noise for the
simulation.

An ideal \gls{snr} is set to be in the range $\sqrt{10^3}$ to
$\sqrt{10^7}$ for our simulations, though we note: the observed
\gls{snr} is reduced as the condition number increases, and thus the
actual \gls{snr} is a function of \gls{ixr} as shown in Figure
\ref{fig:actual_snr_vs_ixr}.  The estimated Stokes vector
$\hat{\mathbfit{e}}^S$ of the observed pulsar is
\begin{equation}
\label{eq:est_pulsar_stokes}
\hat{\mathbfit{e}}^S = \hat{\mathbfss{M}}_{\text{sys}}^{-1} \left ( \mathrm{SN}_{\text{I}} \times \mathbfss{M}_{\text{sys}} \, \mathbfit{e}^S + \mathbfit{n}^{S} \right )
\end{equation}
The inversion of $\hat{\mathbfss{M}}_{\text{sys}}$ is the calibration
stage (Eq. \ref{invert_jones_rime}). We have simulated two types of
calibration. The first is a `gain' calibration where we are only
interested in solving the bandpass and electronic gain of the system,
that is $\hat{\mathbfss{J}}_{\text{sys}} = \mathbfss{BGC} + \Delta
\mathbfss{J} = \mathbfss{I} + \Delta \mathbfss{J}$, where
$\mathbfss{I}$ is the identity matrix. This is an ideal calibration
solution where the gain terms can be solved for independently of any
$\mathbfss{D}$ effects, such as if using a known noise reference in a
single dish system. The second type of calibration is a `full' calibration where the
$\mathbfss{D}$ term is included, $\hat{\mathbfss{J}}_{\text{sys}} = \mathbfss{BGCD} + \Delta
\mathbfss{J} = \mathbfss{D} + \Delta \mathbfss{J}$. This is a more realistic approach, as the
\gls{ixr} will affect any calibration solution. And, when performing timing
with an array, complex gain solutions are required to combine signals in
phase by using a sky calibrator source or self-calibration. The calibratability of
an array is a topic which should be studied in further work. Figure \ref{fig:actual_snr_vs_ixr}
shows the effect of these two methods on observed \gls{snr}, independent
of pulsar and ideal \gls{snr}. Examples
of these types of calibration on the observed profile are shown in
Figures \ref{fig:sim3} and \ref{fig:sim4}. After a simulated
$\hat{\mathbfit{e}}^S$ is produced, a \gls{toa} is then determined
with standard pulsar timing software. The effect of these calibration
methods on timing will be shown in the following section. The
simulation code is available as a git
repository\footnote{https://github.com/griffinfoster/pulsar-polarization-sims}.

\subsection{Methods for TOA determination}
\label{section:timing}

\begin{figure*}

\begin{tabular}{rcc}

    \includegraphics[width=55mm]{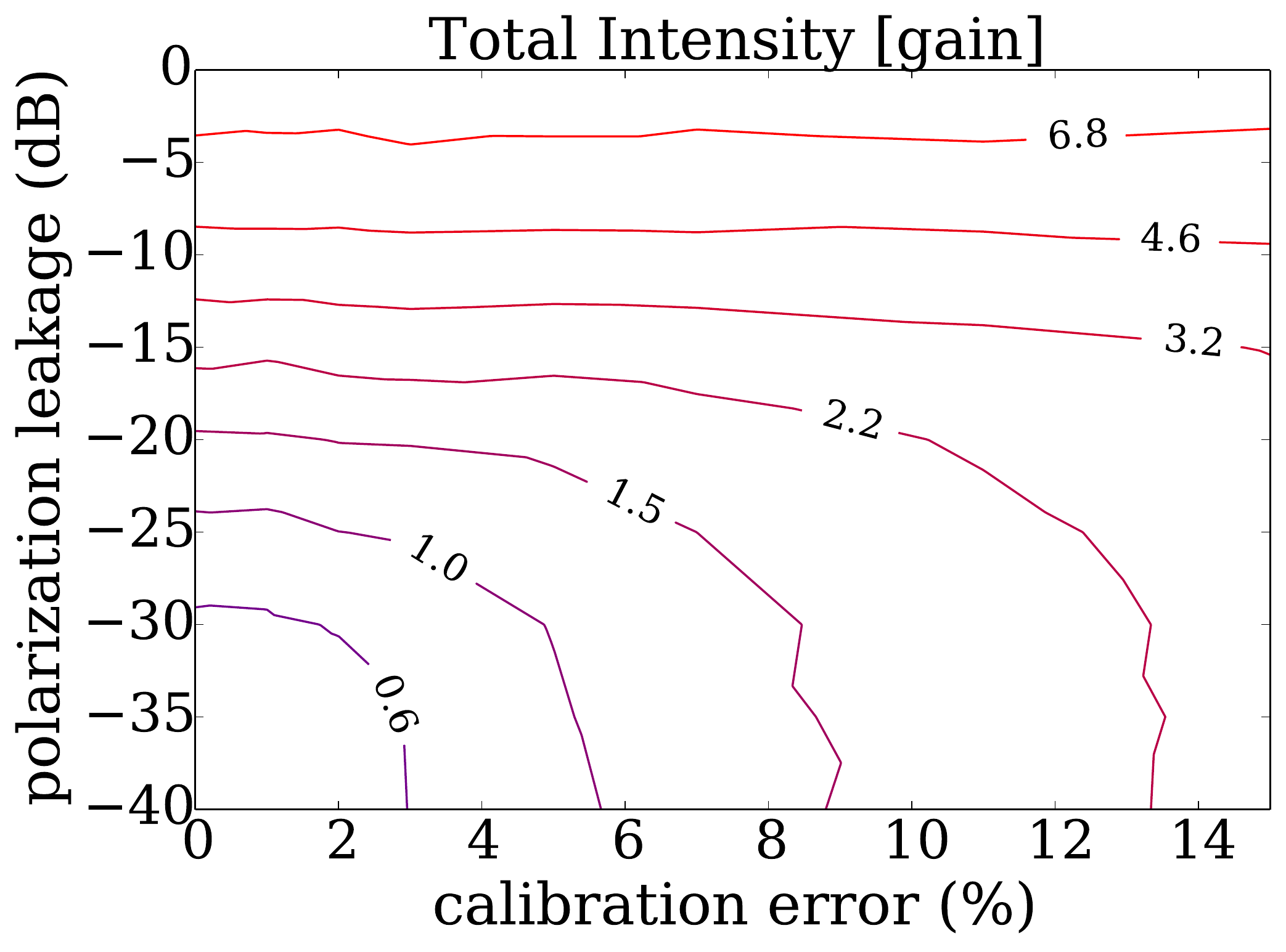} &
    \includegraphics[width=55mm]{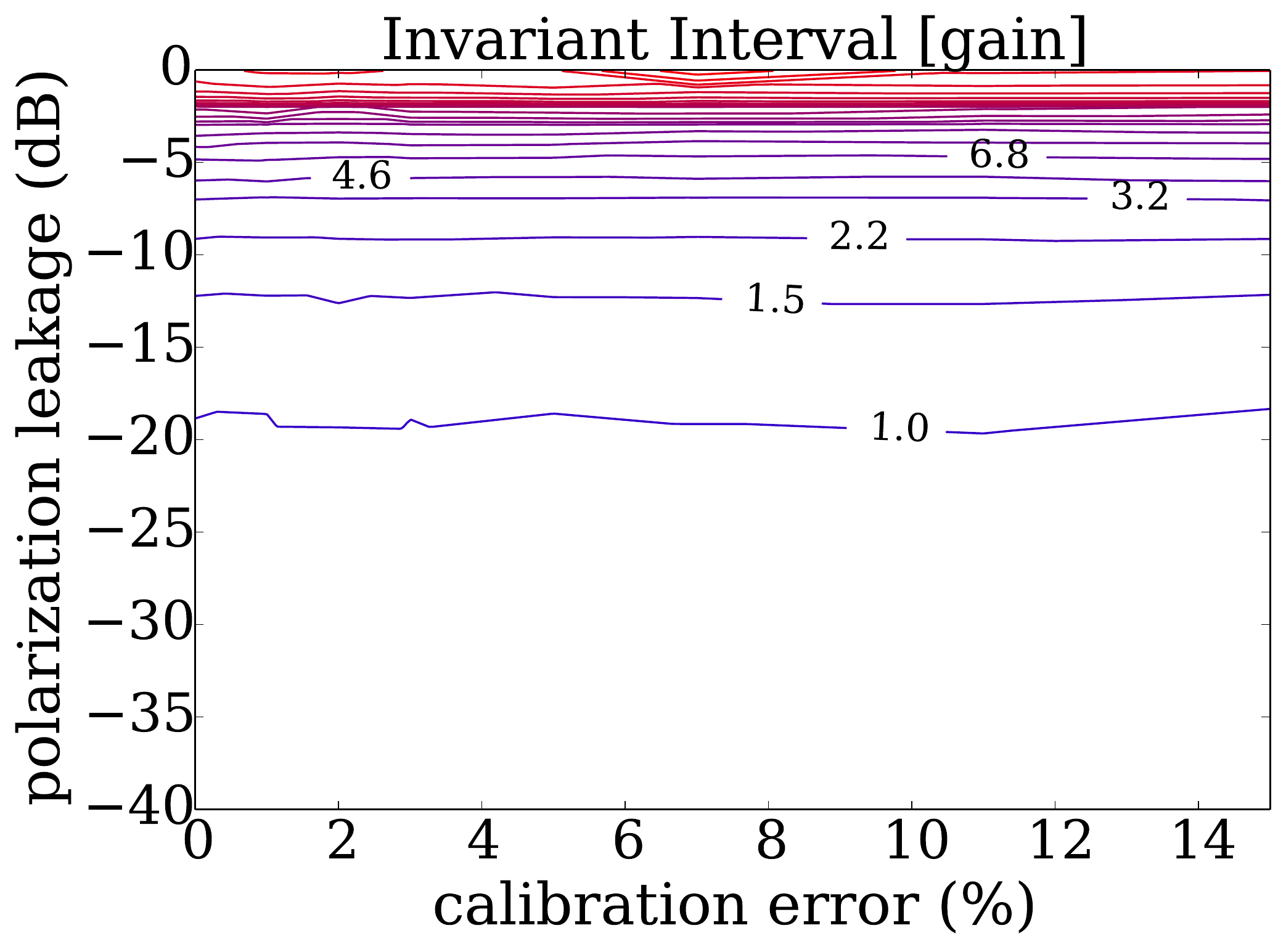} &
    \includegraphics[width=55mm]{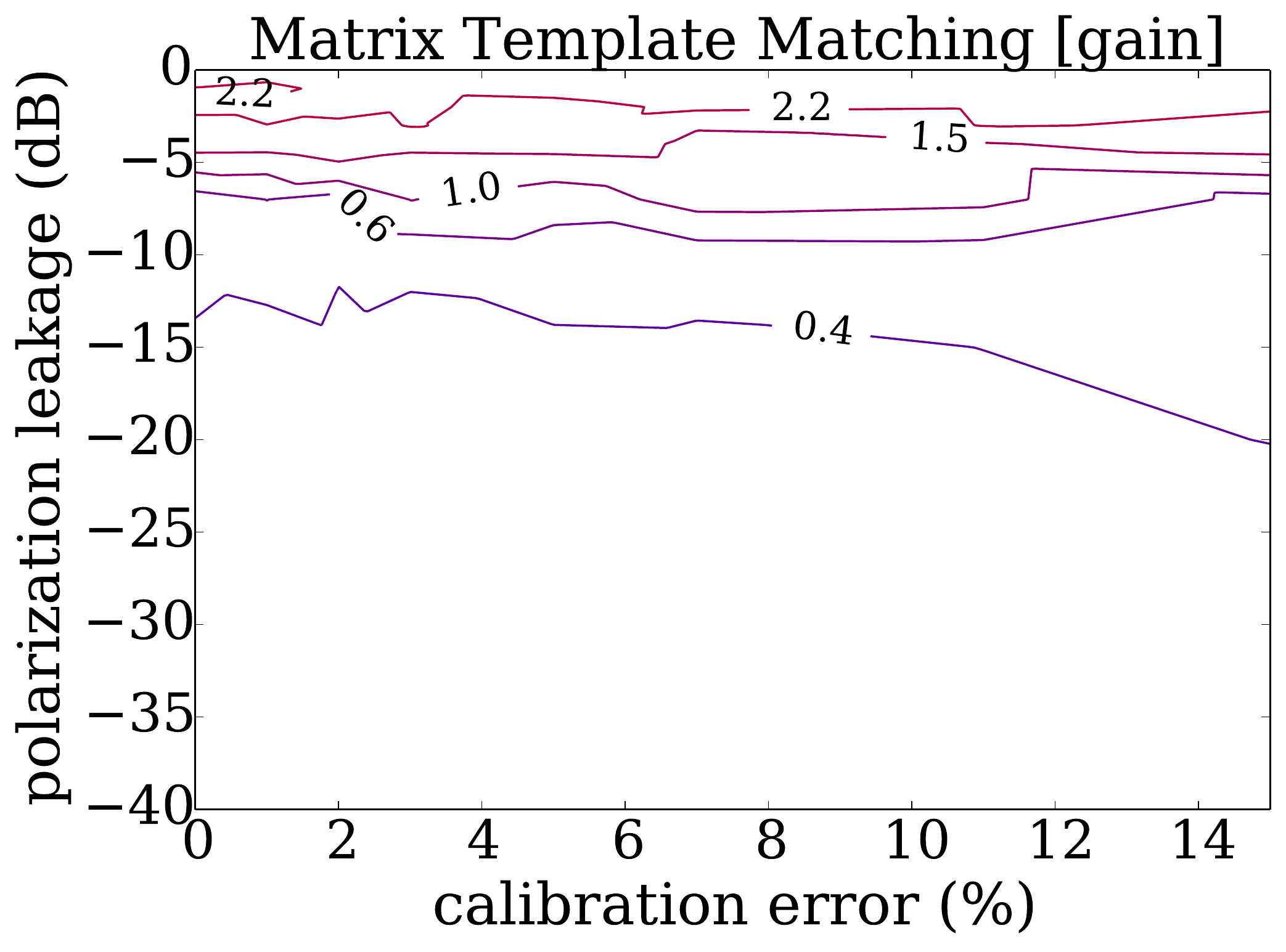} \\

    \includegraphics[width=55mm]{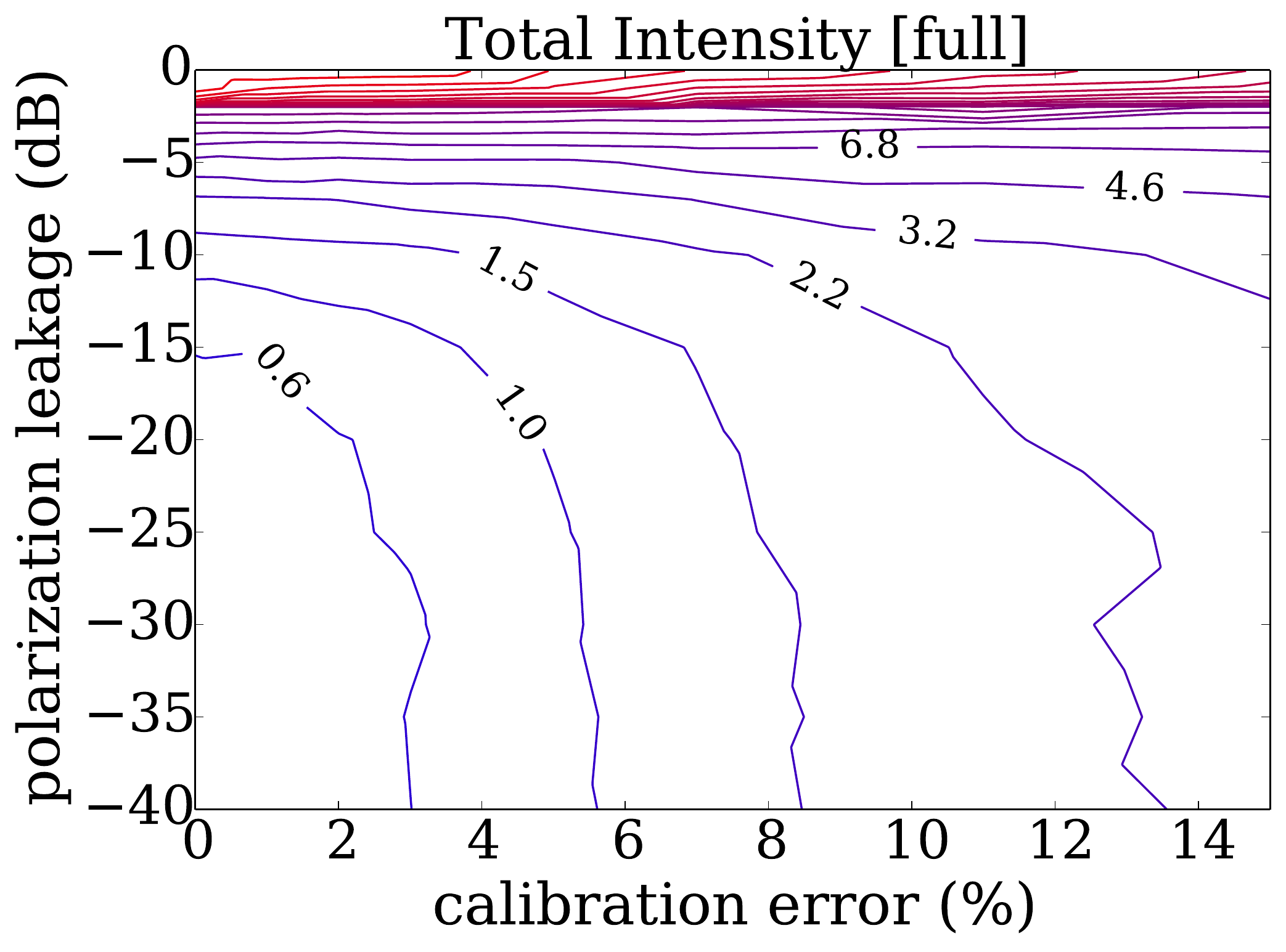} &
    \includegraphics[width=55mm]{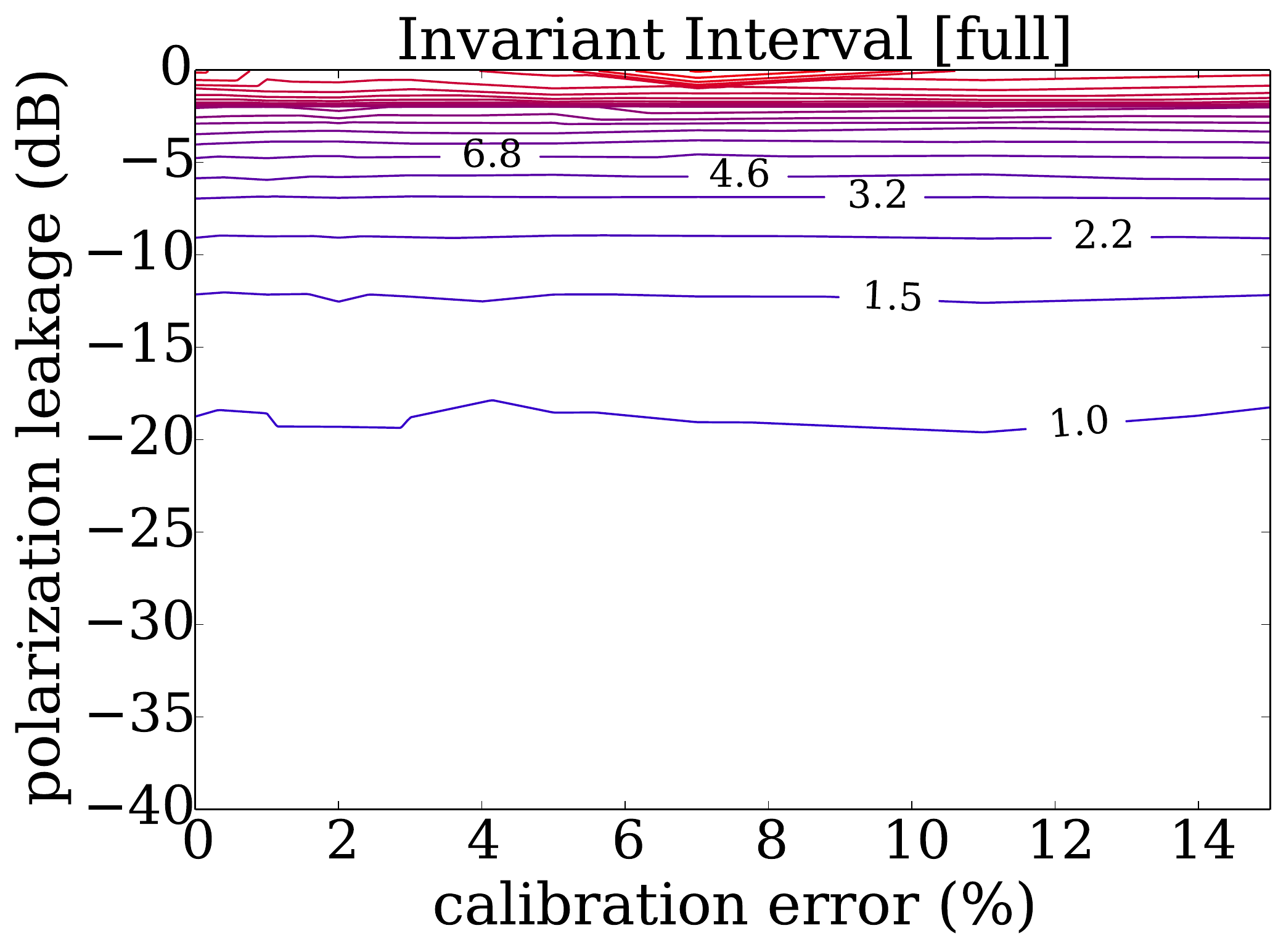} &
    \includegraphics[width=55mm]{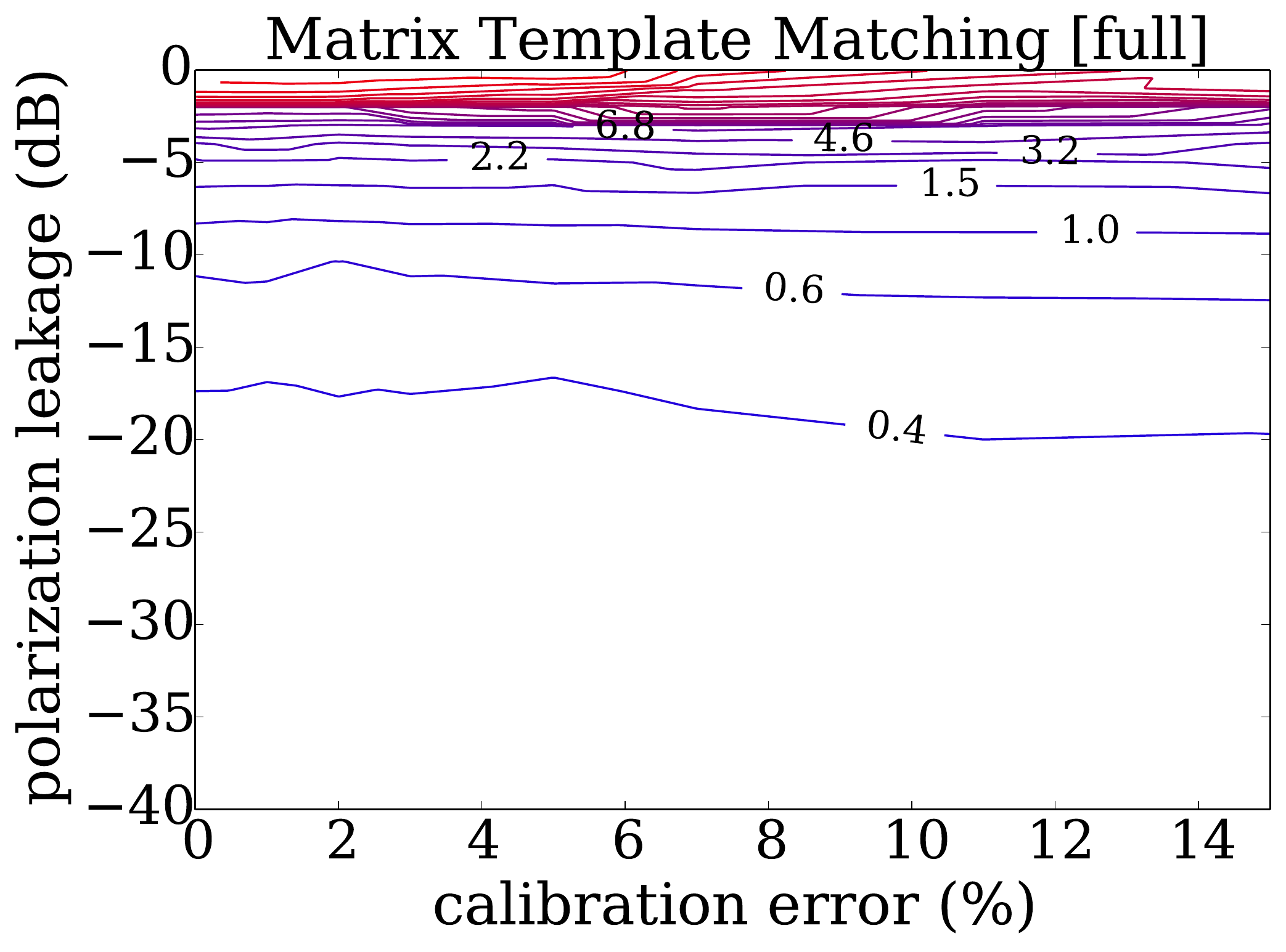} \\

\end{tabular}

\caption{
    Contour plots of time of arrival rms noise ($\mu$s) from simulation of
    PSR J1603$-$7202 as a function of calibration error and intrinsic polarization
    leakage. The top row shows the effect of intrinsic polarization leakage on timing
    for the three methods when `gain' calibration is applied. The bottom row is
    the same simulations but with `full' calibration applied. There is a strong
    dependence on polarization calibration error when using the total intensity
    method. The solutions when using the invariant interval are effectively identical
    for both types of calibration. Matrix template matching outperforms the invariant
    interval method in both the calibrated and uncalibrated cases. An ideal SNR
    of 100 was used for these figures.
    }
\label{fig:timing_results}

\end{figure*}

For \gls{toa} measurements, three standard methods are used: total
intensity \citep{1992RSPTA.341..117T}, invariant interval
\citep{2000ApJ...532.1240B}, and matrix template matching
\citep{2006ApJ...642.1004V}. These methods are included in
\texttt{PAT}, from the \texttt{PSRCHIVE} package.  For the total
intensity method, each observed Stokes $I$ profile is cross-correlated
with the ideal template.  The invariant interval technique uses
all Stokes parameters in the form of a Lorentz 4-vector
$(I^2-Q^2-U^2-V^2)^{1/2}$.  By including all Stokes parameters
complete information is used. However using the invariant form, the
\gls{snr} decreases, leading to a less precise \gls{toa} determination
compared to the intensity fitting, when the source is highly polarized
or intrinsically weak (see \cite{2013ApJS..204...13V} for more information
on this effect).  Matrix template matching represents the
profile in Jones notation and produces a \gls{toa} measurement while
simultaneously solving for a calibration solution by transforming the
template profile to the observed profile. In practice, the \gls{snr}
requirements for timing often ``force'' a pre-processing step of
frequency and time averaging. This requires some prior level of
polarization calibration, as described in the previous section,
rendering the assumptions for matrix template matching no longer
strictly valid, by introducing covariances between the Stokes parameters,
especially fir large polarization leakage.

Figure \ref{fig:timing_results} shows the \gls{toa} \gls{rms} derived
from the three timing methods used in simulations of the well studied
\gls{msp} J1603$-$7202.  There is a strong dependence on polarization
calibration error when using the total intensity timing
method. Applying the `full' calibration results in better timing
solutions except in the high intrinsic polarization leakage region ($>
-5$ dB), compared to using only the `gain' calibration. We will use
the `full' calibration simulations for our total intensity
results. The timing error, when using the invariant interval method,
is effectively the same for both types of calibration.

For the matrix template matching method there is a significant change
in the \gls{toa} \gls{rms} at high intrinsic polarization leakage
between the two types of calibration. This is due to `full'
calibration introducing covariances between Stokes parameters, compared to just `gain'
calibration. Simulations using `gain' calibration and timed with
matrix template matching produce the best timing results for a given
calibration error and intrinsic polarization leakage. We will present
matrix template matching results using both types of calibration, but
will focus our analysis on the `full' calibration as is represents a
more pragmatic result in terms of system calibration, especially in when
using an array for timing which requires multiple signals to be combined
in phase.

In practice, total intensity and invariant interval are commonly used
methods, where as the matrix template matching method is more rarely
used due to the difficulties in practical polarization
calibration. However, as timing limits are pushed, matrix template
matching will become necessary \citep{2013ApJS..204...13V}.

\section{Results}
\label{section:results}

\begin{figure}
    \includegraphics[width=1.0\linewidth]{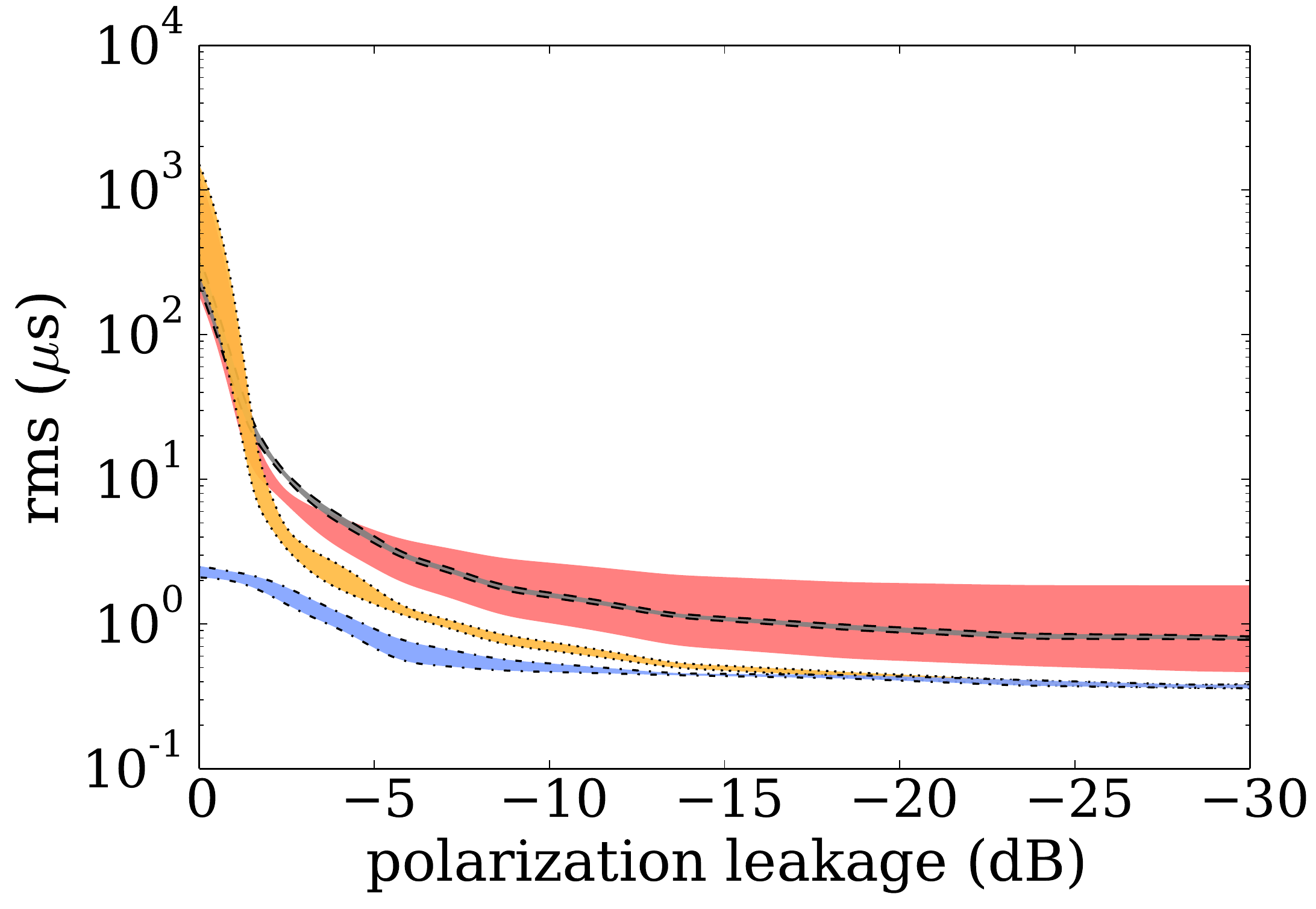}
    \caption{Time of arrival rms noise for J1603$-$7202 as a function of intrinsic
    polarization leakage using the three timing methods: total intensity (red/no border), 
    invariant interval (gray/dashed border), matrix template matching (full calibration: orange/dotted border,
    gain calibration: blue/dot dash border). The width of the lines show the effect polarization
    calibration error has on the ToA rms noise, a polarization calibration error
    from 0\% to 15\% was used. These simulations use a fractional integration time
    of 0.01 which would produce an ideal signal-to-noise ratio of 100.
    }
    \label{fig:rms_vs_ixr}
\end{figure}

As polarization calibration error is not the focus of these
simulations, we can compress the contour plots in Figure
\ref{fig:timing_results} into a more concise and useful figure by
collapsing the calibration error axis. Figure \ref{fig:rms_vs_ixr}
shows the bound range for each method on a single plot. The narrowness
of the bounds for the invariant interval and matrix template matching
show their independence from polarization calibration error. Plots of
this style for all the simulated \glspl{msp} are shown at the end in
Figures \ref{fig:psr_plots0} -- \ref{fig:psr_plots1}. These figures
show that as the intrinsic polarization leakage decrease, so too do
the timing residuals using all methods.

We have chosen to use the \gls{toa} \gls{rms} noise as the metric for
these results.  This is a measure of the time of arrival scatter for
the simulations. Using the average \gls{toa} uncertainty $\sigma_{\text{ToA}}$
underestimates the error. As per Eq. 8.2 of \cite{lorimer2005handbook}
$\sigma_{\text{ToA}} \simeq W/\textrm{SN}_{\textrm{o}}$, where $W$ is
the pulse width and $\textrm{SN}_{\textrm{o}}$ the observed \gls{snr}.
In our simulation results we see the intrinsic polarization leakage
affects timing results beyond a simple reduction of the \gls{snr},
producing poorer rms timing residuals compared to the average
$\sigma_{\text{ToA}}$ of the same simulated observations. In the limit
the $\textrm{IXR}_{\text{J}} \rightarrow \infty$, the \gls{rms} noise
will reach $\sigma_{\text{ToA}}$.

We computed the reduced $\chi^2$ solution to measure the goodness of
fit for each sample of the parameter space. When computing the
$\chi^2$ of the measured time of arrival against the expected time of
arrival, the low intrinsic polarization leakage case results in a good
fit, as would be expected. As the intrinsic polarization leakage
increases the fit degrades. Paradoxically, as the intrinsic
polarization leakage increases to the highest values, the reduced $\chi^2$ fit
approaches 1.  This is because $\sigma_{\text{ToA}}$ grows
exponentially large, and the error of the fit is within the limits of the \gls{toa} variance.

Although the diverse set of profiles we have simulated produce
different results, seen in Figures \ref{fig:psr_plots0} --
\ref{fig:psr_plots1}, there are general trends we observe. All the
simulations with the `full' calibration profiles show exponentially
increasing \gls{rms} timing solutions as the intrinsic polarization
leakage increases. Using only `gain' calibration produces good timing
results in this region, but is not as realistic as the `full'
calibration simulations.

Generally, using matrix template matching, with `gain' or `full'
calibration, produces better timing solutions compared to other
methods at all intrinsic polarization leakage values. For the majority
of the `gain' calibrated profiles, matrix template matching produces
timing solutions that are only weakly dependent on intrinsic
polarization leakage. It is worth noting that, for a few profiles,
such as J1022+1001, matrix template matching of the `full' calibration
profile produces lower \gls{rms} timing residuals than the `gain'
calibrated profile. This profile has highly polarized components,
which are distorted by high intrinsic polarization leakage. At high
intrinsic polarization leakage, the profiles with `full' calibration
result in high \gls{rms} timing residuals and dependence on
polarization calibration error when using matrix template
matching. However, at low intrinsic polarization leakage both types of
calibration converge to a similar timing solution.

The results from using the invariant interval method are consistent
across all profiles.  In a few cases, for example J1744$-$1134 and
J1824$-$2452, the invariant interval method significantly
underperforms compared to the other methods.  Both these pulsars are
highly polarized; therefore, in the invariant interval, the power in
Stokes $I$ is almost entirely cancelled out by the other Stokes
parameters.

\begin{figure}
    \includegraphics[width=1.0\linewidth]{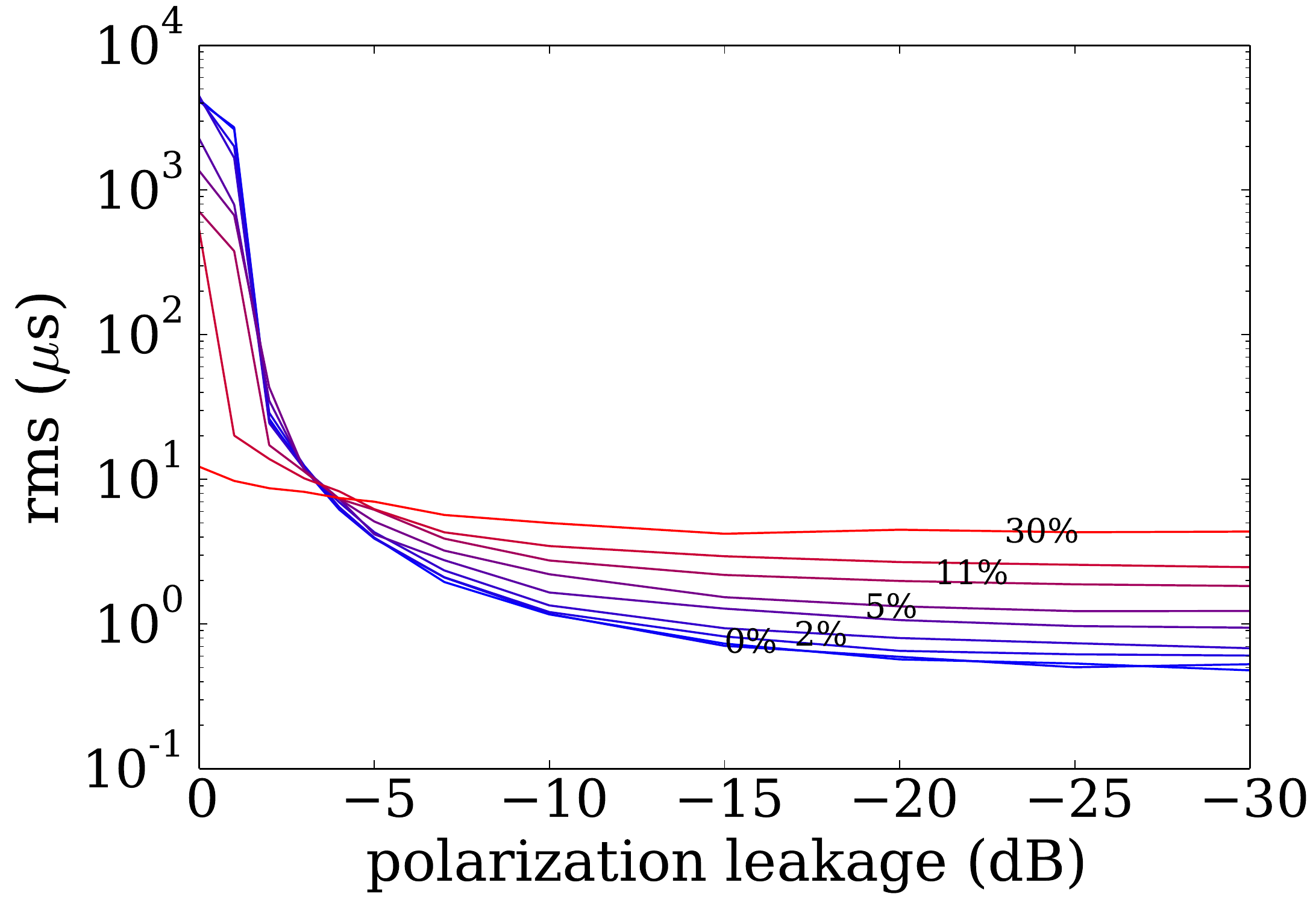}
    \caption{
    Time of arrival rms noise using the total intensity method as a function of intrinsic polarization leakage for J1603$-$7202.
    Lines cover a range of polarization calibration errors from $0\%$ (blue) to $27\%$ (red).
    A normalized integration time of $0.01$ was used for all the lines.
    A counter-intuitive effect occurs when the intrinsic polarization leakage is high, where the rms noise becomes inversely proportional to the calibration error because calibration errors improve the condition of the calibration matrix $\hat{\mathbfss{M}}_{\text{sys}}$.
    }
    \label{fig:sigma_ixr}
\end{figure}

Timing solutions with the total intensity method have a notable
dependence on the polarization calibration error. Plotting the bound
regions as individual lines for different polarization calibration
errors in Figure \ref{fig:sigma_ixr}, there is a polarization leakage
point (e.g. for PSR J1603$-$7202 it is around $-4$ dB), after which
systems with higher polarization calibration error produce better
timing solutions than an ideally calibrated system. This is true for
all \gls{msp} profiles used in our simulation, though transition
points vary between $-3$ dB and $-7$ dB.
%
In this region of the parameter space, this counter-intuitive effect
comes about because the error-free calibrator transformation also has
high intrinsic leakage and therefore further reduces the observed
\gls{snr} of the calibrated profile.  Adding polarization calibration
error reduces the intrinsic polarization leakage, leading to an
increase in the observed \gls{snr}. There is potential use for the
total intensity method at high intrinsic polarization leakage when
only using a `gain' calibration. Needless to say, this regime should
be avoided and this sets a maximum limit on the allowable intrinsic
polarization leakage to around $-5$~dB.

\begin{figure}
    \includegraphics[width=1.0\linewidth]{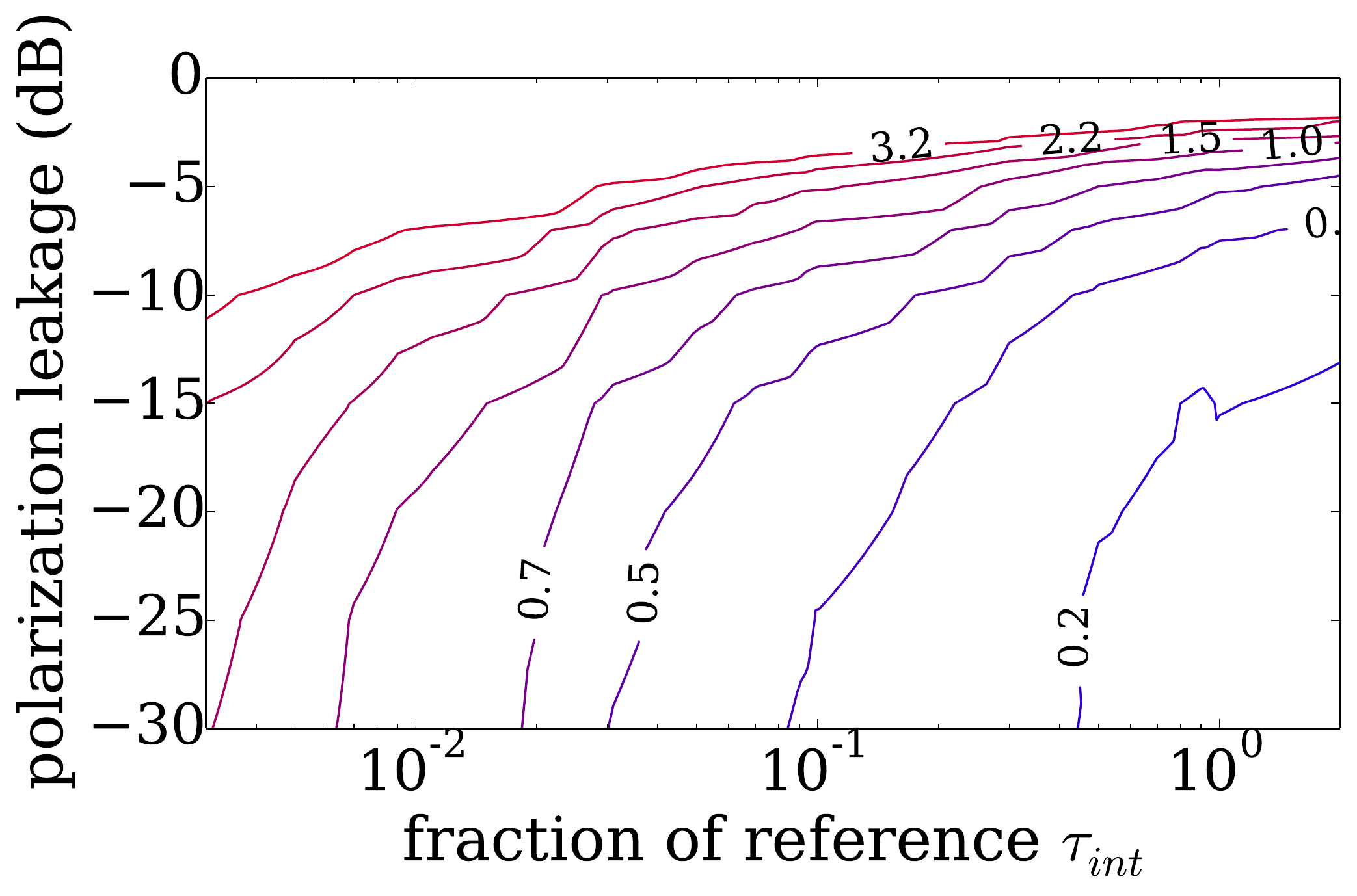}
    \caption{Contour plot of time of arrival rms noise ($\mu$s)
      using matrix template matching as a function of intrinsic polarization
      leakage and integration time for PSR J1603$-$7202. Integration time is a
      fraction of the reference integration time, see \S \ref{sec:ixr_to_snr}.
    }
    \label{fig:tint_plot}
\end{figure}

\begin{figure}
    \includegraphics[width=1.0\linewidth]{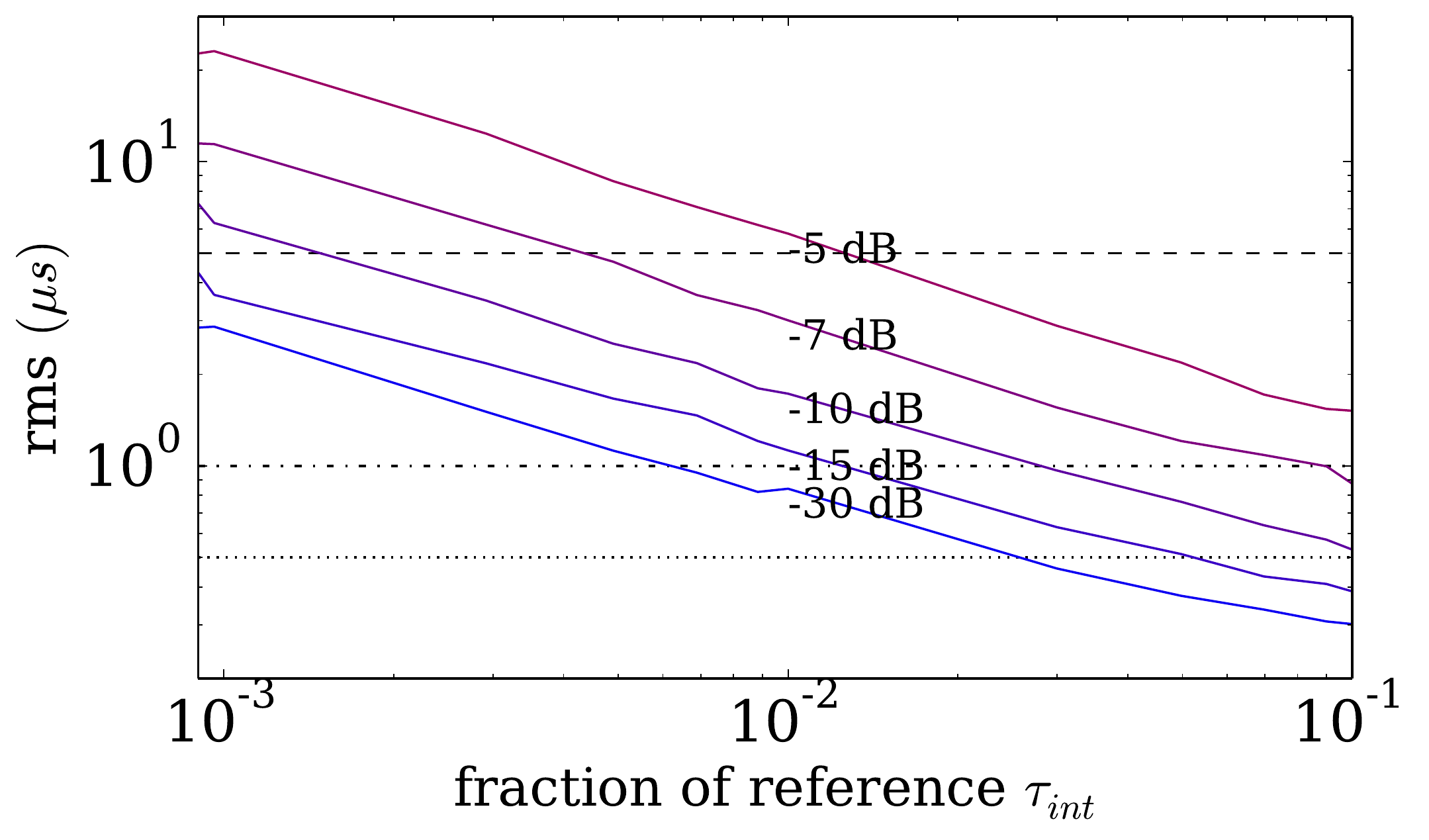}
    \caption{Time of arrival rms noise of PSR J1603$-$7202, using the matrix
      template matching method, as a function of integration lengths for a
      range of intrinsic polarization leakage values. Horizontal lines indicate
      rms thresholds: $0.5~\mu$s (dotted), $1~\mu$s (dot dash), $5~\mu$s (dashed).
    }
    \label{fig:sigma_vs_snr}
\end{figure}

Figure \ref{fig:tint_plot} shows the residual timing \gls{rms} of
J1603$-$7202 as a function of \gls{snr} and $\textrm{IXR}_{\text{J}}$
using a calibration error of $5\%$. For a given integration time, the
achievable time of arrival \gls{rms} noise is dependent on the system
polarization value, e.g. in simulation of J1603$-$7202 for
$\tau_{\text{int}}=0.1$ a system with $-25$ dB intrinsic polarization
leakage can achieve a \gls{toa} \gls{rms} in the timing residuals of
around $300$~ns compared to $3000$~ns with a $-5$~dB intrinsic
polarization leakage system. This variation due to intrinsic
polarization leakage can be better seen in Figure
\ref{fig:sigma_vs_snr}, which shows the residual \gls{rms} as a
function of integration length, for a range of intrinsic polarization
leakage values. The \gls{rms} will continue to decrease as the
integration time increases. The system intrinsic polarization leakage
sets the required integration time to achieve a desired \gls{rms}. In
\S \ref{section:discussion}, we discuss the importance of Figure
\ref{fig:sigma_vs_snr} in optimizing science return when there is
limited available observing time.

\section{Discussion}
\label{section:discussion}

Gravitational wave detection using high precision timing of
\glspl{msp} constitutes a key science project for the \gls{ska}.  This
imposes a requirement on the polarization specifications.  In
\cite{2004NewAR..48.1413C}, a case is presented which sets the
required polarization purity level to $-40$ dB to accomplish the
\gls{ska} pulsar key science goals.  This polarization purity value is
different from what we have defined as polarization purity; it is a
measure of the final, calibrated Stokes data and not a specification
of the front-end design as we have considered here.

Figure \ref{fig:sigma_vs_snr} shows that for a given intrinsic
polarization leakage, a desired timing residual \gls{rms} can be
achieved with sufficient observing time.  This, of course, ignores the
other systematic effects that are part of a timing observation and does
not include the additional sensitivity modulation of the primary beam shape.
We have only focused on intrinsic polarization leakage, which is an
effect on any feed design. The main issue is that the intrinsic
polarization leakage has a strong effect on the required observation
time, which is a limited commodity.  For our simulation of \gls{msp}
J1603$-$7202 in Figure \ref{fig:sigma_vs_snr} the difference in
required observing time to achieve a desired \gls{rms} noise at
$-15$~dB compared to $-30$~dB is a factor of $1.5$. With limited
available observing time, we would like to set an upper limit on the
intrinsic polarization leakage above which it is no longer optimal to
be using observation time on a measurement.
\cite{2009wska.confE..17C} and \cite{2013ITAP...61.2852S} show the
$\textrm{IXR}_{\text{J,dB}}$ for typical feeds to be somewhere between
$30$ dB and $66$ dB at boresight. We see that observing with an
intrinsic polarization leakage of $-15$ dB implies at least a $50\%$
increase in observing time compared to that of a typical feed. A high
intrinsic polarization leakage is not a design issue for a `classic'
single pixel dish system, where a low intrinsic polarization leakage
can be achieved when observing a source on axis. This is not the case
with aperture arrays, \glspl{paf}, and multi-beaming with single pixel
dishes. In these cases, the source will likely not be located in
the optimal intrinsic polarization leakage region of the beam. For
aperture arrays, a source will rarely, if ever, be on axis. Returning
to the example beams in \cite{2009wska.confE..17C} and
\cite{2013ITAP...61.2852S}, we see that the polarization leakage
values can quickly increase to above $-10$ dB away from zenith. This
effectively limits the declination range of sources, depending on the
array latitude, for pulsar timing. 

From our simulation we see there is an intrinsic polarization leakage
lower limit on feed design at which point there is minimal return in
terms of reducing the timing residual \gls{rms} with further
engineering investment, for a given integration time. As there is a cost
to every incremental improvement in \gls{ixr}, we would like to present
our simulation results in terms scientific return for marginal improvements in
engineering specifications. In an effort to create
a meaningful engineering intrinsic polarization leakage lower limit,
Table \ref{tbl:ixr_metric} lists the fractional improvement in
\gls{toa} \gls{rms} noise for different \gls{ixr} values. We have
picked a typical \gls{snr} ($\tau_{\text{int}}=0.01$) for a timing
observation. Columns 2, 4, 6, and 8 are the timing residual \gls{rms}
for each \gls{msp} at $\textrm{IXR}_{\text{J}}=10,20,30,40$ dB
respectively.  Columns 3, 5, and 7 are the percentage change in the
\gls{rms} with the changes in $\textrm{IXR}$. This table shows the
diminishing marginal utility of improving $\textrm{IXR}$ for the
benefit of decreasing the time of arrival \gls{rms} noise. There is,
on average, a $29\%$ improvement in the \gls{rms} when improving the
$\textrm{IXR}_{\text{J}}$ from $10$ dB to $20$ dB, but can vary significantly
with profile shape. For example, timing of J0711--6830 is largely uneffected
by improvements in \gls{ixr}, while timing of J1603--7202 improves with each
increase in \gls{ixr}. Going from $20$ dB
to $30$ dB there is a small improvement, but going above $30$ dB
provides essentially no improvement. This indicates that for a feed
with $\textrm{IXR}_{\text{J,dB}}>30$ dB there is limited fractional
improvement in pulsar timing capabilities. It may be worth considering
that low intrinsic polarization leakage across the field of view may
be preferable to optimizing for minimal intrinsic polarization leakage
around boresight.

% ...
\begin{table*}
    \centering
    \begin{tabular}{| l | r | r | r | r | r | r | r |}
    \hline
                & rms (ns)                          & $\Delta_{\text{ToA}} (\%)$                    & rms (ns)                          & $\Delta_{\text{ToA}} (\%)$ & rms (ns)                 & $\Delta_{\text{ToA}} (\%)$                 & rms (ns) \\
    Pulsar      & $\textrm{IXR}_{\text{J}}\!=\!10\,$dB   & $10\,\textrm{dB}\!\rightarrow\!20\,\textrm{dB}$    & $\textrm{IXR}_{\text{J}}\!=\!20\,$dB & $20\,\textrm{dB}\!\rightarrow\!30\,\textrm{dB}$ & $\textrm{IXR}_{\text{J}}\!=\!30\,$dB & $30\,\textrm{dB}\!\rightarrow\!40\,\textrm{dB}$ & $\textrm{IXR}_{\text{J}}\!=\!40\,$dB \\
    \hline  
    J0437--4715 & 545 & 43.1\% & 310 &  9.4\% & 281 & 3.2\% & 272 \\
    J0613--0200 & 256 & 19.9\% & 205 &  2.5\% & 200 & 0.5\% & 199 \\
    J0711--6830 & 202 &  7.9\% & 186 &  0.5\% & 185 & 0.0\% & 185 \\
    J1022+1001  & 294 & 24.5\% & 222 &  5.0\% & 211 & 0.0\% & 211 \\
    J1024--0719 & 217 & 11.1\% & 193 &  1.6\% & 190 & 1.1\% & 188 \\
    J1045--4509 & 657 & 41.6\% & 384 & 11.2\% & 341 & 4.1\% & 327 \\
    J1600--3053 & 445 & 38.7\% & 273 &  9.2\% & 248 & 2.0\% & 243 \\
    J1603--7202 & 755 & 43.7\% & 425 & 11.1\% & 378 & 4.5\% & 361 \\
    J1643--1224 & 704 & 44.0\% & 394 & 10.9\% & 351 & 1.7\% & 345 \\
    J1713+0747  & 419 & 37.5\% & 262 &  9.2\% & 238 & 1.7\% & 234 \\
    J1730--2304 & 336 & 40.2\% & 201 &  3.5\% & 194 & 0.5\% & 193 \\
    J1732--5049 & 247 & 18.2\% & 202 &  4.0\% & 194 & 0.5\% & 193 \\
    J1744--1134 & 463 & 37.1\% & 291 &  9.6\% & 263 & 1.9\% & 258 \\
    J1824--2452 & 377 & 33.7\% & 250 &  6.8\% & 233 & 0.9\% & 231 \\
    J1857+0943  & 211 &  8.1\% & 194 &  1.5\% & 191 & 0.5\% & 190 \\
    J1909--3744 & 356 & 35.1\% & 231 &  5.6\% & 218 & 2.8\% & 212 \\
    J1939+2134  & 354 & 33.9\% & 234 &  5.6\% & 221 & 1.8\% & 217 \\
    J2124--3358 & 201 &  7.5\% & 186 &  1.1\% & 184 & 0.5\% & 183 \\
    J2129--5721 & 274 & 22.6\% & 212 &  3.8\% & 204 & 2.0\% & 200 \\
    J2145--0750 & 398 & 35.9\% & 255 &  6.7\% & 238 & 0.8\% & 236 \\
    \hline
    Average     &     & 29.2\% &     &  6.0\% &     & 1.6\% &     \\
    Range       &     & 7\% --- 44\% &     & 1\% --- 11\%  &     & 0\% --- 4\% &     \\
    \hline
    \end{tabular}
    \caption{
    Fractional improvement in time of arrival rms noise between two $\textrm{IXR}_{\text{J}}$
    values for the simulated MSPs using the matrix template method. Percent change
    is computed as $\Delta_{\text{ToA}}=100(\text{rms}_{\text{i}}-\text{rms}_{\text{f}})/\text{rms}_{\text{i}}$,
    where $\text{rms}_{\text{i}}$ is the initial (lower) $\textrm{IXR}_{\text{J}}$
    ToA rms noise, and $\text{rms}_{\text{f}}$ is the final (higher) $\textrm{IXR}_{\text{J}}$
    ToA rms noise. These simulations used a fractional integration length of $0.01$,
    which is an ideal SNR of 100. The reported rms values have an uncertainty of $\sim1\%$ set
    by the number of ToA simulations ($n=5000$). The last rows of the table are the average ToA rms
    noise percent improvement and minimum/maximum range.
    }
    \label{tbl:ixr_metric}
\end{table*}

\section{Conclusion}

On the pathway towards the \gls{ska} a number of aperture arrays,
dishes, and \glspl{paf} are being developed as precursor
instruments. As pulsar timing is a key science project, design of
these instruments should take into account the intrinsic polarization
leakage specification we have presented in this paper.

There is a relative increase in required integration time as a
function of the feed \gls{ixr}, as seen in Figure
\ref{fig:sigma_vs_snr}. At high intrinsic polarization leakage this
can make the required integration time inefficiently long.

We have shown that there are diminishing returns (Table
\ref{tbl:ixr_metric}) on building feed systems which have intrinsic
polarization leakage below $-30$ dB in the direction of
observation. Achieving this intrinsic polarization leakage limit
should be easily affordable for single pixel dishes on axis, where the
leakage is at a minimum. However, aperture arrays, \glspl{paf}, and
multi-beam systems, where observations are not always made in the
direction of minimum leakage, could have limited use for pulsar timing
if intrinsic polarization leakage is not taken into account while
developing the feed system. The complex aperture array and \gls{paf}
beams with sharp, frequency dependent structure lead to regions of
high intrinsic polarization leakage.

Given the effect of intrinsic polarization leakage on pulsar timing
and a costing model for a feed design, a desirable optimization could
be to maximize the average \gls{ixr} across the intended usable field
of view for the element and not just in the direction of boresight.

The calibratability of an array has a key effect on timing as we have
shown with the idealized `gain' calibration technique against the more
realistic `full' calibration.  Beyond this work, there is scope for
additional work on effect of calibration on timing. Additionally,
the matrix template matching method should be extended to account for
the covariances between the Stokes parameters induced by poorly conditioned
calibration matrices.

Ideally, we can further refine these values on precursor instruments. The MeerKAT,
KAT-7, ASKAP, and APERTIF arrays will provide a platform to study dish array
polarization effects with multi-beam feeds and \glspl{paf}. LOFAR and MWA, though
not ideal for pulsar timing experiments due to the low observing frequencies, will
be useful to study instrumental polarization in aperture arrays.  A study of these
array polarization properties is necessary to assure the \gls{ska} science
specifications can be met for pulsar timing.

\section*{Acknowledgments}

This work is based upon research supported by the South African Research Chairs
Initiative of the Department of Science and Technology and National Research
Foundation. C.B.

\bibliography{pp_article}

\begin{thebibliography}{}

\bibitem[\protect\citeauthoryear{{Britton}}{{Britton}}{2000}]{2000ApJ...532.1240B}
{Britton} M.~C.,  2000, \apj, 532, 1240

\bibitem[\protect\citeauthoryear{{Carozzi} \& {Woan}}{{Carozzi} \&
  {Woan}}{2011}]{2011ITAP...59.2058C}
{Carozzi} T.~D.,  {Woan} G.,  2011, IEEE Transactions on Antennas and
  Propagation, 59, 2058

\bibitem[\protect\citeauthoryear{{Carozzi}, {Woan} \& {Maaskant}}{{Carozzi}
  et~al.}{2009}]{2009wska.confE..17C}
{Carozzi} T.~D.,  {Woan} G.,    {Maaskant} R.,  2009, in Wide Field Astronomy
  \& Technology for the Square Kilometre Array {Polarization Diversity for SKA
  Wide-field Polarimetry}.
p.~17

\bibitem[\protect\citeauthoryear{{Cordes}, {Kramer}, {Lazio}, {Stappers},
  {Backer} \& {Johnston}}{{Cordes} et~al.}{2004}]{2004NewAR..48.1413C}
{Cordes} J.~M.,  {Kramer} M.,  {Lazio} T.~J.~W.,  {Stappers} B.~W.,  {Backer}
  D.~C.,    {Johnston} S.,  2004, \nar, 48, 1413

\bibitem[\protect\citeauthoryear{{Hamaker}}{{Hamaker}}{2000}]{2000A&AS..143..515H}
{Hamaker} J.~P.,  2000, \aaps, 143, 515

\bibitem[\protect\citeauthoryear{{Hamaker}, {Bregman} \& {Sault}}{{Hamaker}
  et~al.}{1996}]{1996A&AS..117..137H}
{Hamaker} J.~P.,  {Bregman} J.~D.,    {Sault} R.~J.,  1996, \aaps, 117, 137

\bibitem[\protect\citeauthoryear{{Han}, {Demorest}, {van Straten} \&
  {Lyne}}{{Han} et~al.}{2009}]{2009ApJS..181..557H}
{Han} J.~L.,  {Demorest} P.~B.,  {van Straten} W.,    {Lyne} A.~G.,  2009,
  \apjs, 181, 557

\bibitem[\protect\citeauthoryear{{Hobbs}, {Edwards} \& {Manchester}}{{Hobbs}
  et~al.}{2006}]{2006MNRAS.369..655H}
{Hobbs} G.~B.,  {Edwards} R.~T.,    {Manchester} R.~N.,  2006, \mnras, 369, 655

\bibitem[\protect\citeauthoryear{{Hotan}, {van Straten} \&
  {Manchester}}{{Hotan} et~al.}{2004}]{2004PASA...21..302H}
{Hotan} A.~W.,  {van Straten} W.,    {Manchester} R.~N.,  2004, \pasa, 21, 302

\bibitem[\protect\citeauthoryear{IEEE}{IEEE}{1998}]{705931}
IEEE 1998, IEEE Standard Definitions of Terms for Radio Wave Propagation.
IEEE

\bibitem[\protect\citeauthoryear{{Janssen}, {Hobbs}, {McLaughlin}, {Bassa},
  {Deller}, {Kramer}, {Lee}, {Mingarelli}, {Rosado}, {Sanidas}, {Sesana},
  {Shao}, {Stairs}, {Stappers} \& {Verbiest}}{{Janssen}
  et~al.}{2015}]{2015arXiv150100127J}
{Janssen} G.~H.,  {Hobbs} G.,  {McLaughlin} M.,  {Bassa} C.~G.,  {Deller}
  A.~T.,  {Kramer} M.,  {Lee} K.~J.,  {Mingarelli} C.~M.~F.,  {Rosado} P.~A.,
  {Sanidas} S.,  {Sesana} A.,  {Shao} L.,  {Stairs} I.~H.,  {Stappers} B.~W.,
   {Verbiest} J.~P.~W.,  2015, ArXiv e-prints

\bibitem[\protect\citeauthoryear{Lorimer}{Lorimer}{2005}]{lorimer2005handbook}
Lorimer D.,  2005, Handbook of Pulsar Astronomy.
Cambridge Observing Handbooks for Research Astronomers, Cambridge University
  Press

\bibitem[\protect\citeauthoryear{{Manchester} et~al.,}{{Manchester}
  et~al.}{2013}]{2013PASA...30...17M}
{Manchester} R.~N.,  et~al., 2013, \pasa, 30, 17

\bibitem[\protect\citeauthoryear{{Smirnov}}{{Smirnov}}{2011a}]{2011A&A...527A.106S}
{Smirnov} O.~M.,  2011a, \aap, 527, A106

\bibitem[\protect\citeauthoryear{{Smirnov}}{{Smirnov}}{2011b}]{2011A&A...527A.107S}
{Smirnov} O.~M.,  2011b, \aap, 527, A107

\bibitem[\protect\citeauthoryear{{Sutinjo} \& {Hall}}{{Sutinjo} \&
  {Hall}}{2013}]{2013ITAP...61.2852S}
{Sutinjo} A.~T.,  {Hall} P.~J.,  2013, IEEE Transactions on Antennas and
  Propagation, 61, 2852

\bibitem[\protect\citeauthoryear{{Taylor}}{{Taylor}}{1992}]{1992RSPTA.341..117T}
{Taylor} J.~H.,  1992, Royal Society of London Philosophical Transactions
  Series A, 341, 117

\bibitem[\protect\citeauthoryear{{van Straten}}{{van
  Straten}}{2006}]{2006ApJ...642.1004V}
{van Straten} W.,  2006, \apj, 642, 1004

\bibitem[\protect\citeauthoryear{{van Straten}}{{van
  Straten}}{2013}]{2013ApJS..204...13V}
{van Straten} W.,  2013, \apjs, 204, 13

\bibitem[\protect\citeauthoryear{{van Straten}, {Demorest} \& {Oslowski}}{{van
  Straten} et~al.}{2012}]{2012AR&T....9..237V}
{van Straten} W.,  {Demorest} P.,    {Oslowski} S.,  2012, Astronomical
  Research and Technology, 9, 237

\end{thebibliography}

\section*{Simulation Plots}
\label{sec:sim_plots}

\begin{figure*}
\strut

\noindent\null\hfill
\includegraphics[width=.2\linewidth]{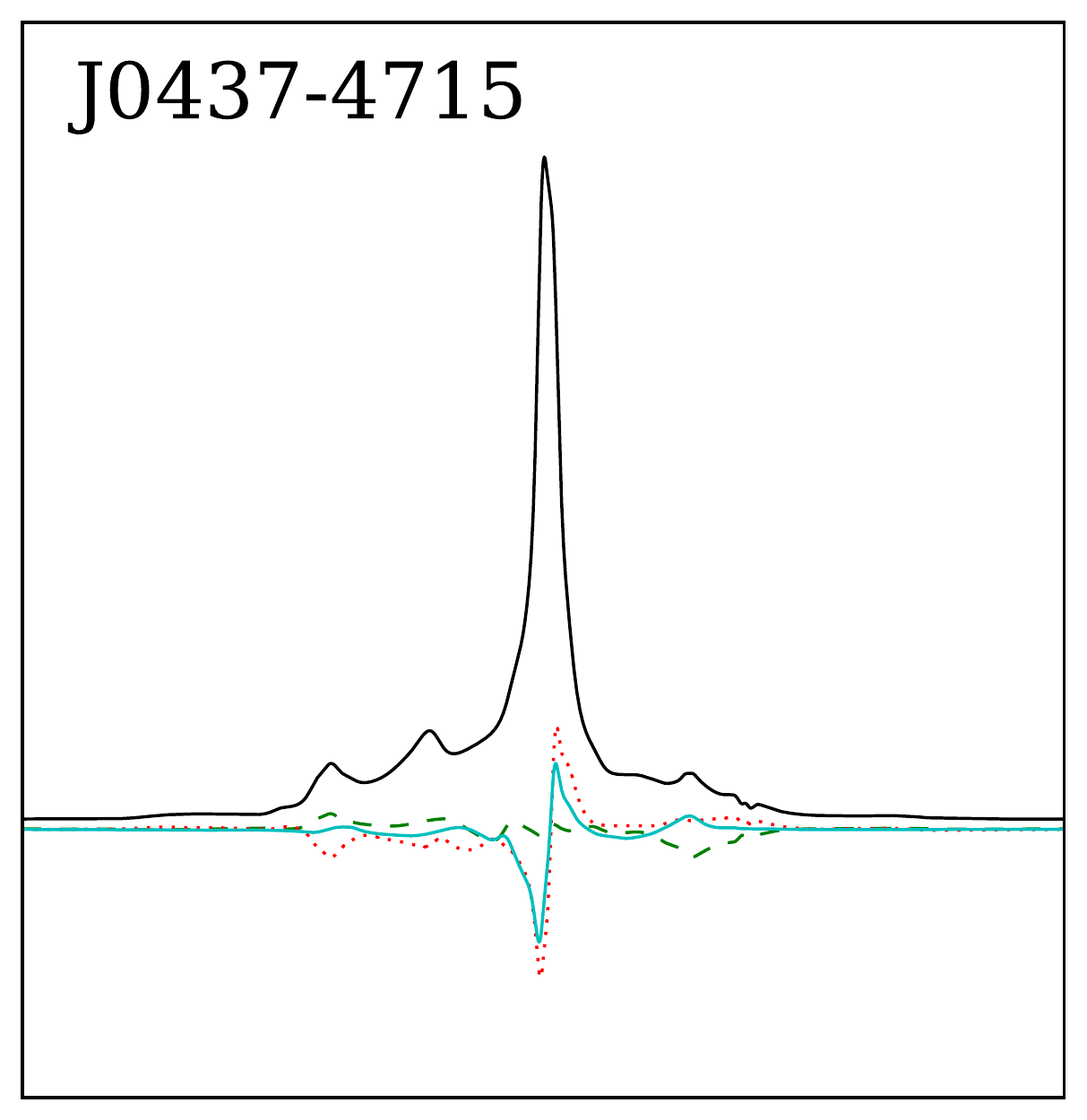} \hfill
\includegraphics[width=.28\linewidth]{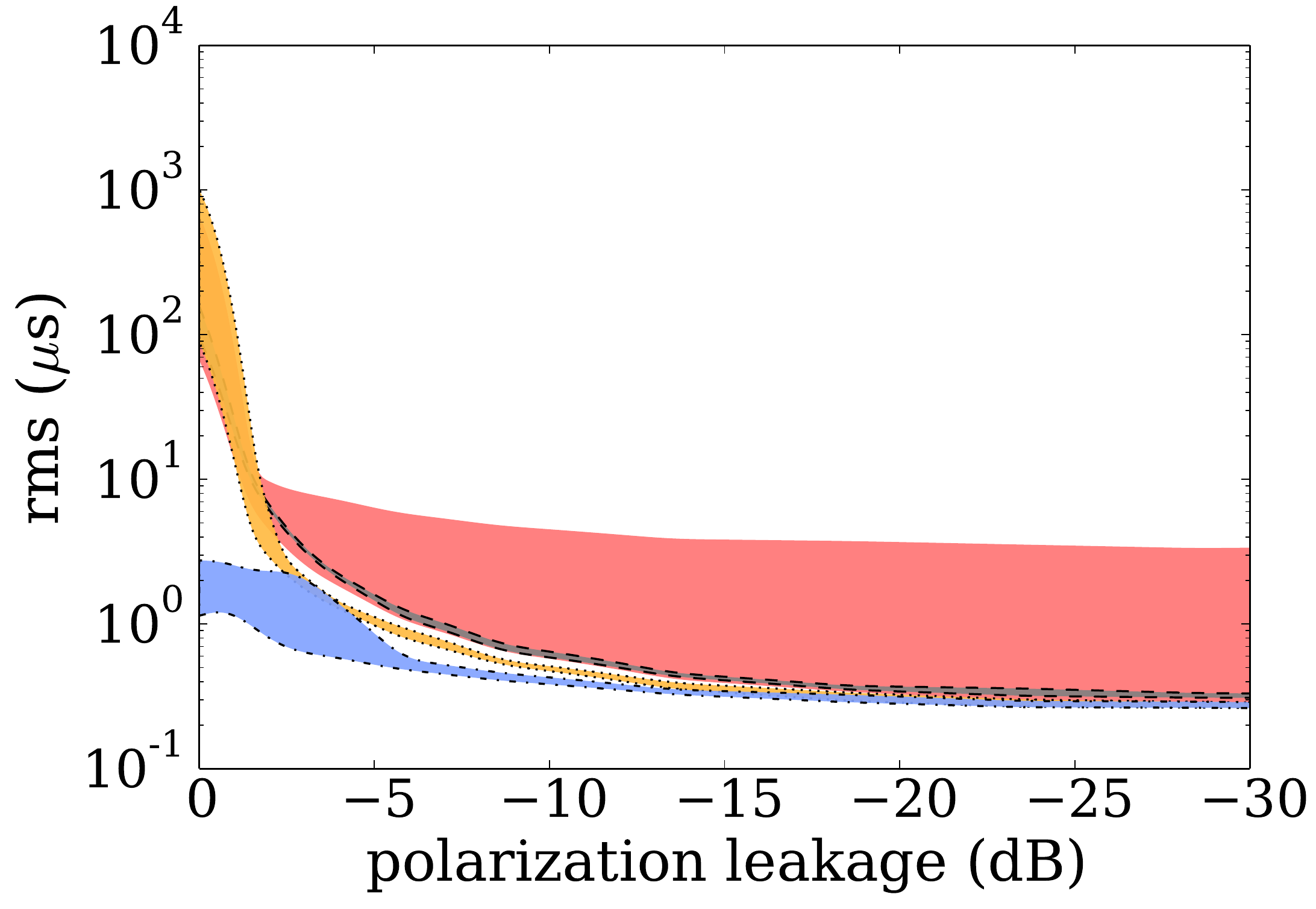} \hfill
\includegraphics[width=.2\linewidth]{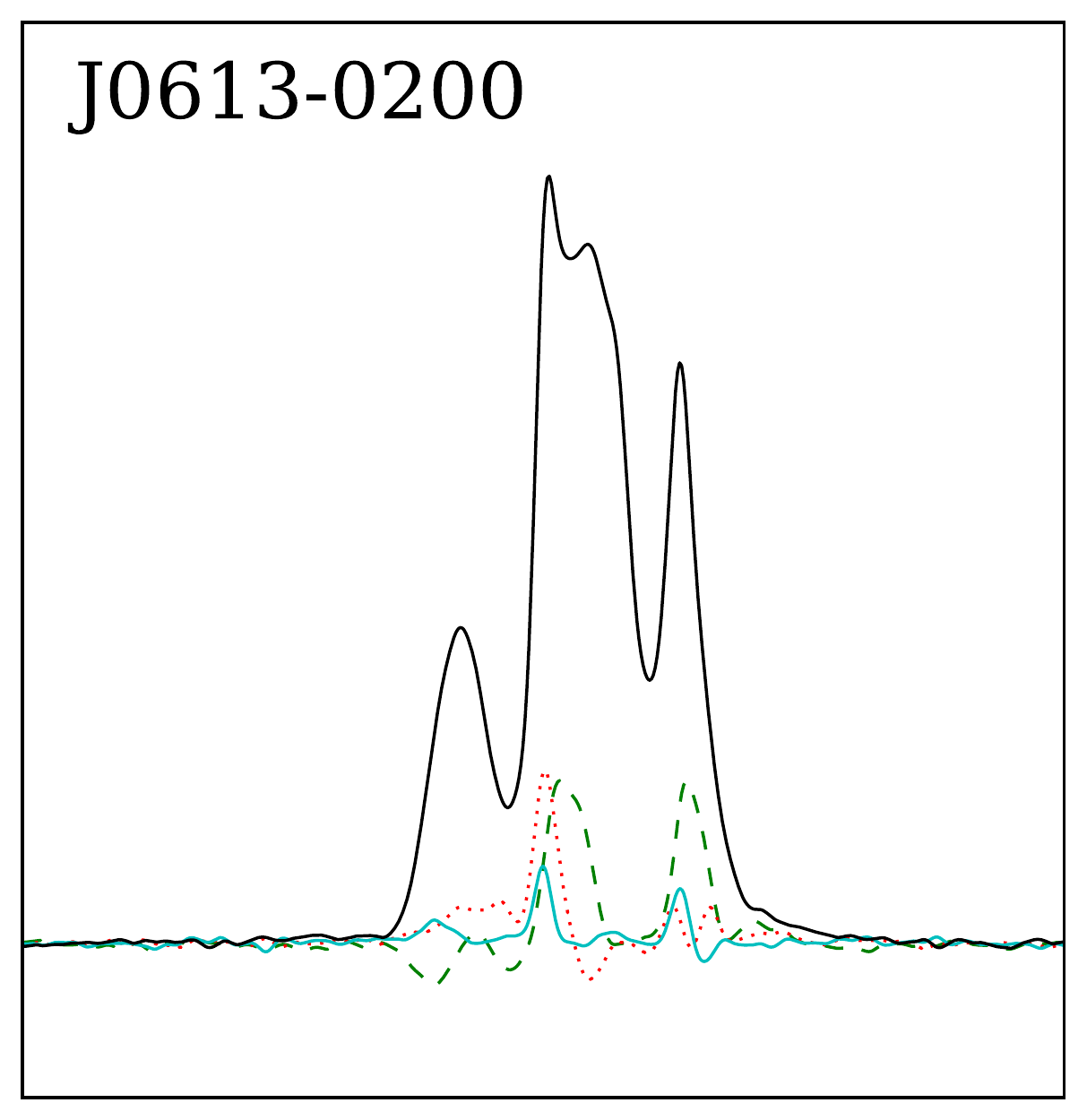} \hfill
\includegraphics[width=.28\linewidth]{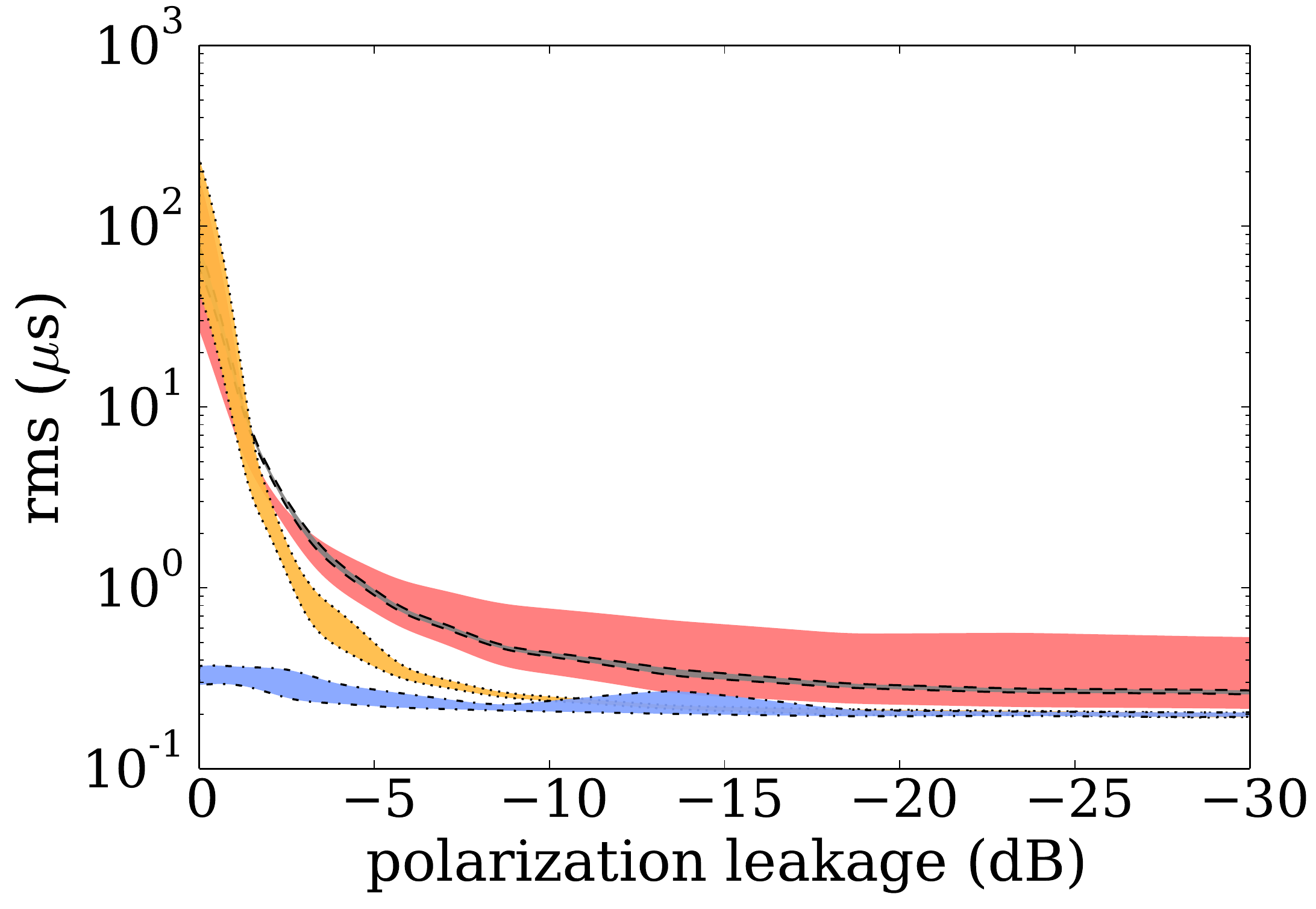} \hfill
\null

\noindent\null\hfill
\includegraphics[width=.2\linewidth]{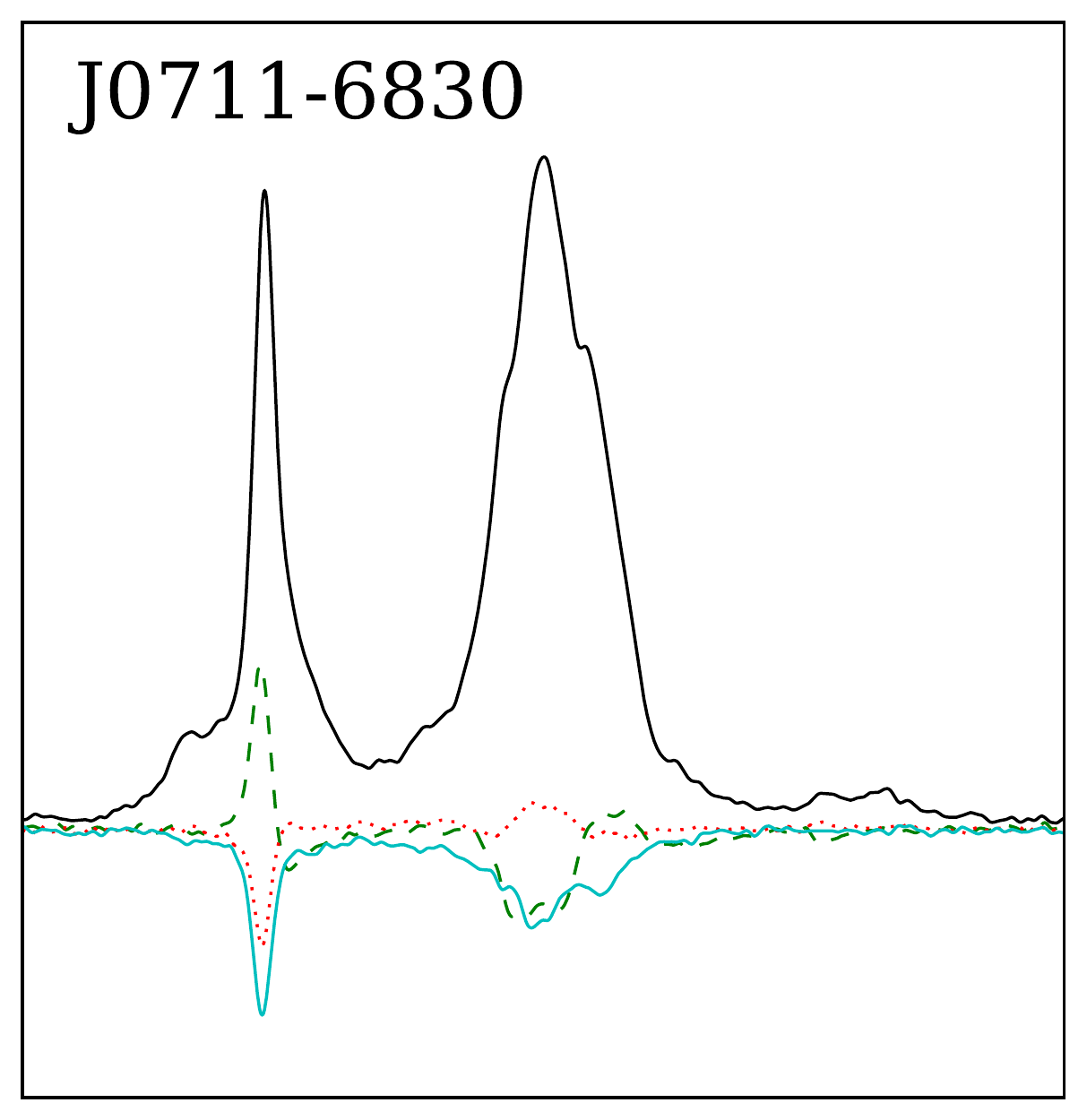} \hfill
\includegraphics[width=.28\linewidth]{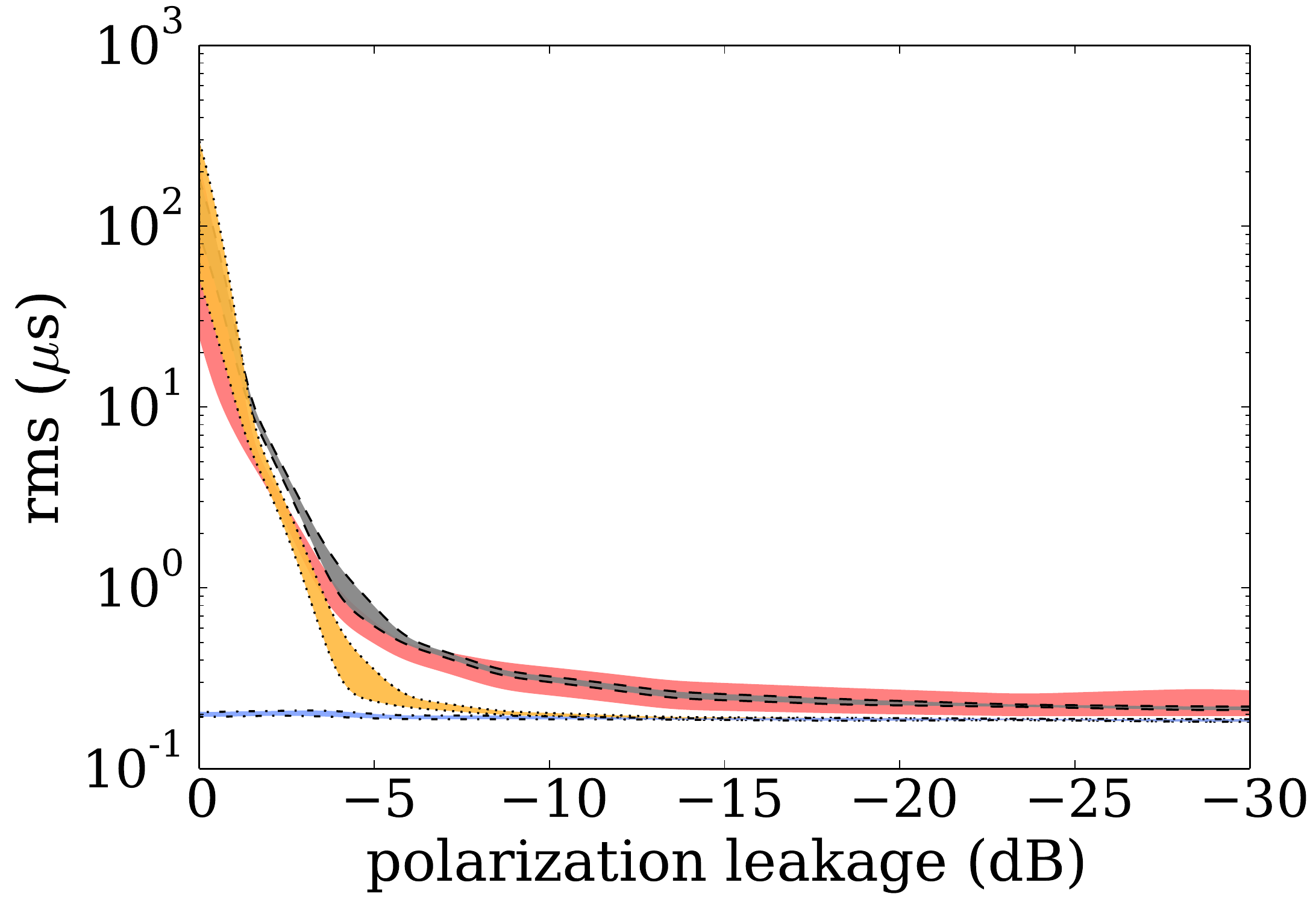} \hfill
\includegraphics[width=.2\linewidth]{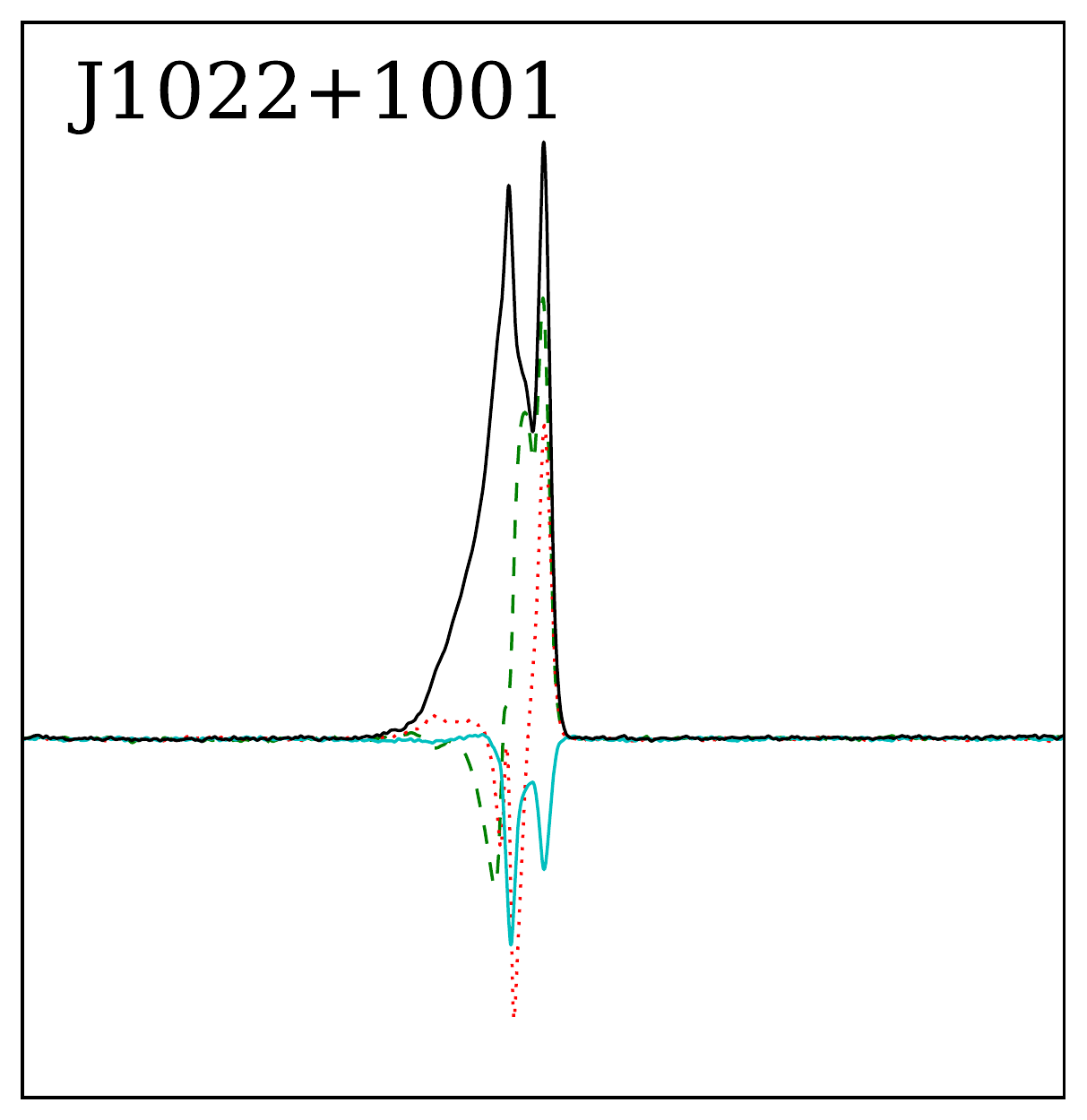} \hfill
\includegraphics[width=.28\linewidth]{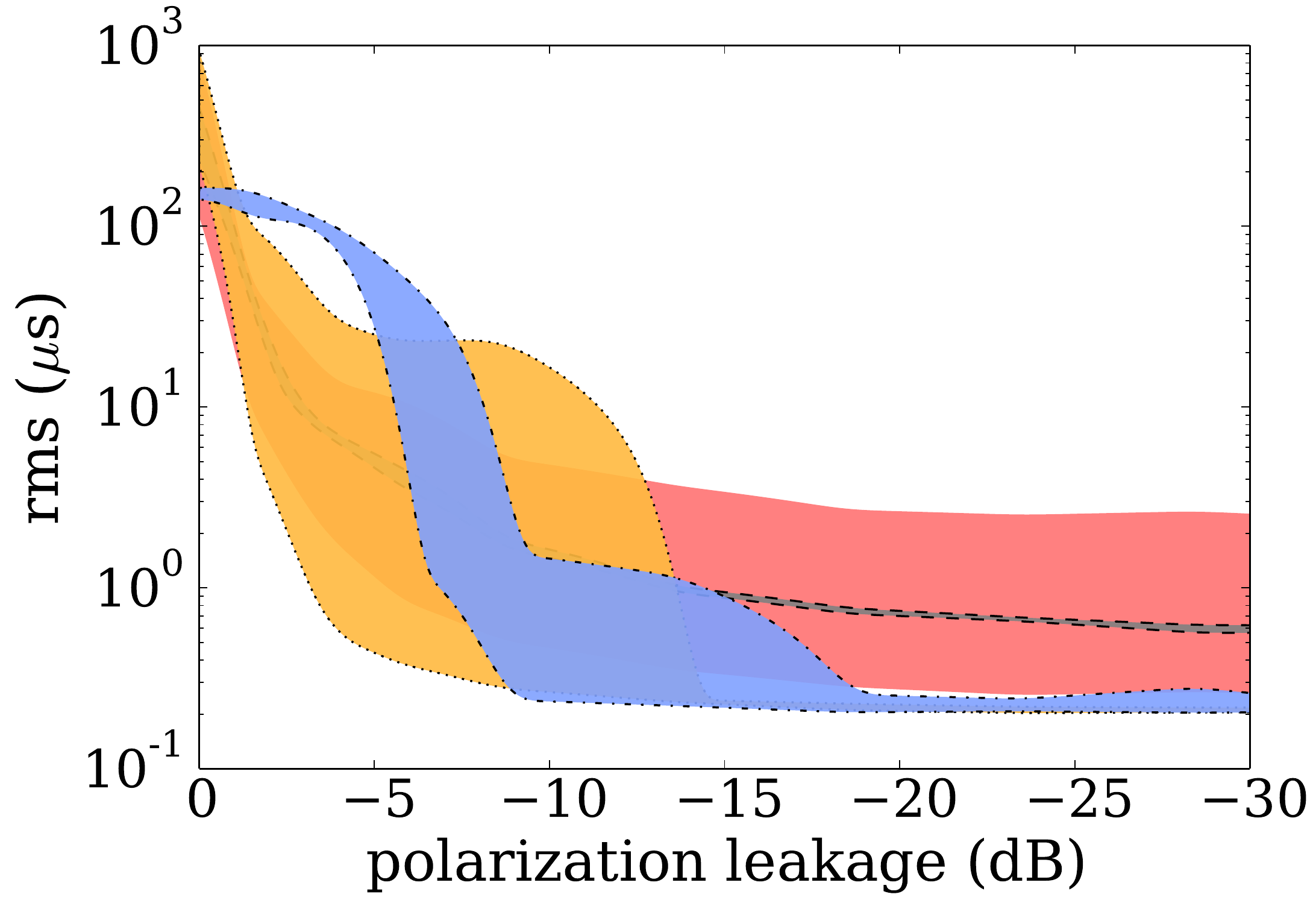} \hfill
\null

\noindent\null\hfill
\includegraphics[width=.2\linewidth]{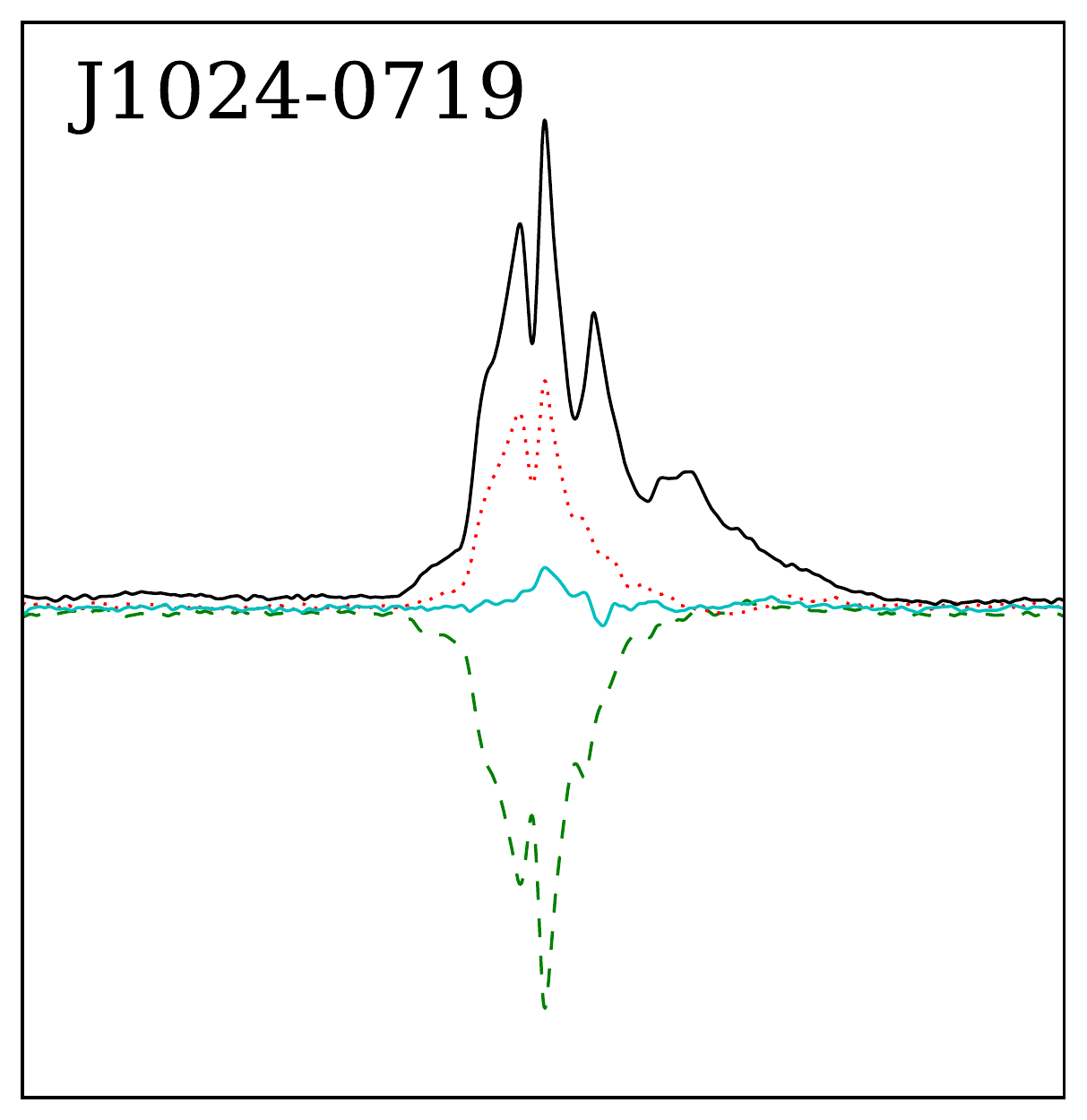} \hfill
\includegraphics[width=.28\linewidth]{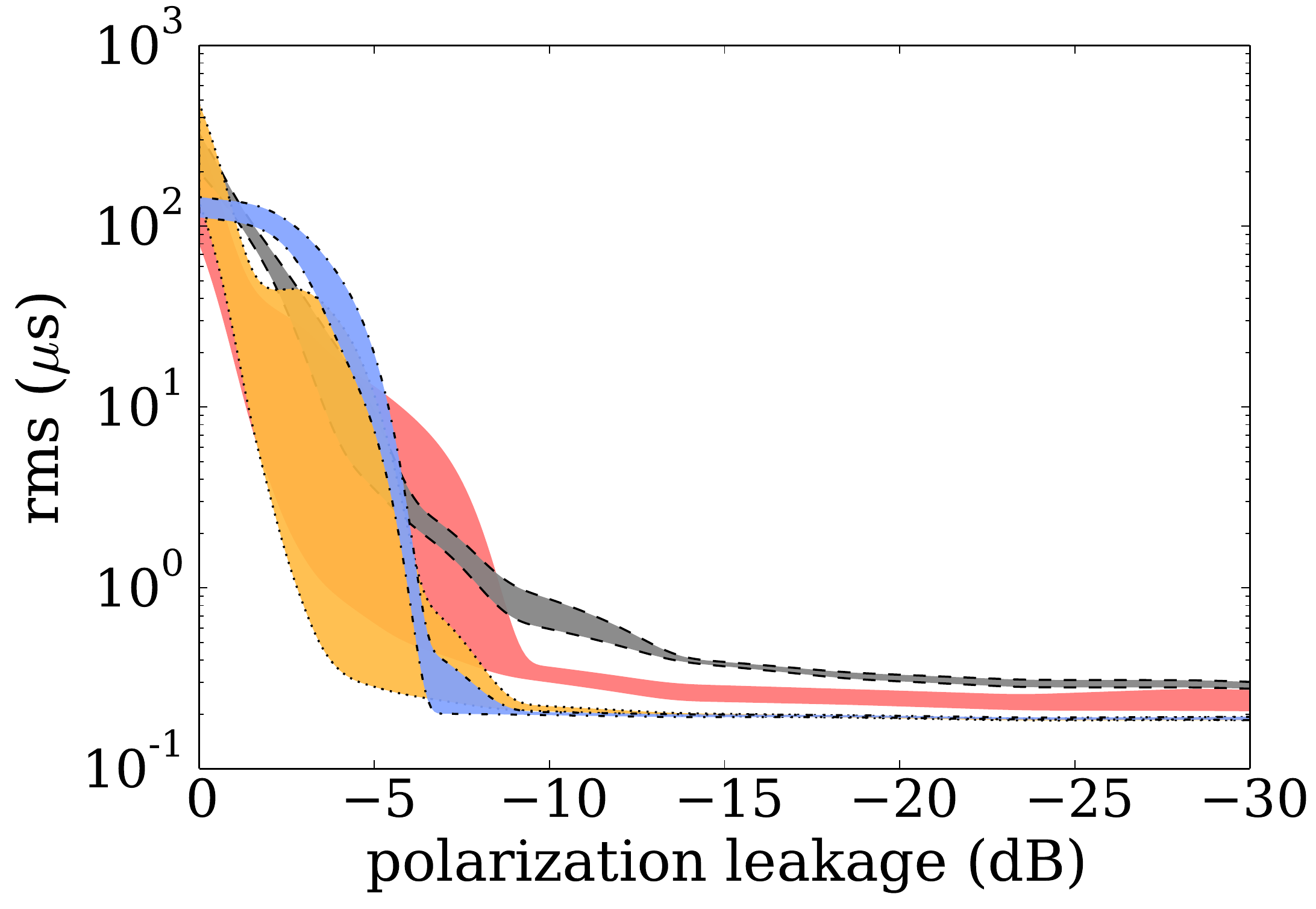} \hfill
\includegraphics[width=.2\linewidth]{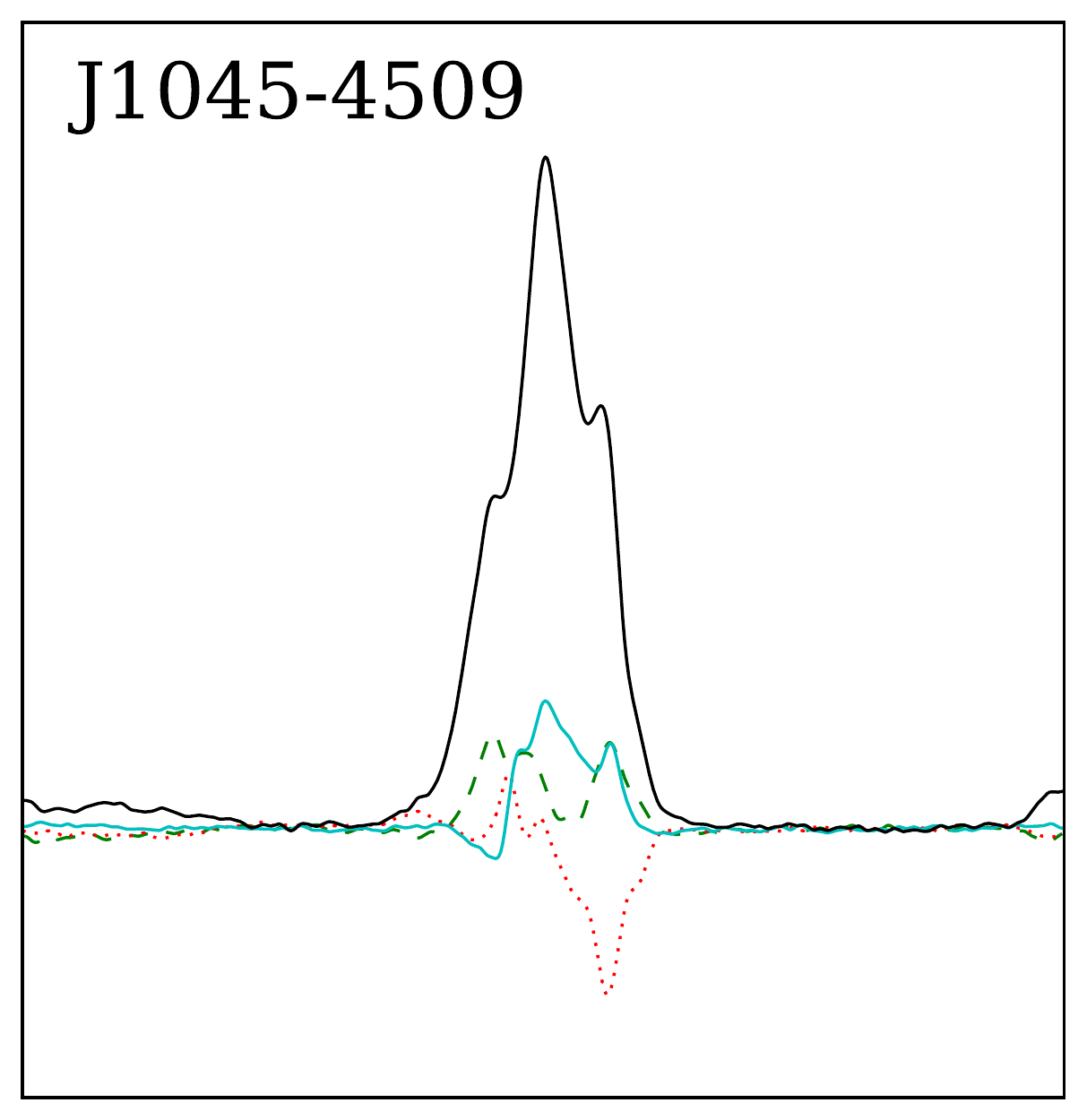} \hfill
\includegraphics[width=.28\linewidth]{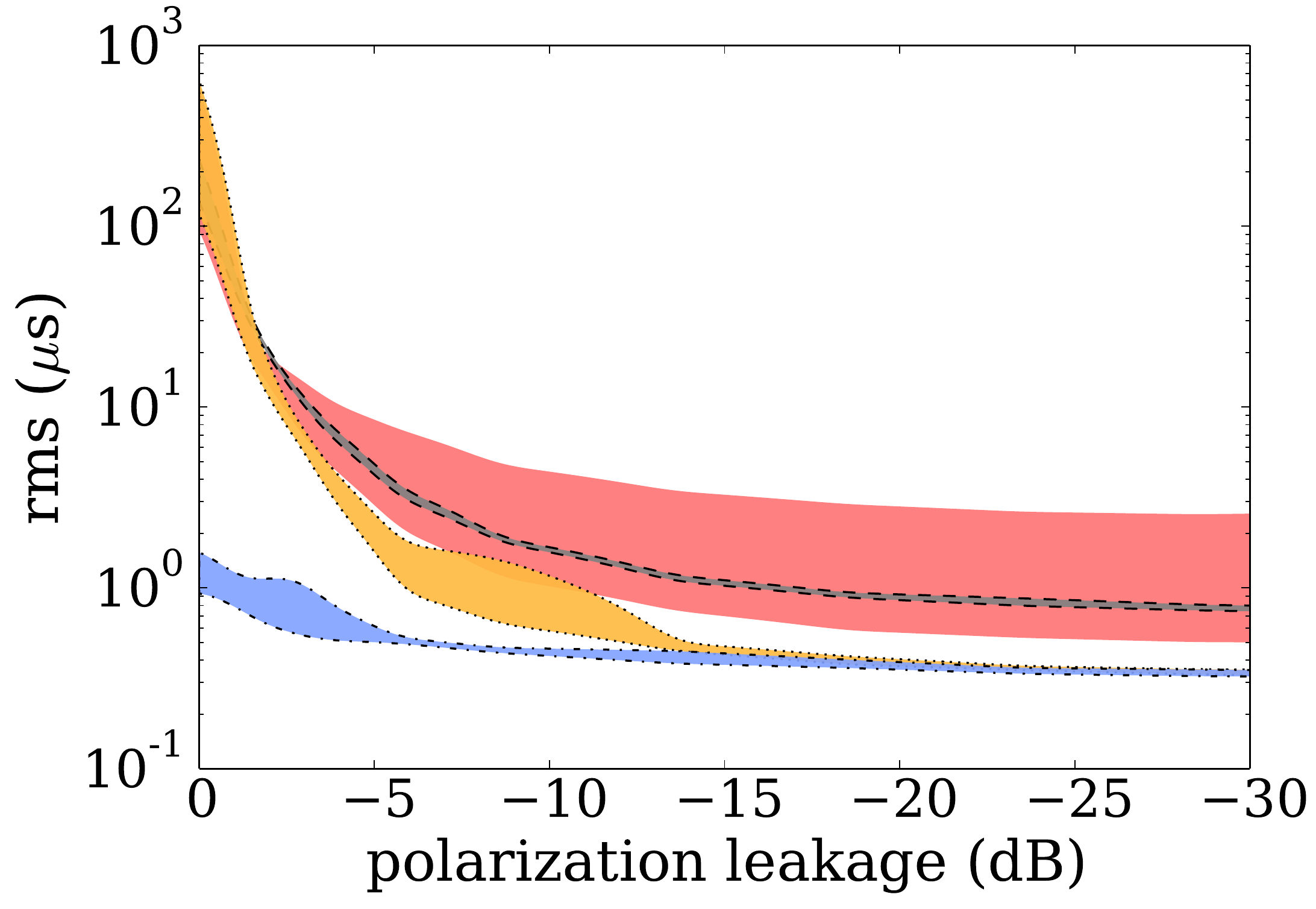} \hfill
\null

\noindent\null\hfill
\includegraphics[width=.2\linewidth]{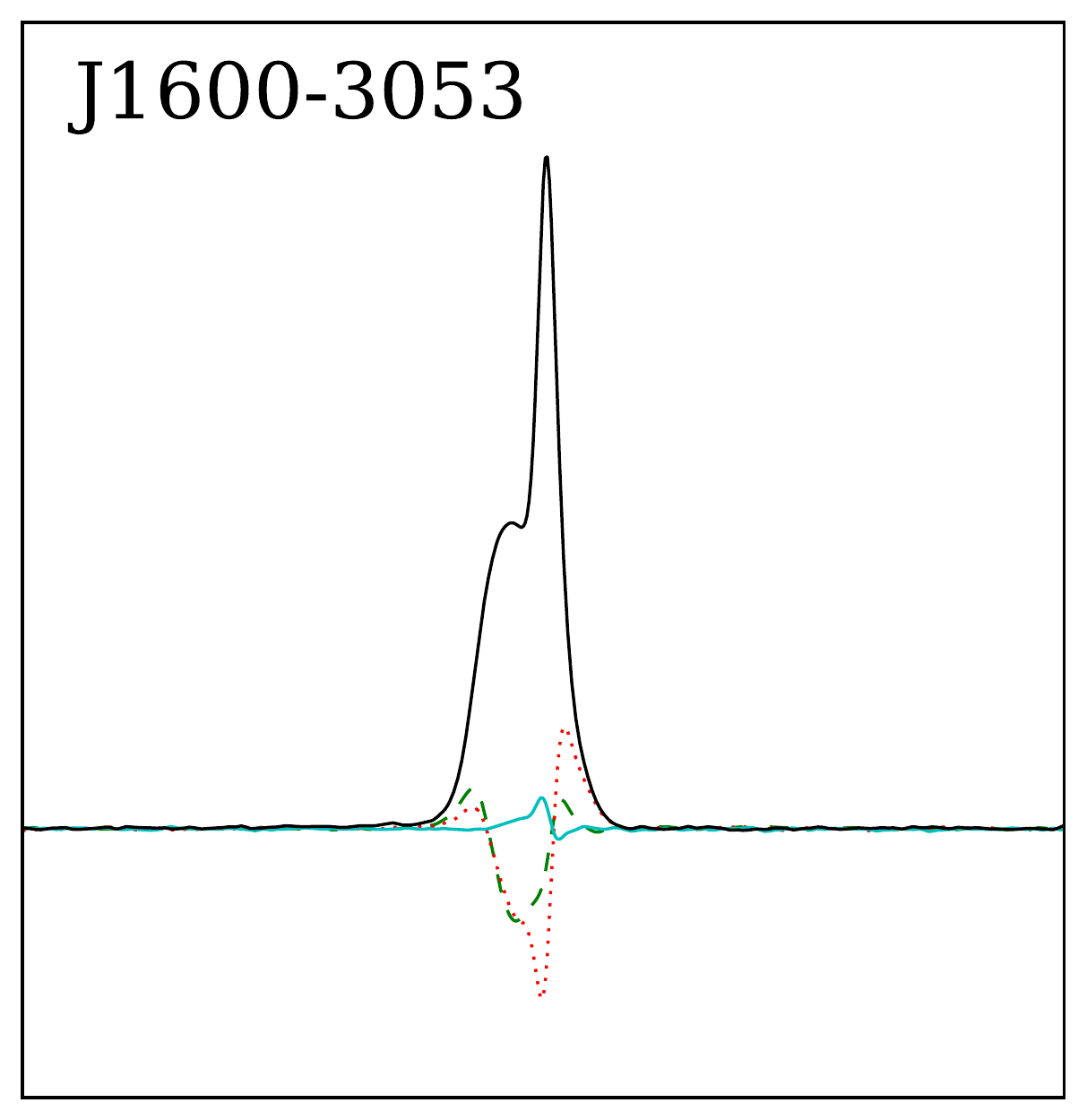} \hfill
\includegraphics[width=.28\linewidth]{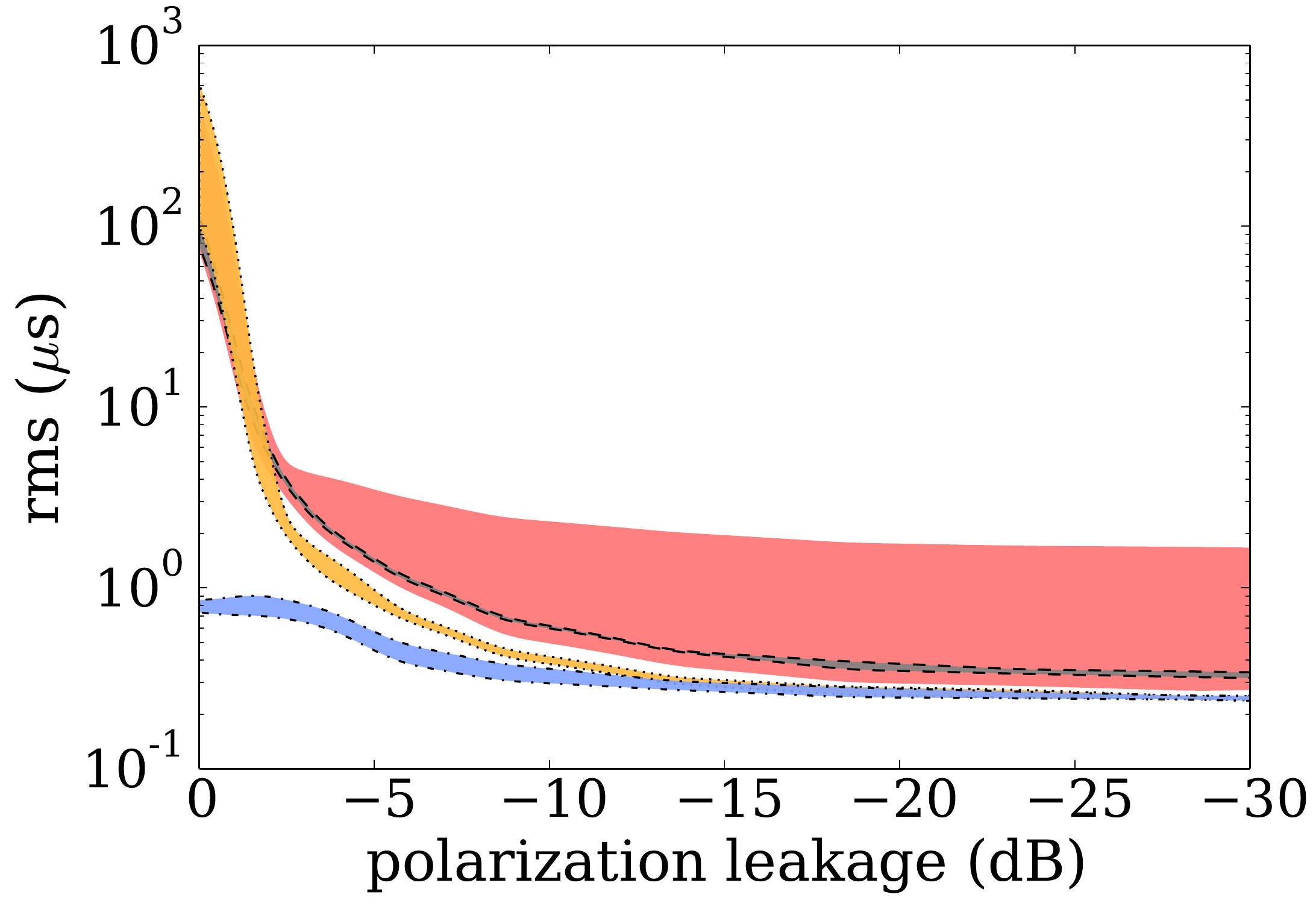} \hfill
\includegraphics[width=.2\linewidth]{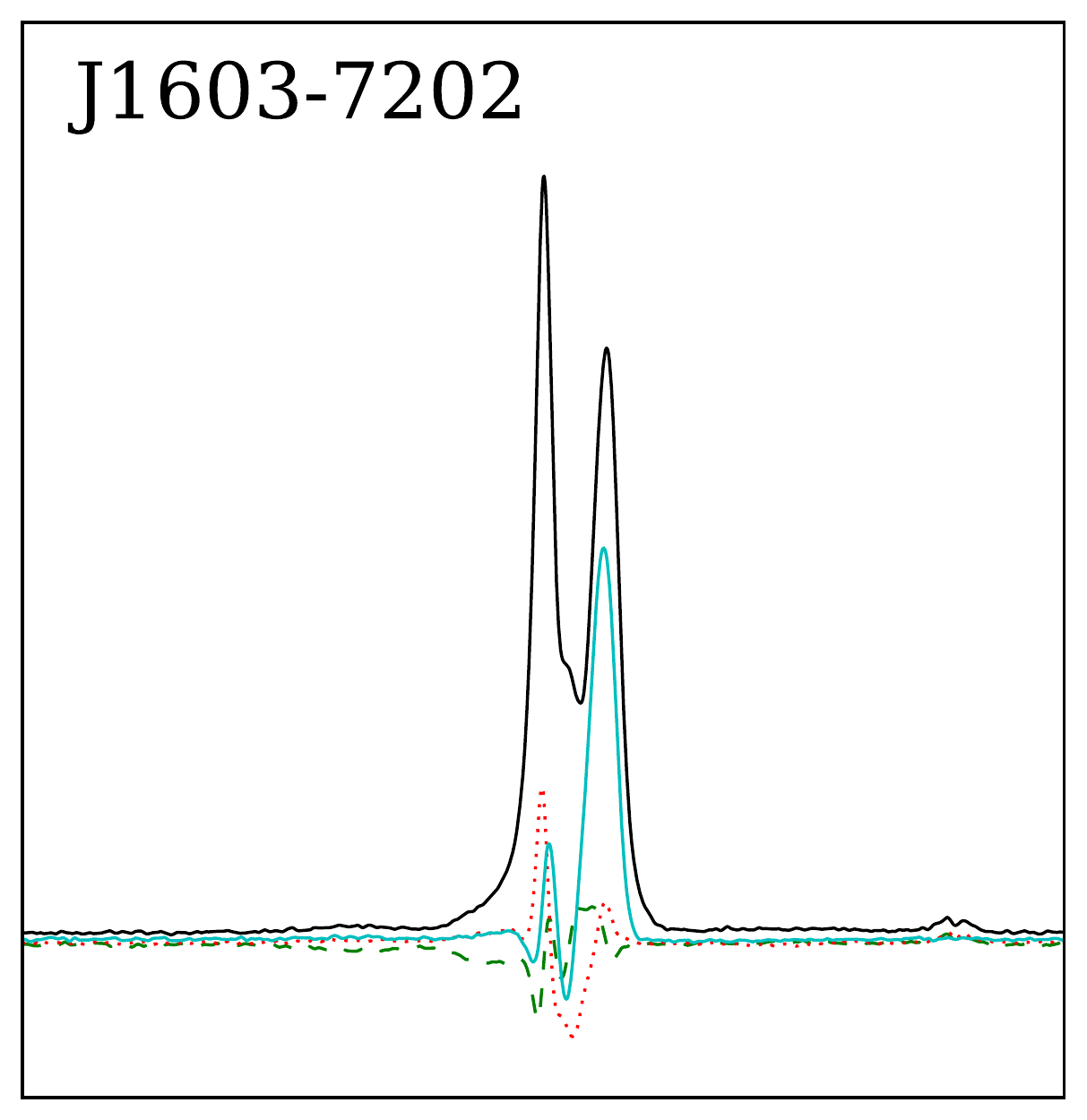} \hfill
\includegraphics[width=.28\linewidth]{graphics/fill/fill_rms_ixr_snr100_J1603.pdf} \hfill
\null

\noindent\null\hfill
\includegraphics[width=.2\linewidth]{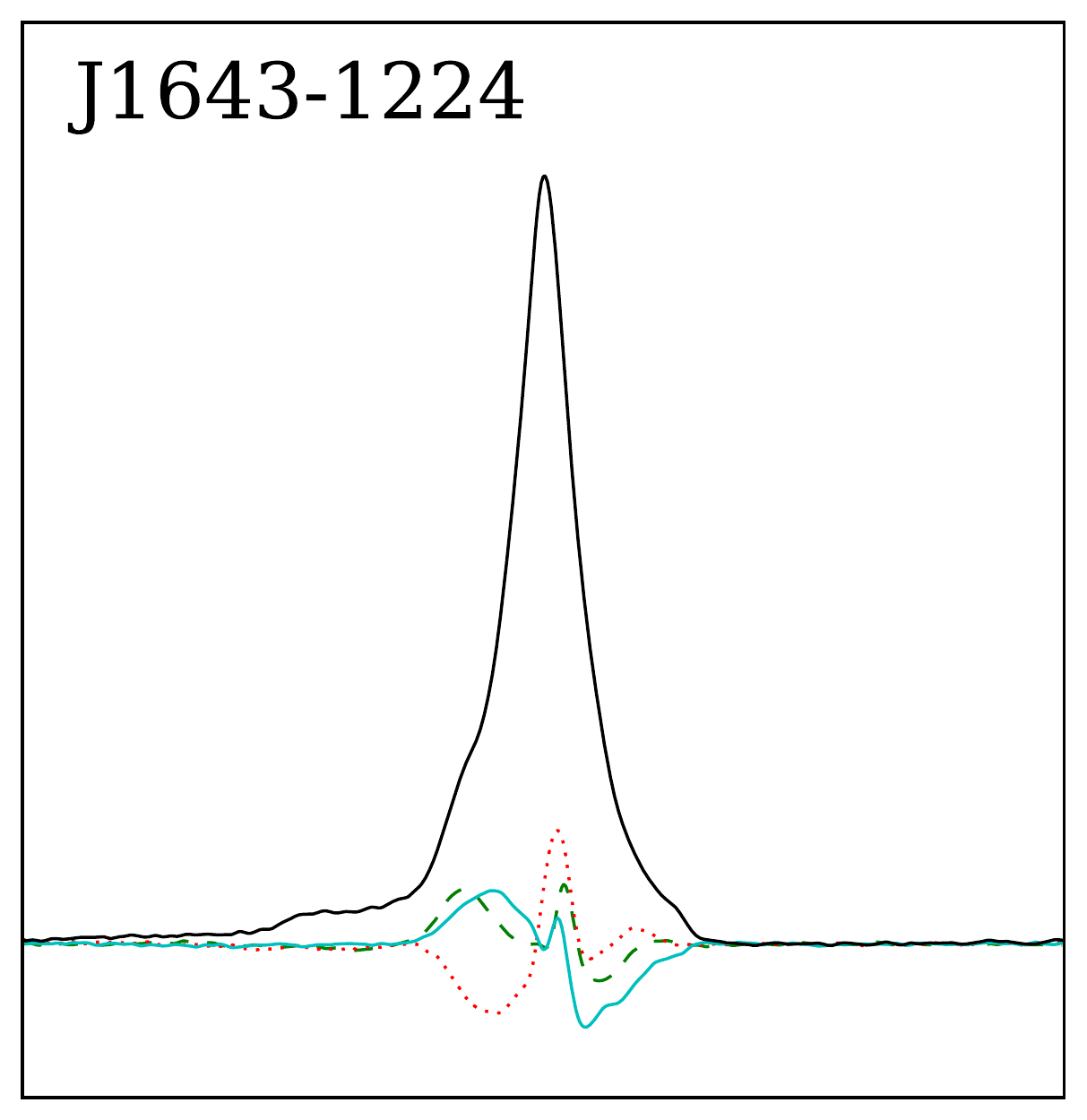} \hfill
\includegraphics[width=.28\linewidth]{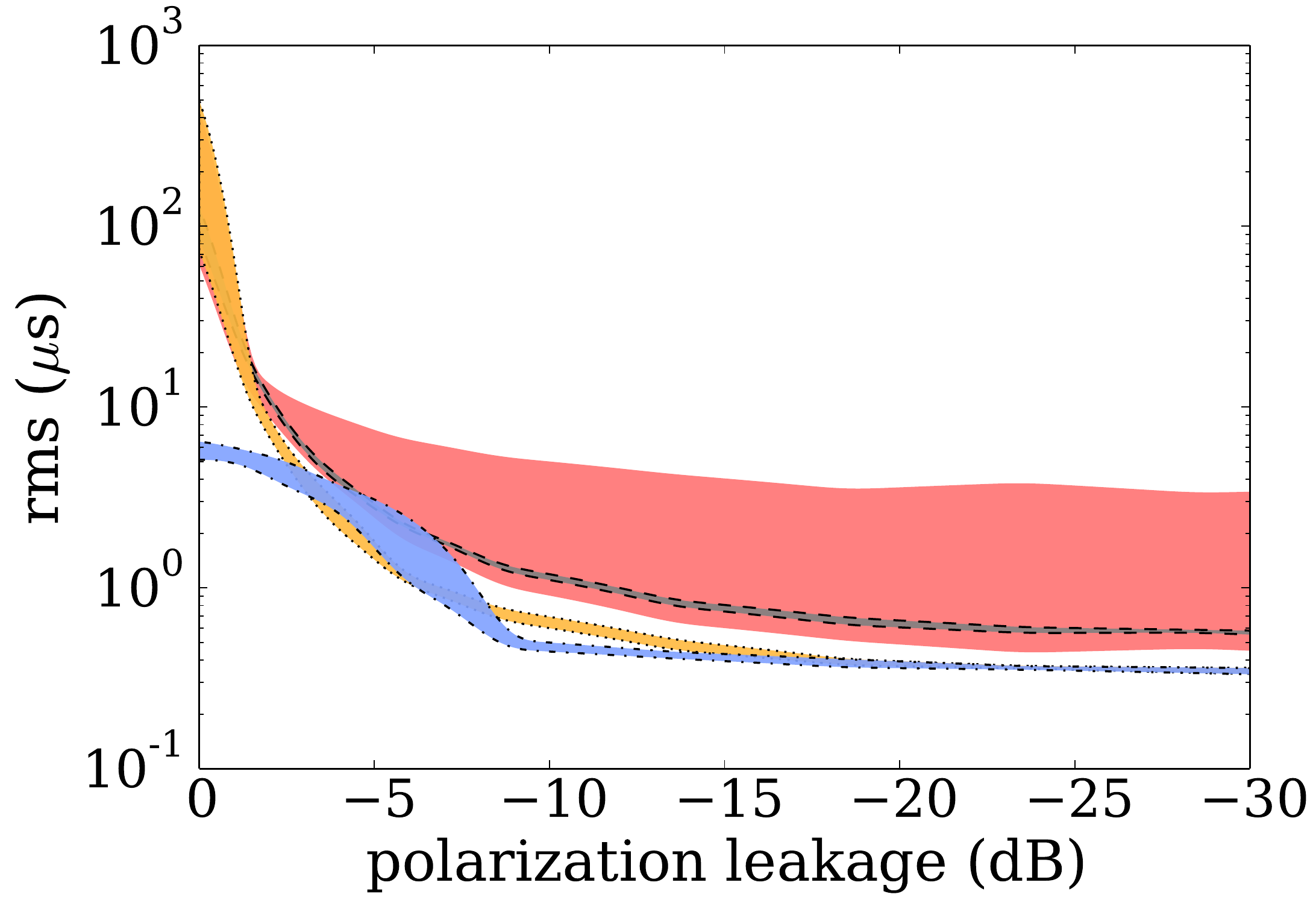} \hfill
\includegraphics[width=.2\linewidth]{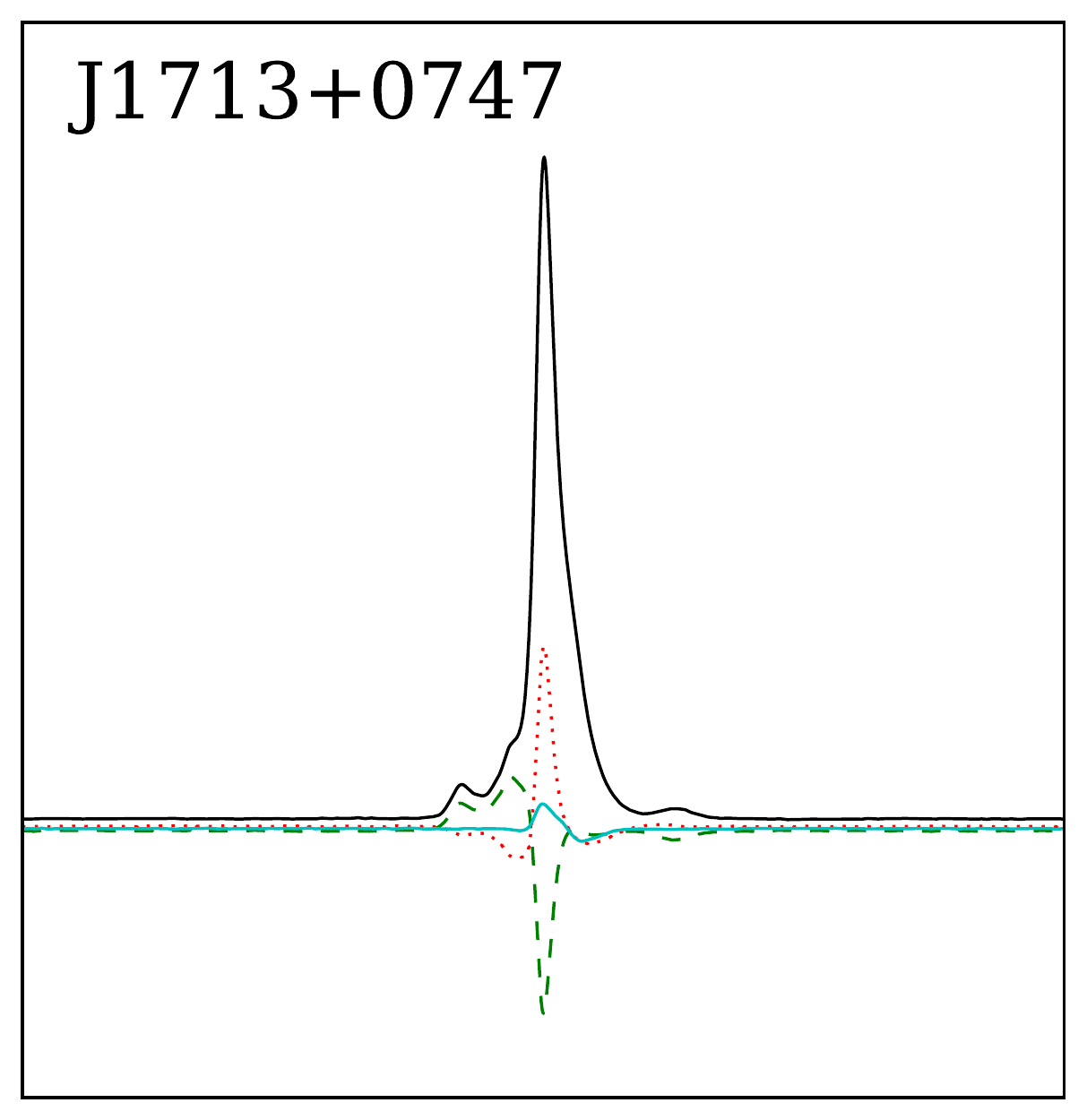} \hfill
\includegraphics[width=.28\linewidth]{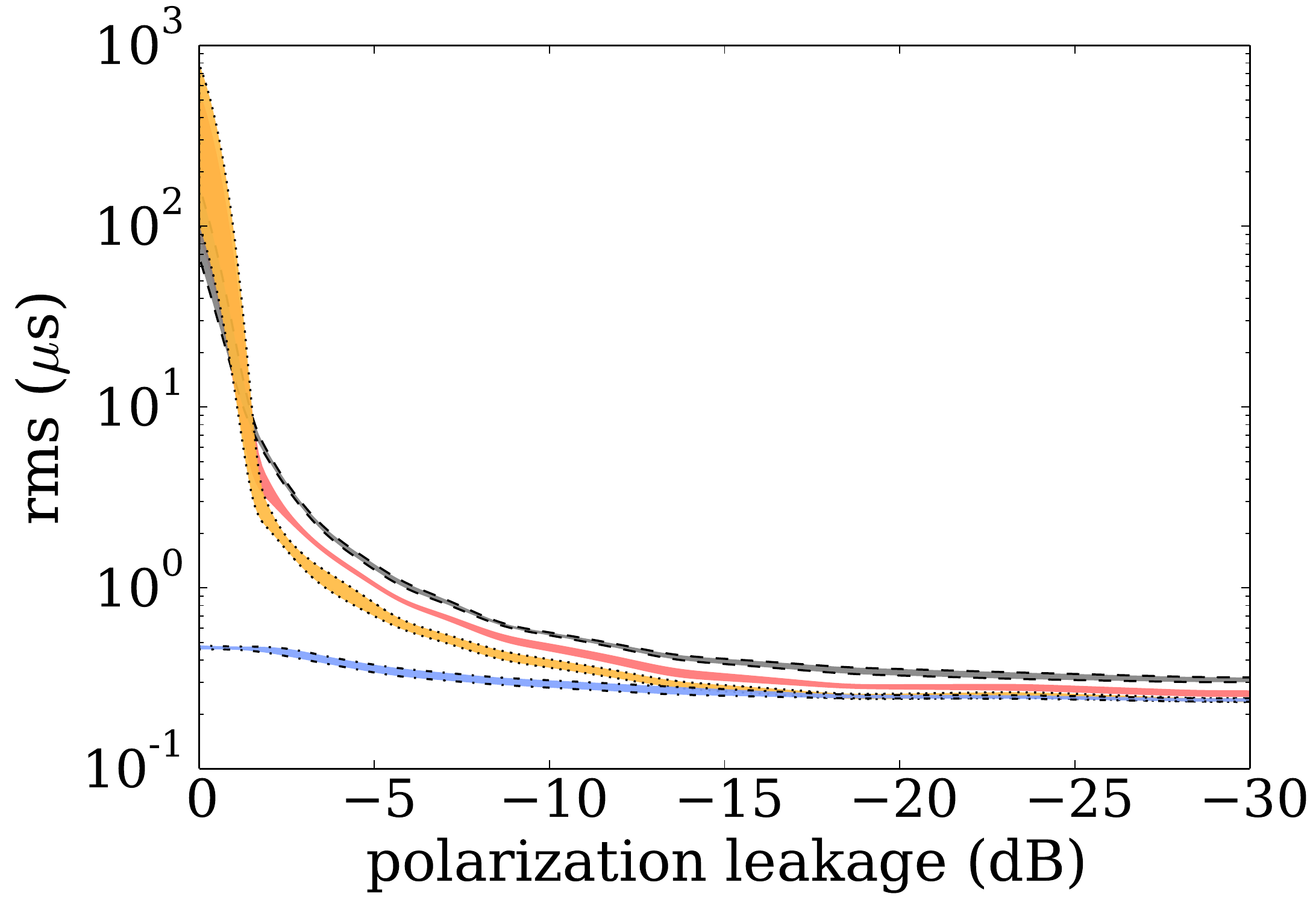} \hfill
\null

\caption{
    Time of arrival rms noise from simulation of various MSPs. The first and third
    columns are the Stokes parameters of the ideal profile: I (black/solid), Q
    (green/dashed), U (red/dotted), and V (cyan/solid). The second and fourth columns
    show ToA rms noise ($\mu$s) as a function of intrinsic polarization leakage when
    using total intensity (red/no border), invariant interval (gray/dashed border), and matrix template
    matching (full calibration: orange/dotted border, gain calibration: blue/dot-dash border) methods on the profile.
    The width of the lines show the effect polarization calibration error has on the
    rms, a polarization calibration error from 0\% to 15\% was used. These simulations
    use a fractional integration time of $0.01$ which would produce an ideal signal-to-noise
    ratio of 100. Plots continue in Figure \ref{fig:psr_plots1}.
    }
\label{fig:psr_plots0}

\end{figure*}

\begin{figure*}
\strut

\noindent\null\hfill
\includegraphics[width=.2\linewidth]{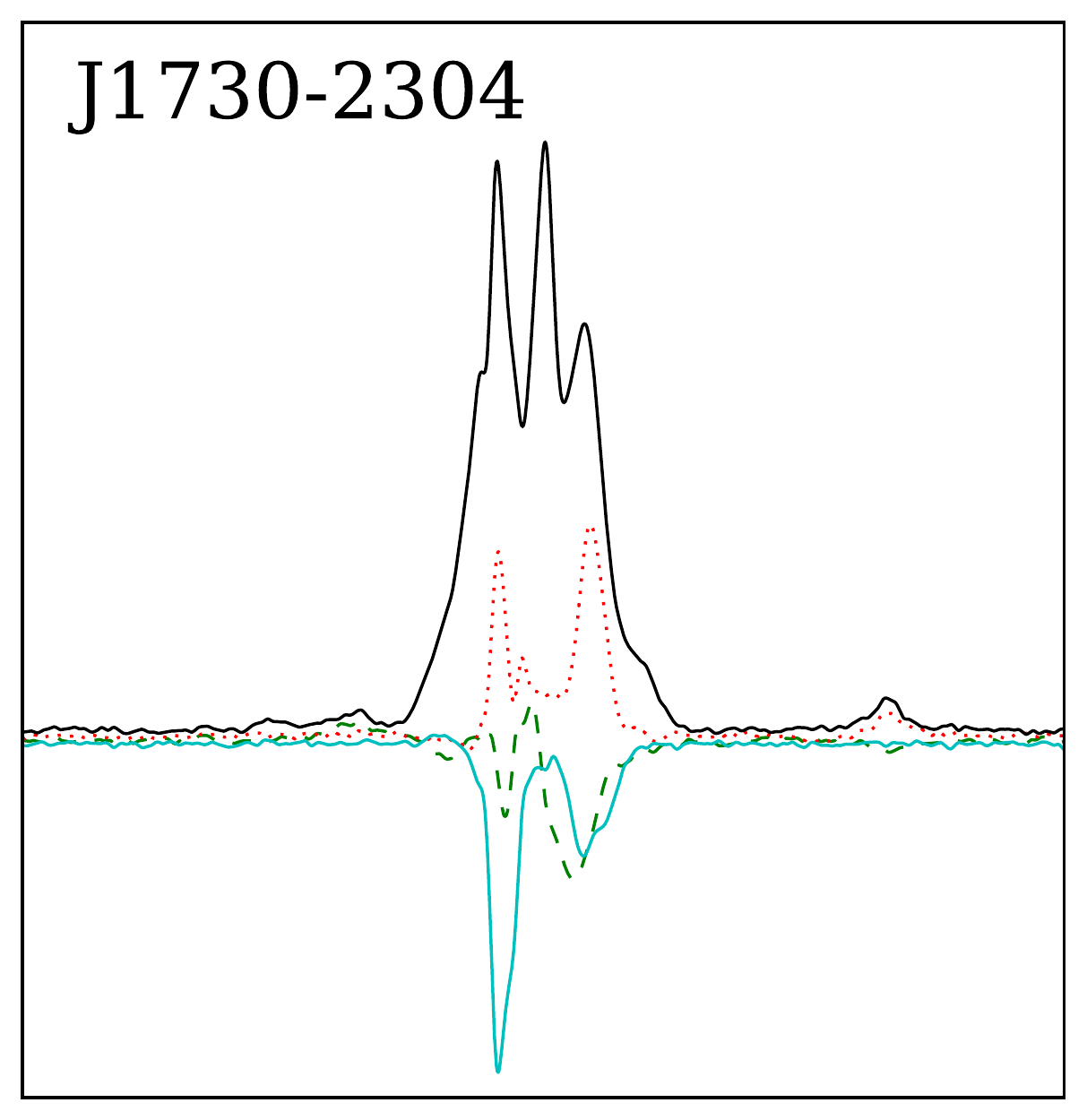} \hfill
\includegraphics[width=.28\linewidth]{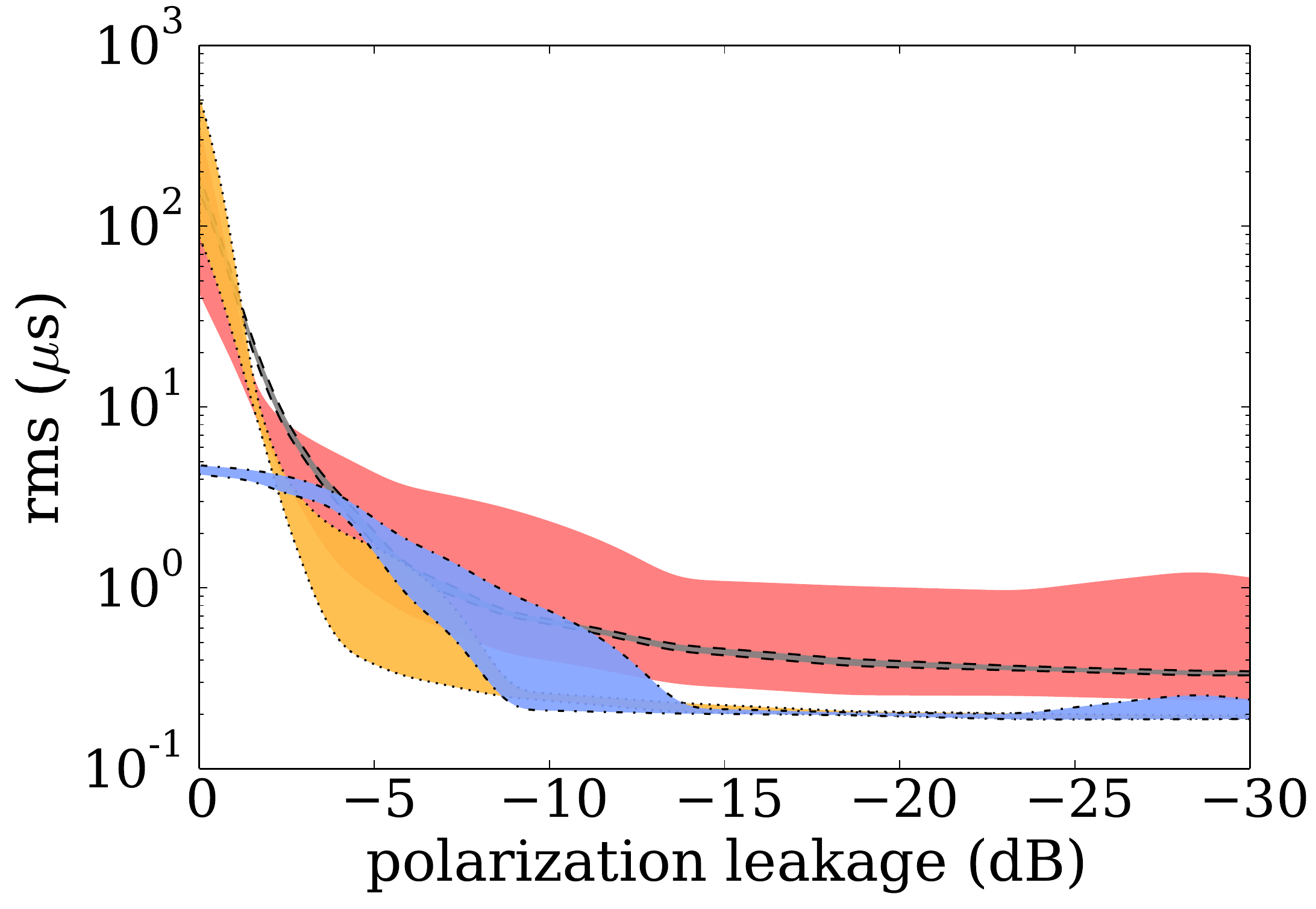} \hfill
\includegraphics[width=.2\linewidth]{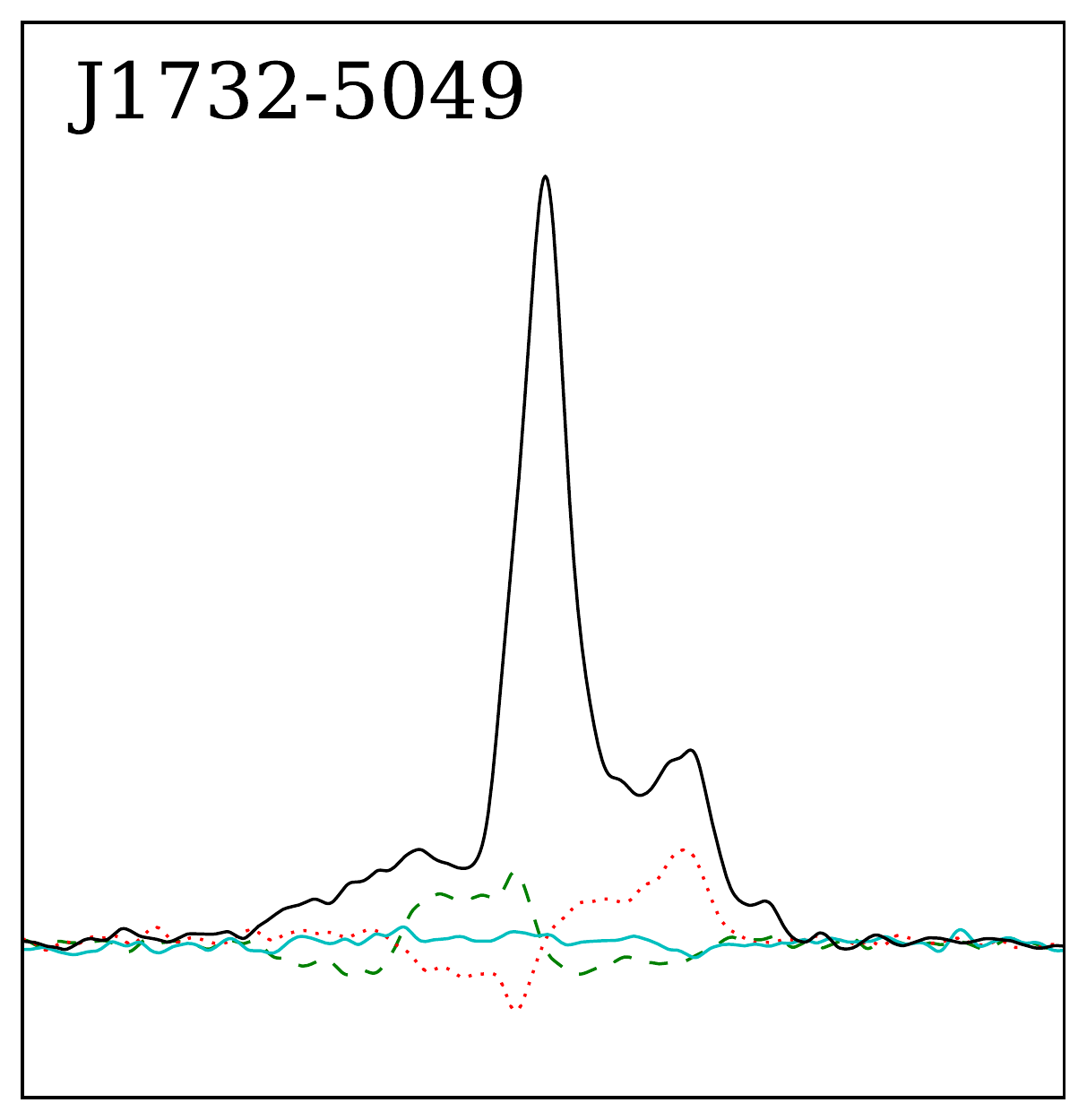} \hfill
\includegraphics[width=.28\linewidth]{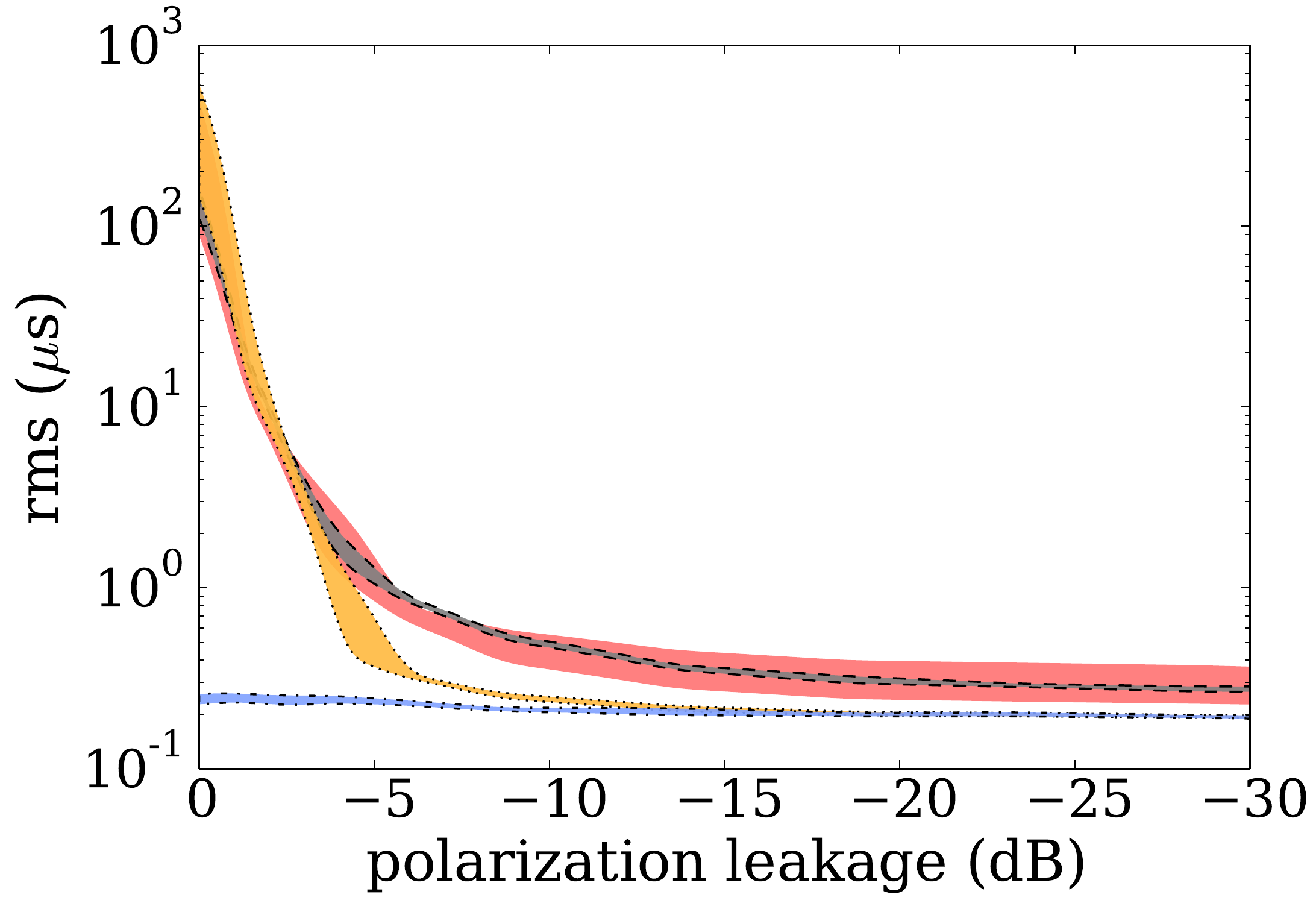} \hfill
\null

\noindent\null\hfill
\includegraphics[width=.2\linewidth]{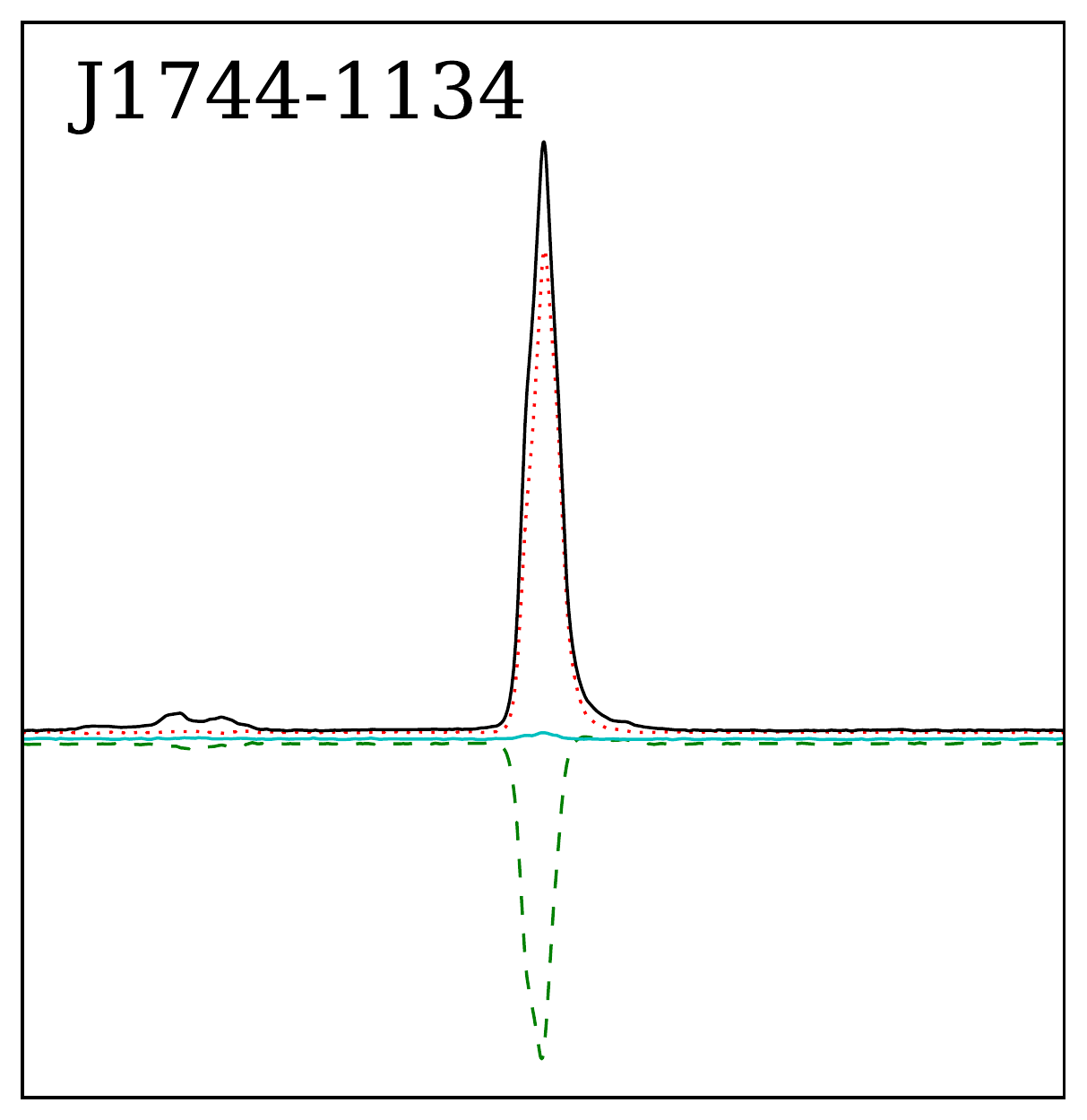} \hfill
\includegraphics[width=.28\linewidth]{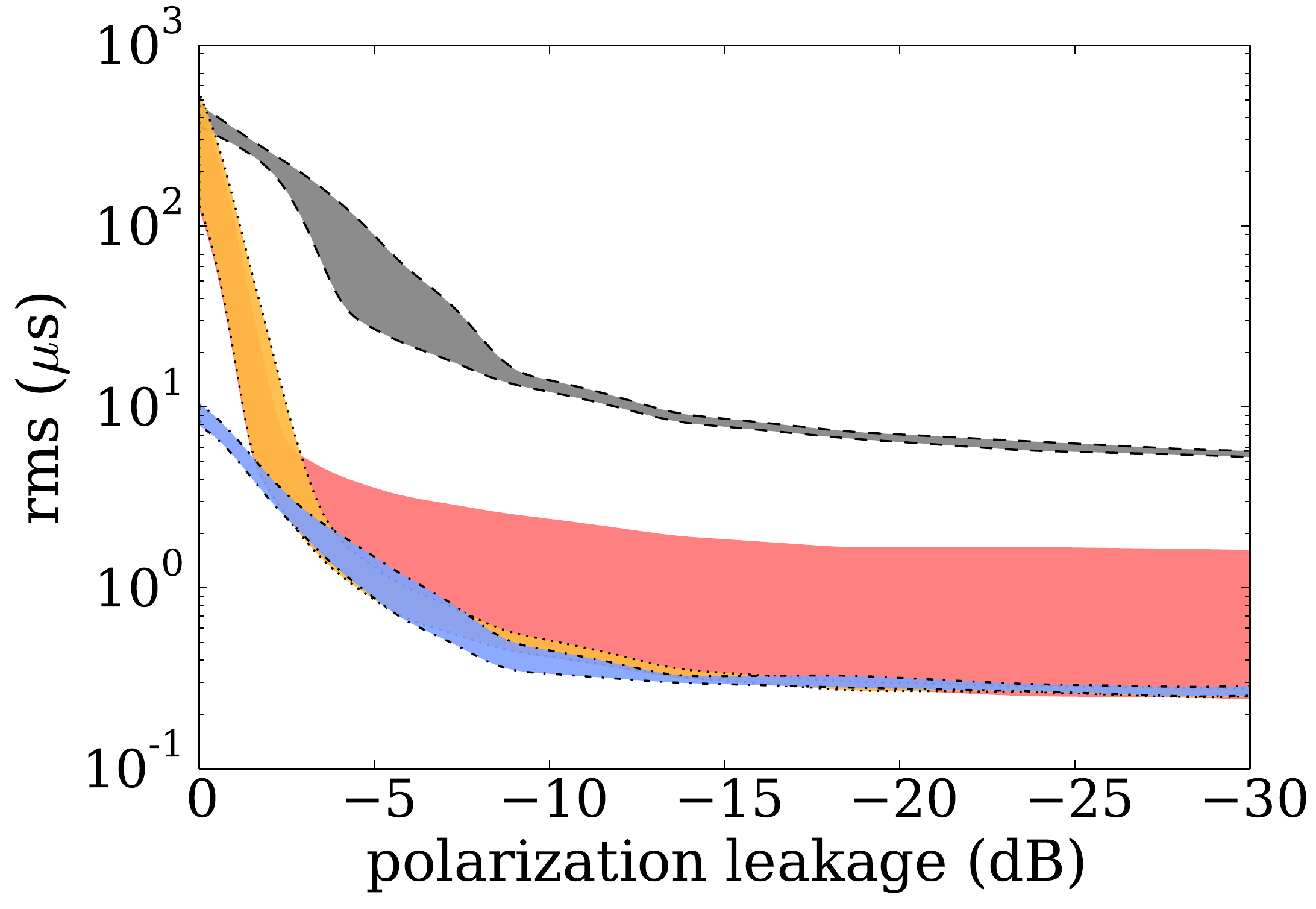} \hfill
\includegraphics[width=.2\linewidth]{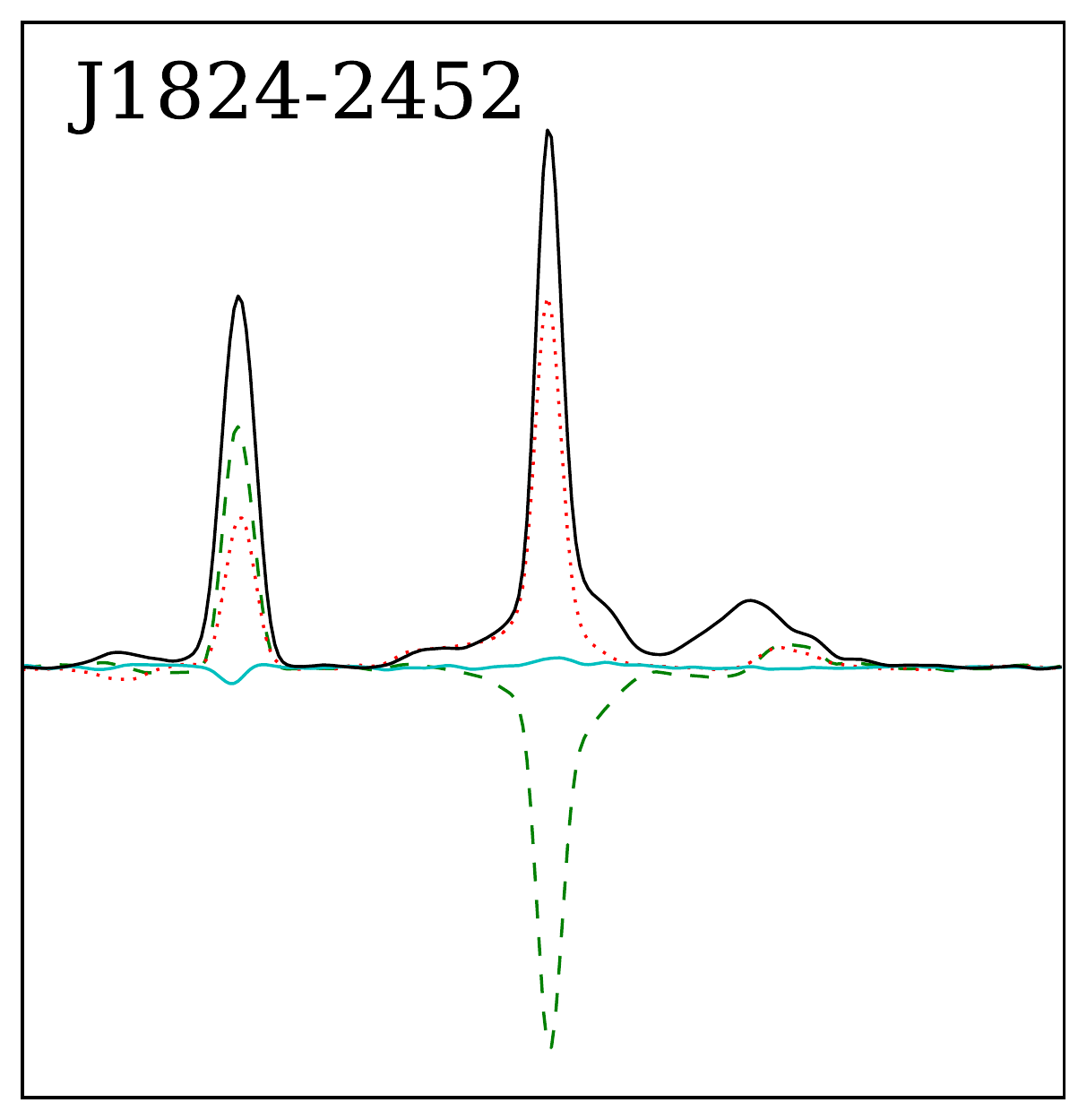} \hfill
\includegraphics[width=.28\linewidth]{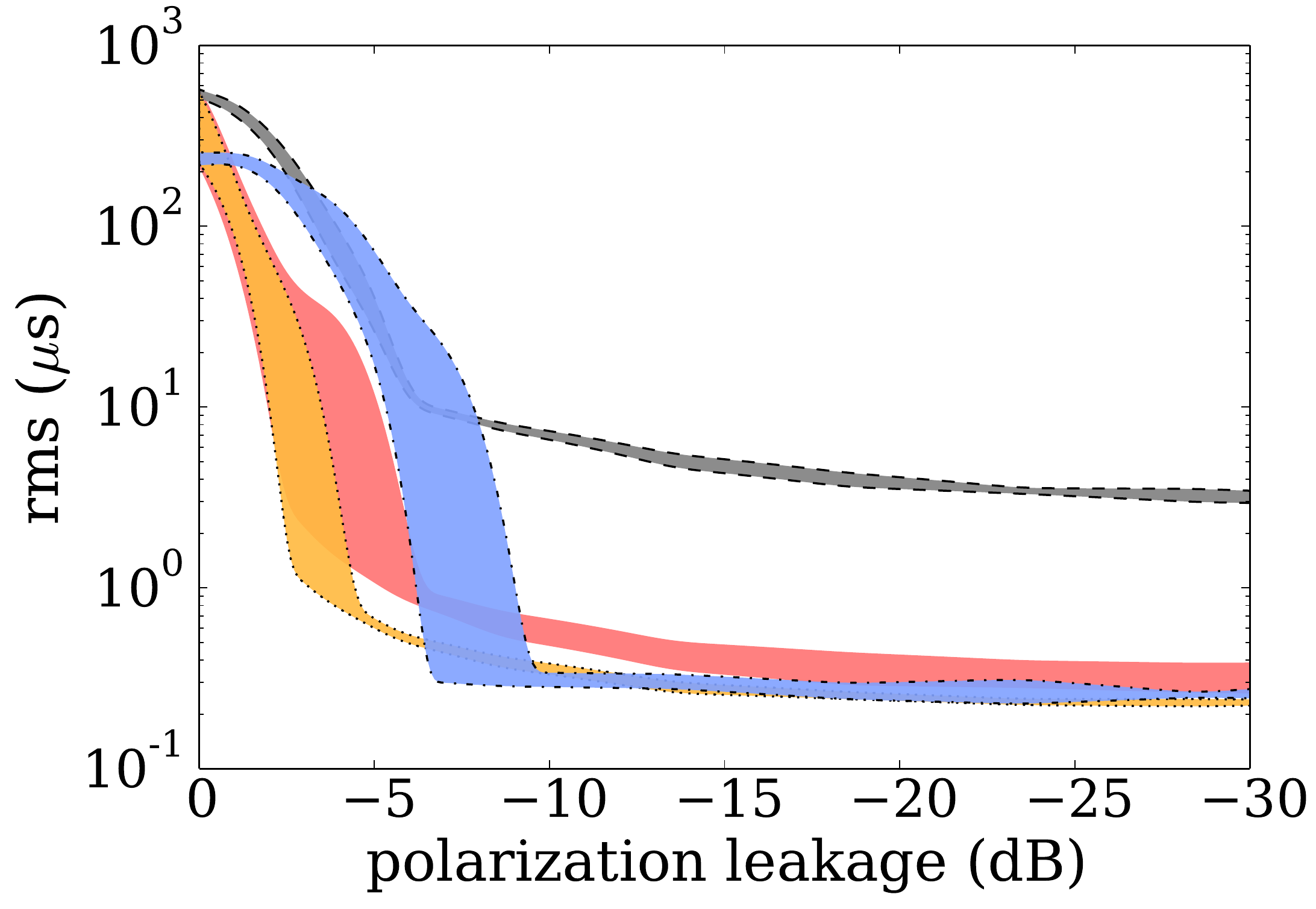} \hfill
\null

\noindent\null\hfill
\includegraphics[width=.2\linewidth]{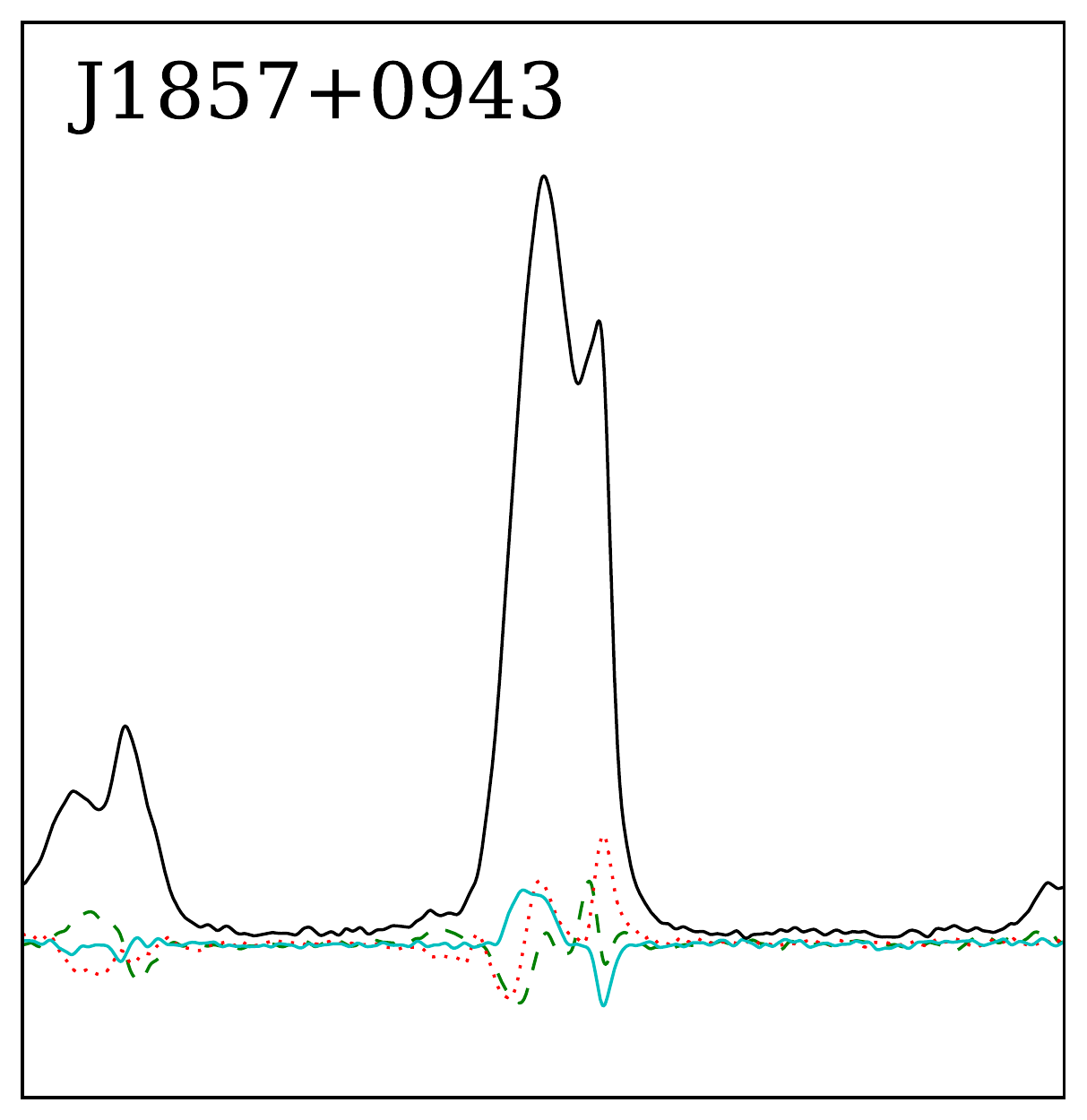} \hfill
\includegraphics[width=.28\linewidth]{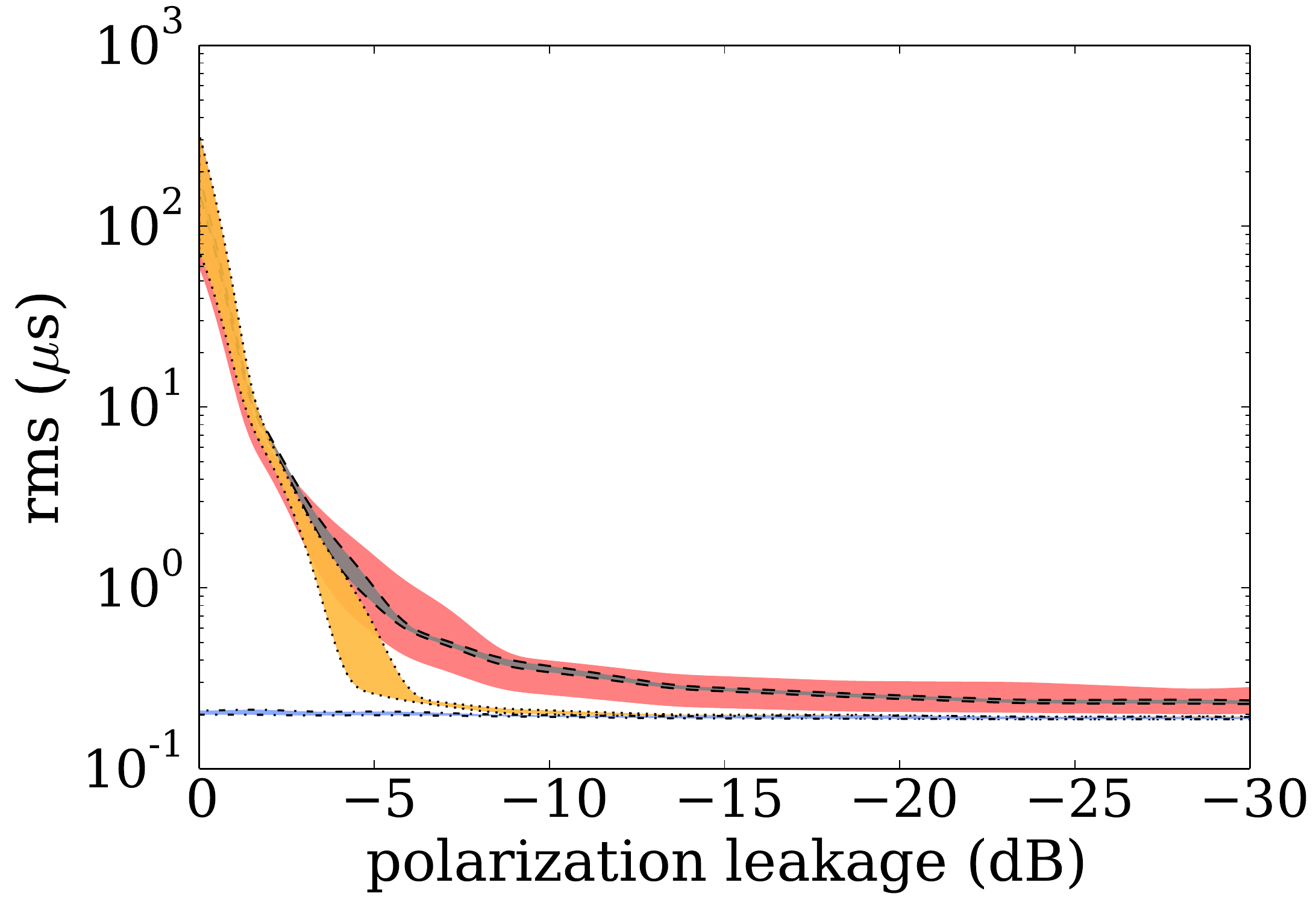} \hfill
\includegraphics[width=.2\linewidth]{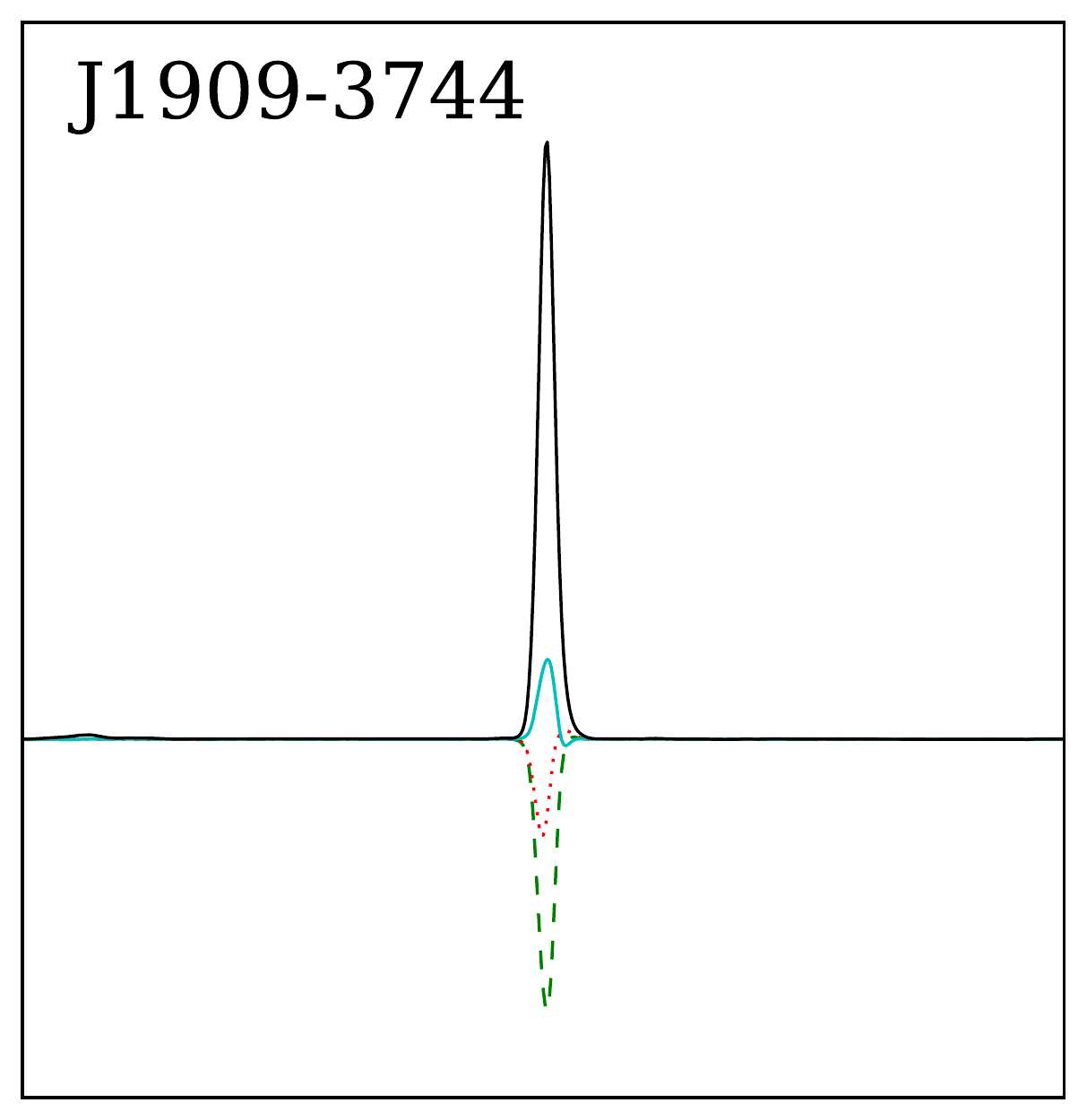} \hfill
\includegraphics[width=.28\linewidth]{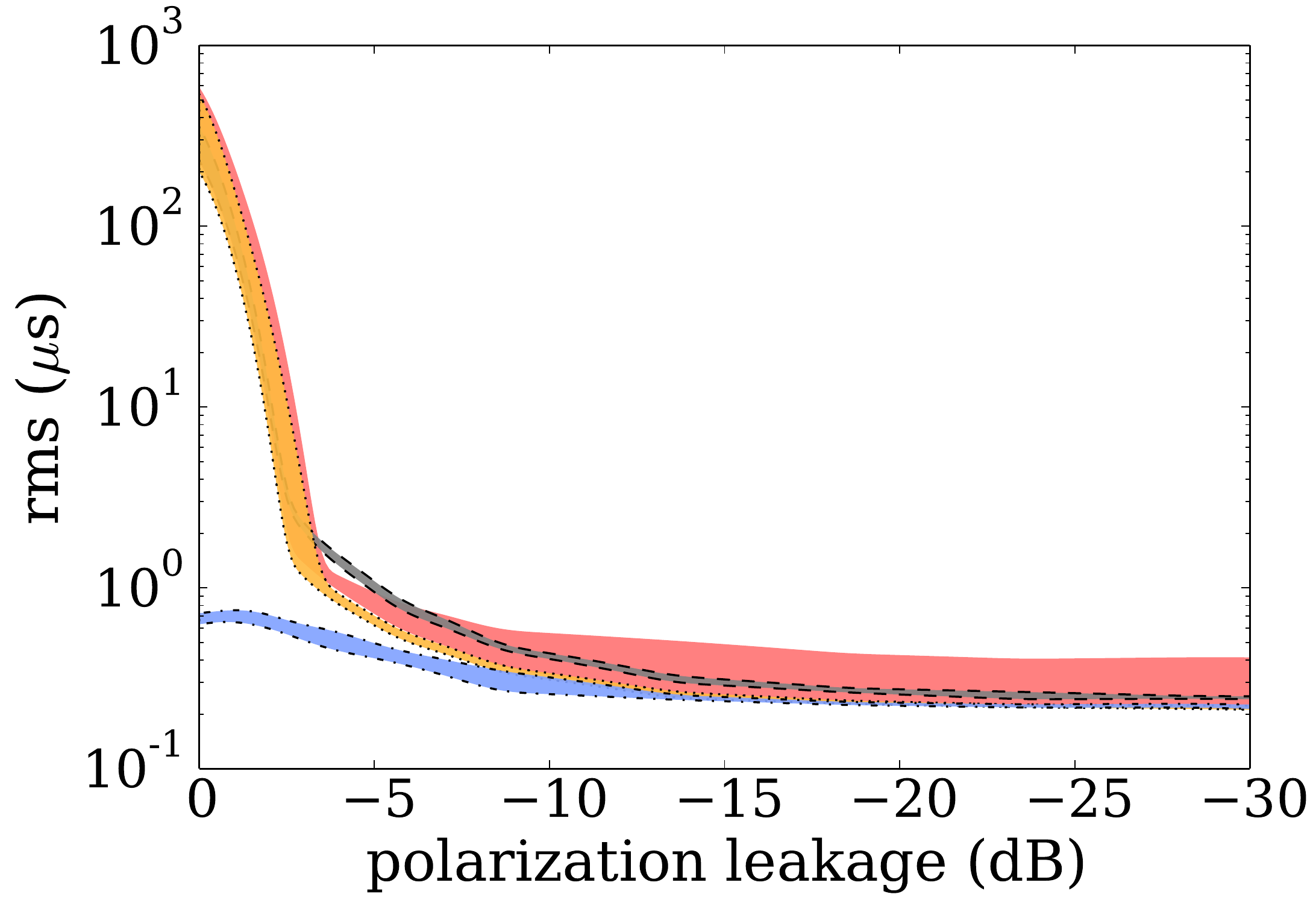} \hfill
\null

\noindent\null\hfill
\includegraphics[width=.2\linewidth]{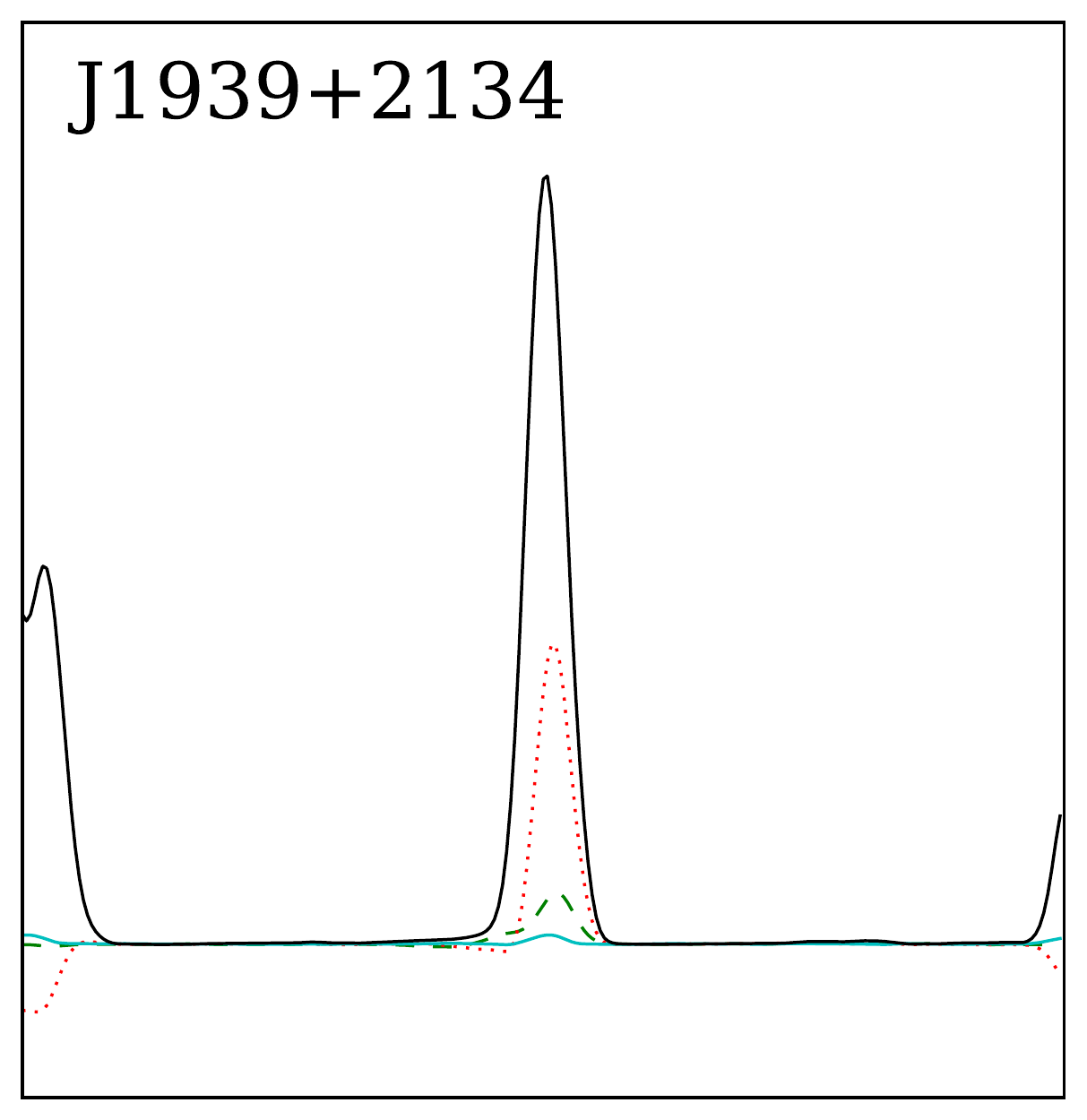} \hfill
\includegraphics[width=.28\linewidth]{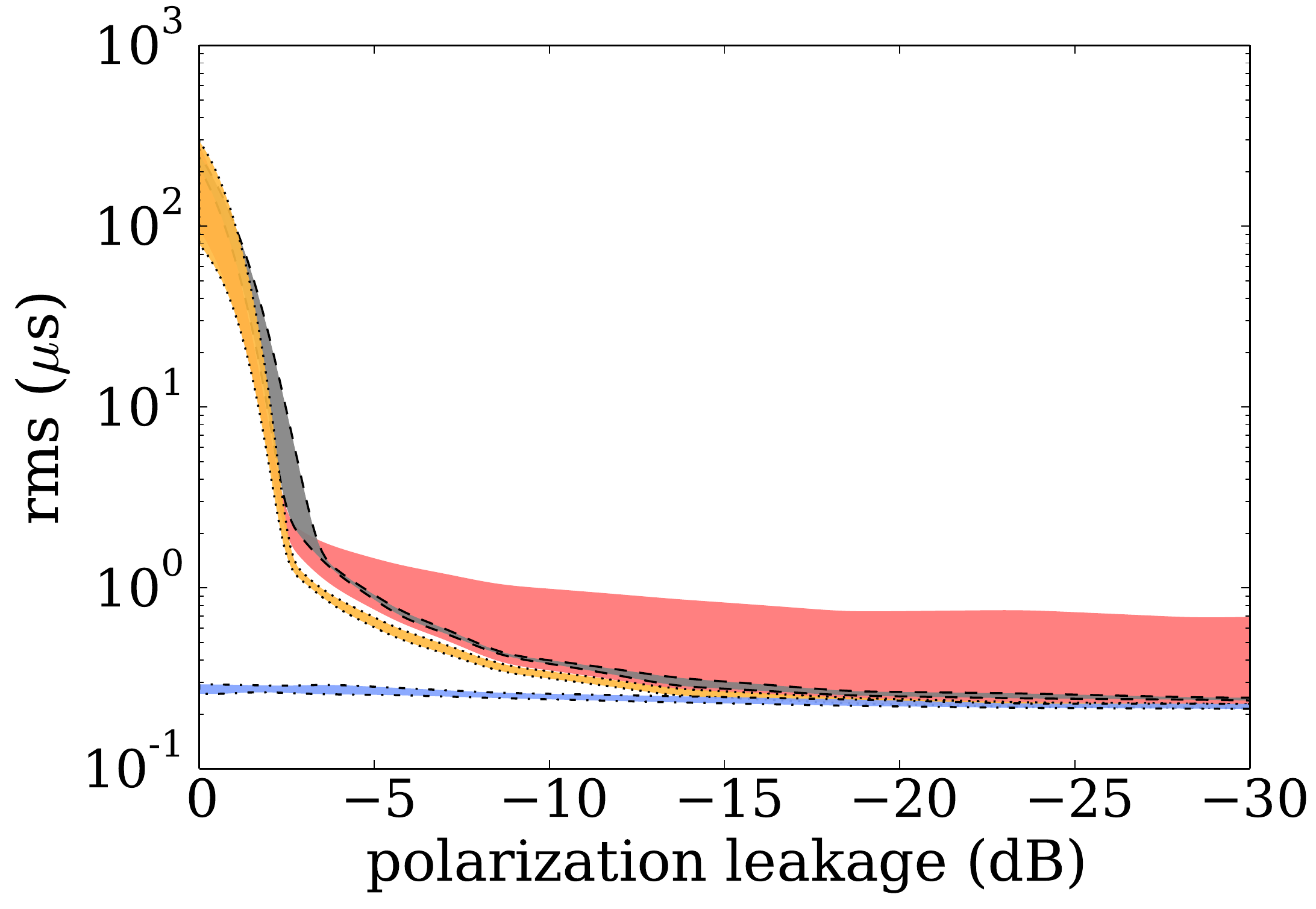} \hfill
\includegraphics[width=.2\linewidth]{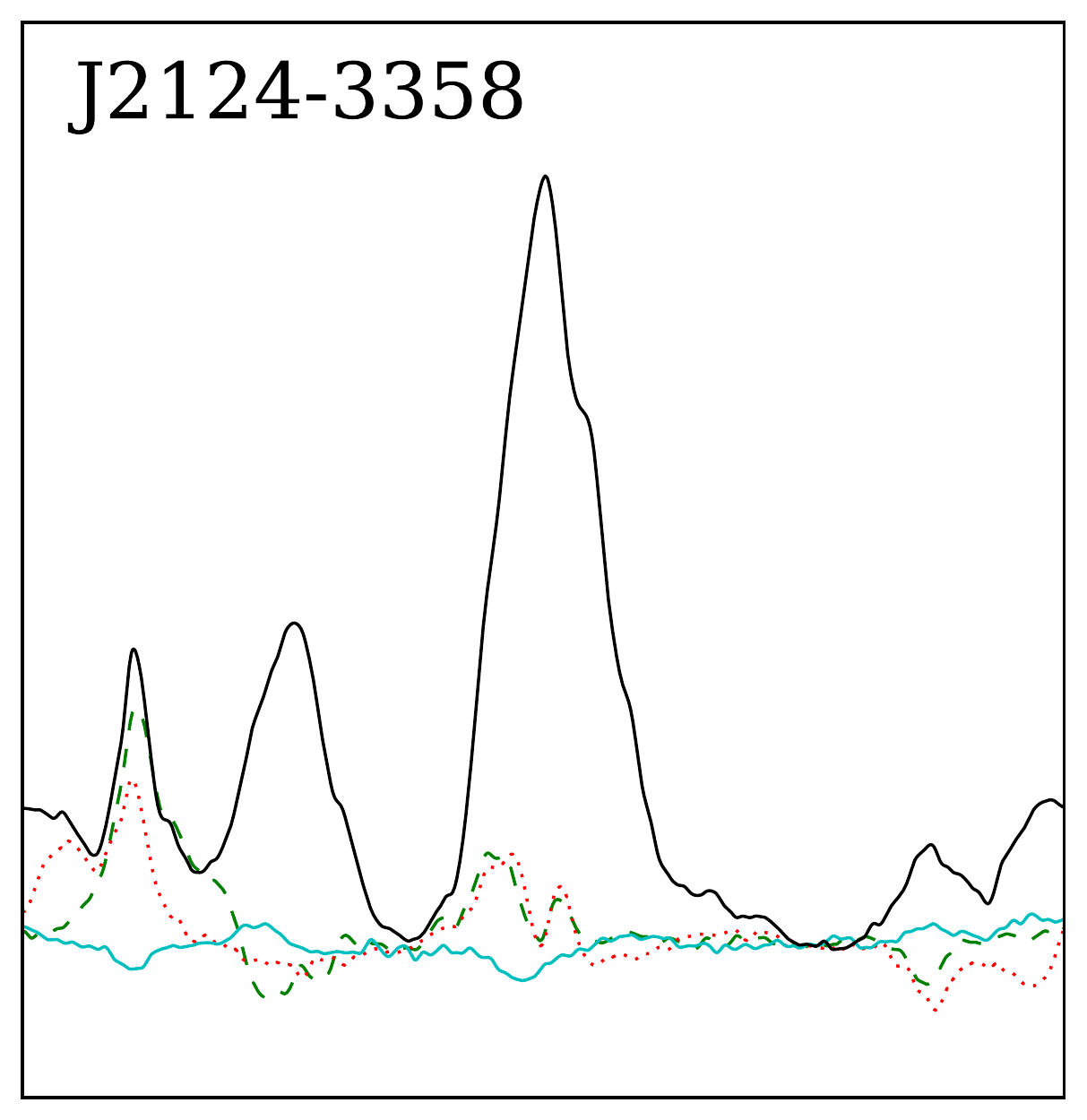} \hfill
\includegraphics[width=.28\linewidth]{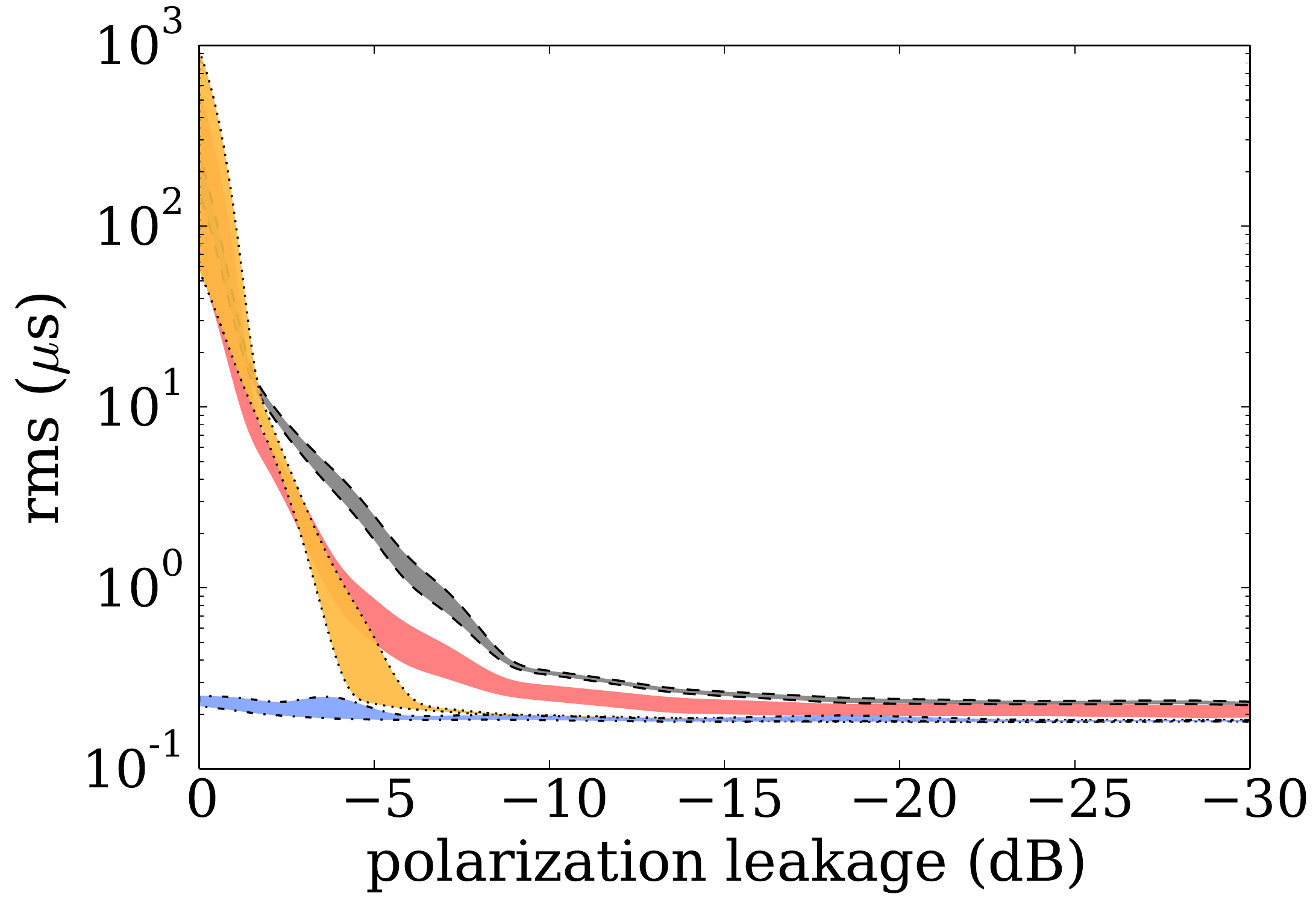} \hfill
\null

\noindent\null\hfill
\includegraphics[width=.2\linewidth]{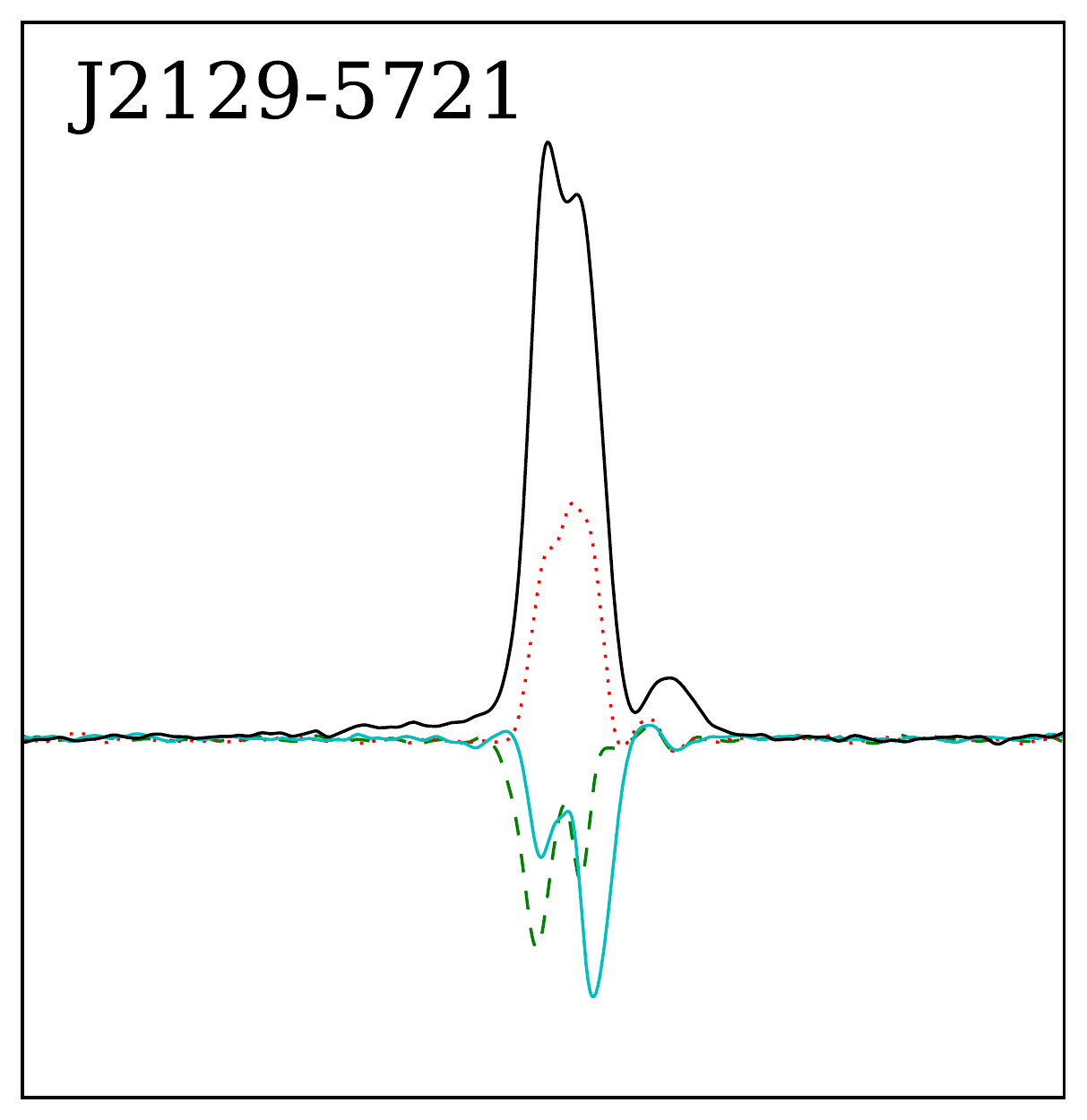} \hfill
\includegraphics[width=.28\linewidth]{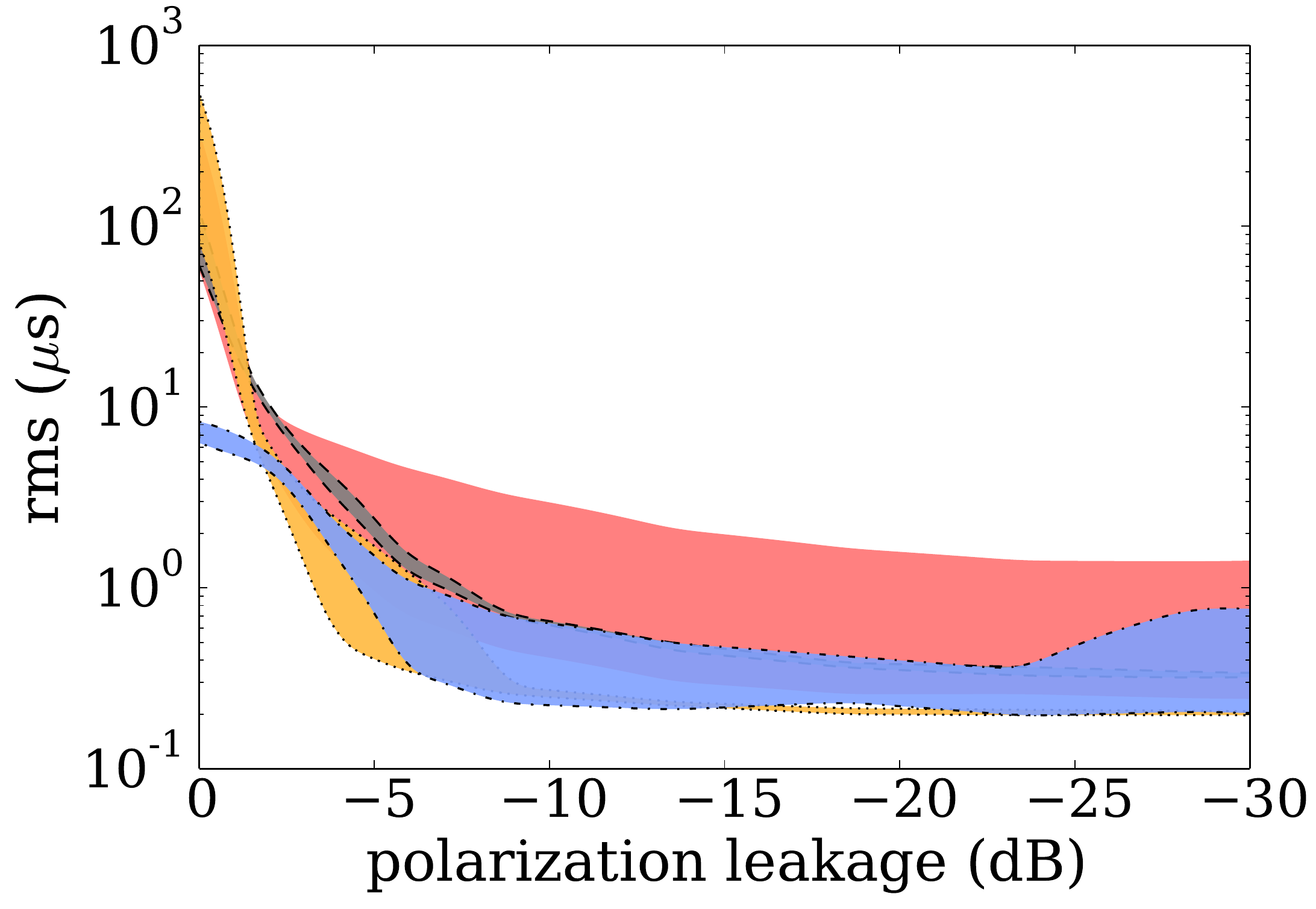} \hfill
\includegraphics[width=.2\linewidth]{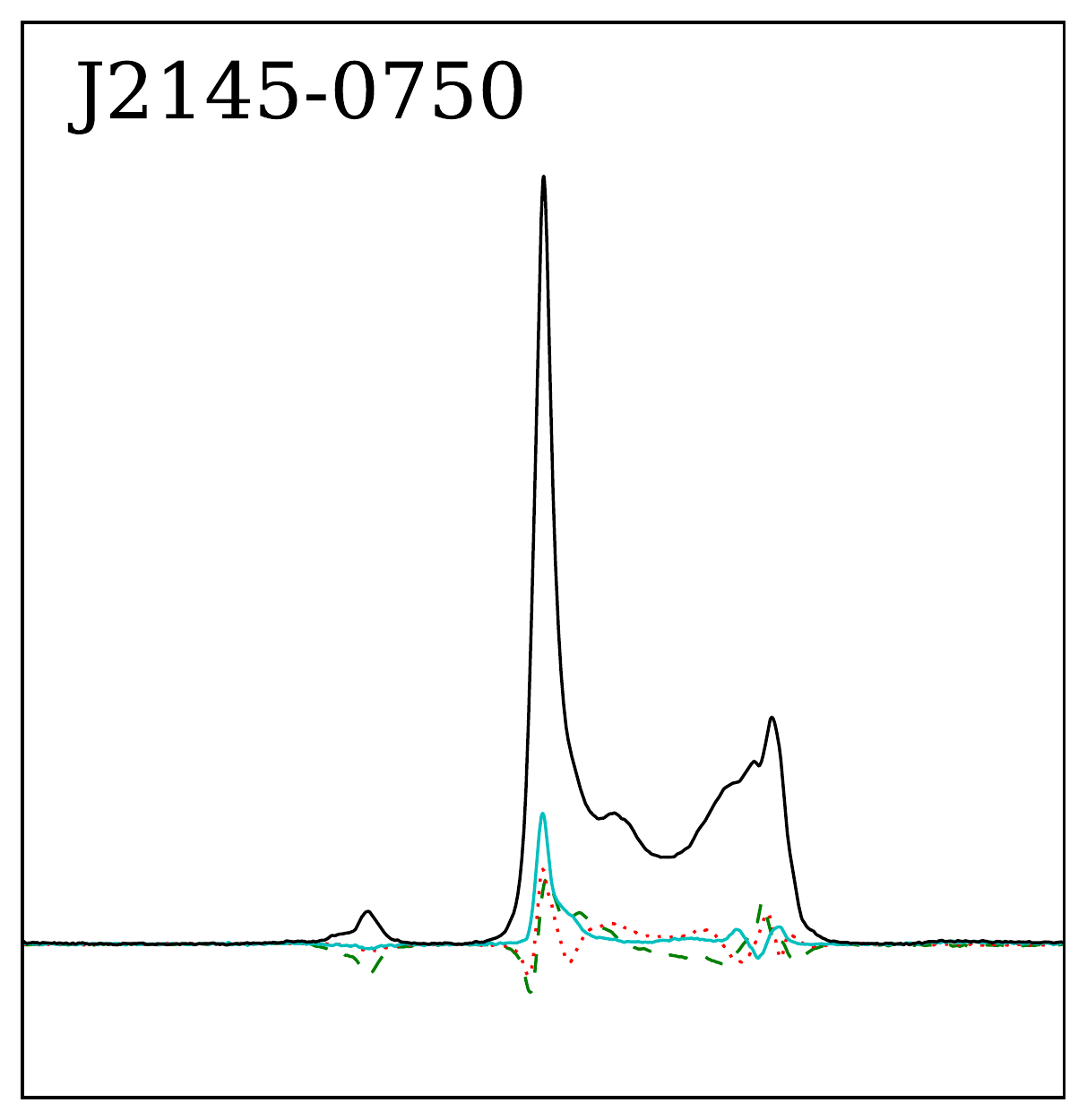} \hfill
\includegraphics[width=.28\linewidth]{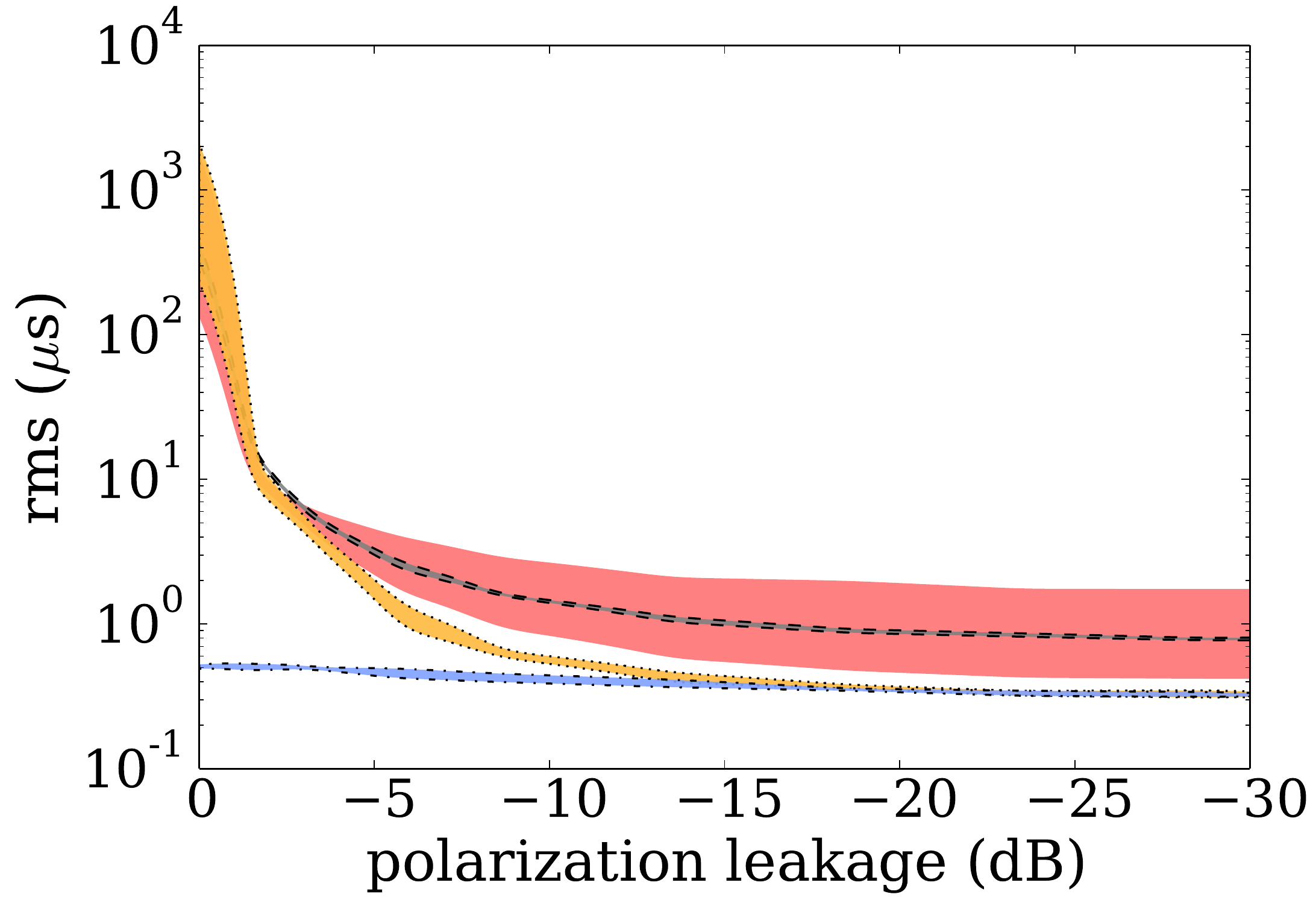} \hfill
\null

\caption{
    Continuation of Figure \ref{fig:psr_plots0}, see that figure for description.
    }
\label{fig:psr_plots1}
\end{figure*}

\label{lastpage}

\end{document}